\documentclass[aps,12pt,eqsecnum,tightenlines,showpacs,nofootinbib,endfloats,amsmath,amssymb]{revtex4}
\usepackage{graphicx}
\usepackage{bm}

\begin{document}

\title{Three dimensional stationary cyclic symmetric
Einstein--Maxwell solutions; energy, mass, momentum, and
algebraic tensors characteristics}

\topmargin
-2cm

\author{Alberto A. Garcia--Diaz\\{Departamento~de~F\'{\i}sica,
~Centro~de~Investigaci\'on~y~de~Estudios~Avanzados~del~IPN, Apdo.
Postal 14-740, 07360 M\'exico DF, M\'exico \\
} \email{aagarcia@fis.cinvestav.mx}}

\begin{abstract}
The main purpose of this contribution is to determine
physical and geometrical characterizations of whole classes
of stationary cyclic symmetric gravitational
fields coupled to Maxwell electromagnetic fields within the
$(2+1)$--dimensional gravity. The physical characterization is based on the
determination of the local and global energy--momentum--mass
quantities using the Brown--York approach. As far as
to the algebraic--geometrical characterization is concerned, the eigenvalue
problem for the electromagnetic field, energy--momentum and Cotton tensors
is solved and their types are established.

The families of Einstein--Maxwell solutions to be considered are: all uniform
electromagnetic solutions possessing electromagnetic fields with
vanishing covariant derivatives (
stationary uniform and spinning Clement classes), all fields having constant
electromagnetic field and energy--momentum tensors' invariants ( Kamata--Koikawa solutions), the whole classes of hybrid
electromagnetic Ayon--Cataldo--Garcia solutions, a new family of stationary electromagnetic
solutions,
the electrostatic and magnetostatic solutions with Peldan limit, the Clement spinning charged
metric, the Martinez--Teitelboim--Zanelli black hole solution, and
Dias--Lemos electromagnetic solution.

\vspace{0.5cm}
\end{abstract}

\date{July 22, 2011}

\maketitle \tableofcontents


\section{Introduction}

During the last two decades three--dimensional gravity has received
some attention, in particular, in topics such as: black hole
physics, search of exact solutions, quantization of fields coupled
to gravity, cosmology, topological aspects, and others. This
interest in part has been motivated by the discovery, in 1992, of
the $2+1$ stationary circularly symmetric black hole solution by
Ba\~nados, Teitelboim and Zanelli~\cite{BTZ}--the BTZ black hole--
see also~\cite{BTZ93,CangemiLM93,Carlip-cqg-95}, which possesses
certain features inherent to $3+1$ black holes. On the other hand,
it is believed that $2+1$ gravity may provide new insights towards a
better understanding of the physics of $3+1$ gravity. In the
framework of $2+1$ gravity the list of references
on exact solutions is extremely vast; one finds works on point masses,
perfect fluid solutions, dilaton and string fields,
electromagnetic fields coupled to gravity, and cosmologies, among others.

The general form of electromagnetic fields for stationary cyclic
symmetric 2+1 spacetimes is given by:$\ast
\bm{F}=a\bm{dt}+b\bm{d\phi}+c{g_{rr}}/{\sqrt{-g}}\bm{dr}$,
which splits into various sub--families: the electric $b\neq0$ fields,
the magnetic $a\neq0$ fields, the
uniform fields characterized by the
vanishing of the covariant derivatives $F_{\alpha\beta;\gamma}=0$,
the class of stationary fields with constant invariant $F_{\mu\nu}\,F^{\mu\nu}$,
and consequently, due to the structure of
the electromagnetic fields, with constant energy--momentum tensor invariants,
the gravitational stationary cyclic solutions for the hybrid
electromagnetic field ${\ast {\bm{F}}}=
c{g_{rr}}/{\sqrt{-g}}\bm{dr}$; the explicit derivation of
the solutions belonging to the quoted branches
can be found in~\cite{GarciaAnnals09}.\\
This report is organized as follows. Section \ref{Cotton} is
devoted to the determination of the
algebraic types of the Cotton tensor for a generic stationary
cyclic symmetric metric. In section~\ref{Energy}
a detailed derivation of the energy, momentum and mass
quantities for this stationary cyclic symmetric metric is accomplished;
these characteristic expressions will be evaluated in the next sections
for each of the spacetimes to be considered. In particular, since the static and
the stationary BTZ solutions are considered as
limits for vanishing electromagnetic
fields, it is convenient to have at hand their energy--momentum characteristics,
this is done in Section~\ref{BTZenergy}. Next sections,
following all a similar pattern, are devoted to the determination
of the energy and momentum
densities as well as of the corresponding integral energy,
momentum and mass. Emphasis is put on the asymptotic
behavior of these quantities at spatial infinity.

The application of the Hayward
black hole dynamics formulation and the
Ashtekar isolated horizon approach to the reported
here static and stationary black hole solutions is straightforward.

\section{Cotton tensor algebraic classification}\label{Cotton}

In $(n+1)$--dimensional space--times, for $n>3$, the invariant decomposition
of the Riemannian curvature tensor gives rise to the conformal Weyl tensor, the traceless
Ricci tensor, and the scalar curvature; for the classification of gravity one classifies the
Weyl tensor, and the classification of matter is achieved through the classification
of the traceless Ricci tensor. For details, in $(3+1)$--dimensional
space--times, see for instance, the book~\cite{Kramer03}.

In $(2+1)$--dimensional space--times there is no room for the conformal Weyl
tensor, the Riemannian curvature tensor decomposes into the Ricci tensor, and
the scalar curvature. The role of the conformal tensor in 2+1 gravity is
played by the Cotton tensor, see~\cite{Kramer03}, which is defined by means of the
Ricci tensor and the scalar curvature through their covariant derivatives
\begin{eqnarray}
C^{\alpha\beta}={\epsilon}^{\alpha\gamma\delta}
( {R^\beta}_{\gamma}-\frac{1}{4}R\,
{\delta^\beta}_{\gamma})_{;\delta},\,{C^{\alpha}}_{\alpha}=0.
\end{eqnarray}
For the standard stationary (static) cyclic symmetric metric
\begin{equation*} ds^2=-N^2dt^2+L^{-2}\,dr^2
+K^2[d\phi+Wdt]^2,
\end{equation*}
the traceless Cotton tensor, in the form ${C^{\alpha}}_{\beta}$,
occurs to be
\begin{eqnarray}\label{cotteq1}
({C^{\alpha}}_{\beta})=\left[ \begin {array}{ccc} {{C^1}_{1}}&0&
{{C^1}_{3}}\\\noalign{\medskip}0&{{C^2}_{2}}
&0\\\noalign{\medskip}{{C^3}_{1}}&0&{{C^3}_{3}}\end {array}
 \right];\,{{C^1}_{1}}+{{C^2}_{2}}+{{C^3}_{3}}=0.
\end{eqnarray}
Determining the eigenvalues and eigenvectors of the
Cotton matrix (\ref{cotteq1}) one establishes
the algebraic Cotton type of the space--time one is dealing with.
Accordingly, the characteristic equation for the eigenvalue
$\lambda$ amounts to
\begin{eqnarray}
({{C^2}_{2}}-{\lambda})\left(({{C^1}_{1}}-{\lambda})
({{C^3}_{3}}-{\lambda})-{{C^3}_{1}}\,{{C^1}_{3}}\right)=0,
\end{eqnarray}
or, in terms of its solutions, as
\begin{eqnarray}
({{C^2}_{2}}-{\lambda})&&\left({\lambda}+\frac{1}{2}{C^2}_{2}
+\frac{1}{2}\sqrt{({C^2}_{2}+2\,{C^1}_{1})^2+4\,{{C^3}_{1}}\,{{C^1}_{3}}}
\right)\times\nonumber\\
&&\left({\lambda}+\frac{1}{2}{C^2}_{2}
-\frac{1}{2}\sqrt{({C^2}_{2}+2\,{C^1}_{1})^2+4\,{{C^3}_{1}}\,{{C^1}_{3}}}
\right)=0
\end{eqnarray}
while the eigenvector
equations are
\begin{eqnarray}
&&({{C^1}_{1}}-{\lambda})V^{1}+{{C^1}_{3}}V^{3}=
0,\nonumber\\&&({{C^2}_{2}}-{\lambda})V^{2}=0,\,
\nonumber\\&&
{{C^3}_{1}}V^{1}+({{C^3}_{3}}-{\lambda})V^{3}=0.
\end{eqnarray}
For each eigenvalue the corresponding solution is:
\begin{eqnarray}
&&{\lambda}_{1}={C^2}_{2},\,\,{\bf V}1=(0,V^{2},0),\nonumber\\&&
{\lambda}_{2}=-\frac{1}{2}{C^2}_{2}
+\frac{1}{2}\sqrt{({C^2}_{2}+2\,{C^1}_{1})^2+4\,{{C^3}_{1}}\,{{C^1}_{3}}},\,\,{\bf V}2
=(V^{1}=-\frac{{C^1}_{3}}{{{C^1}_{1}}-{\lambda}_{2}}V^{3},0,V^{3}),\nonumber\\&&
{\lambda}_{3}=-\frac{1}{2}{C^2}_{2}
-\frac{1}{2}\sqrt{({C^2}_{2}+2\,{C^1}_{1})^2+4\,{{C^3}_{1}}\,{{C^1}_{3}}},\,\,{\bf V}3
=(V^{1}=-\frac{{C^1}_{3}}{{{C^1}_{1}}-{\lambda}_{3}}V^{3},0,V^{3}).\nonumber\\&&
\end{eqnarray}
Thus, the eigenvector ${\bf V}1$--a real one--is oriented in the $\rho$--direction,
the remaining two
vectors ${\bf V}2$ and ${\bf V}3$ might be real vectors lying on the surface
spanned by the $t$ and $\phi$ coordinate
directions or complex eigenvectors
depending, correspondingly, upon whether the value of the radical
$({C^2}_{2}+2\,{C^1}_{1})^2+4\,{{C^3}_{1}}\,{{C^1}_{3}}$ is positive or negative.

The nomenclature to be used for eigenvectors and algebraic types of
tensors is borrowed from Plebanski's monograph~\cite{Plebanski74},
Chapter VI: time-like, space--like, null, and complex vectors are
denoted respectively by ${\bf T}$, ${\bf S}$, ${\bf N}$, and ${\bf
Z}$. For algebraic types are used the symbols:
$\{\lambda_{1}T,\lambda_{2} S_2,\lambda_{3}S_3\}\equiv \{T, S, S\}$,
meaning that the first real eigenvalue $\lambda_{1}$ gives raise to
a time--like eigenvector ${\bf T}$, the second real eigenvalue
$\lambda_{2}$ is associated with a space--like eigenvector ${\bf
S}_2$, finally the third real eigenvalue $\lambda_{3}$ is related to
a space--like eigenvector ${\bf S}_3$; for the sake of simplicity I
use the typing $\{T, S, S\}$. It is clear that $\{N, N, S\}$  stands
for the algebraic type allowing for two different real eigenvalues
giving rise to two null eigenvectors while the third real root is
associated with a space--like eigenvector. When there are a single
and a double real eigenvalues giving rise correspondingly to a
time--like and space --like eigenvectors, the algebraic type is
denoted by $\{T, 2\,S\}$, consequently, for a triple real
eigenvalue, if that were the case, the types could be $\{3T\}$,
$\{3N\}$, or $\{3S\}$. For a complex eigenvalue $\lambda_{Z}$ , in
general, the related eigenvectors occur to be complex and are
denoted by ${\bf Z}$ and ${\bf \bar Z}$ its complex conjugated, the
possible types are $\{T,Z,\bar Z\}$, $\{N,Z,\bar Z\}$, or
$\{S,Z,\bar Z\}$.

In general, the spaces described by the
stationary (static) cyclic symmetric metric above belong to the
Cotton type $I$; if the three eigenvectors
are real the type is $I_{R}$, otherwise the type is $I_{Z}$
with eigenvectors ${\bf S}$, ${\bf N}$, ${\bf T}$,  ${\bf Z}$,
and ${\bf \bar{Z}}$. Following the notation above--proposed,
the algebraic types for the Cotton tensor could
be: $\{S, S, S\}, \{S, N, N\},\{S,Z,\bar Z\}$, and so on.

An alternative treatment of the Cotton tensor and conformal
symmetries for $(2+1)$-dimensional spaces is given
in~\cite{Hall99}, and also in~\cite{Heinicke04}, where also is
developed the analysis on Cotton tensors in n+1--dimensions.

\section{Energy, mass, and momentum for $2+1$ stationary
cyclic symmetric metric}\label{Energy}

In this section is established the general form of the energy
and the momentum functions for metrics with non-flat
but anti--de Sitter asymptotic following the Brown--York
approach \cite{BrownY-prd93}, see also \cite{BrownCM-prd94}.
The $(2+1)$--dimensional stationary cyclic
symmetric metric to be used is given by
\begin{equation} ds^2=-N^2dt^2+L^{-2}\,dr^2
+K^2[d\phi+Wdt]^2,
\end{equation}
where the structural functions $N$, $L$, $K$, and $W$ depend on the
variable $r$. The timelike vector $u_{\mu}$ normal to the hypersurface
$\Sigma:\,t_{\Sigma}=\rm const.$ and the spacelike vector $n_{\mu}$
normal to the surface
${\stackrel{2}{B}}:\,r_{{\stackrel{2}{B}}}=R=\rm const.$ are given
by
\begin{eqnarray}
u_{\mu}&=&-N\delta^{t}_{\mu},\,u^{\mu}
=\frac{1}{N}\delta^{\mu}_{t}
-\frac{W}{N}\delta^{\mu}_{\phi},\nonumber\\
n_{\mu}&=&\frac{1}{L}\delta^{r}_{\mu},\,
n^{\mu}=L\delta_{r}^{\mu}.
\end{eqnarray}
Therefore the projection metrics are:
\begin{eqnarray}
ds^2|_{\Sigma:t=\rm const.}&=&L^{-2}\,dr^2
+K^2\,d\phi^2=h_{ij}dx^i\,dx^j,
\nonumber\\
ds^2|_{\stackrel{2}{B}:r=R=\rm const.}&=&-N^2\,dt^2
+K^2(d\phi+Wdt)^2=\gamma_{AB}dx^A\,dx^B,
\nonumber\\
ds^2|_{B:t=\rm const.,r=R=\rm const.}&=&
K^2\,d\phi^2=\sigma_{\imath\jmath}dx^\imath\,dx^\jmath.
\end{eqnarray}
The components of the projection tensor $h$ are explicitly given by
\begin{eqnarray}
&{}&h_{\mu\nu}=g_{\mu\nu}+u_{\mu}u_{\nu},\nonumber\\
&{}&h_{\mu\nu}=K^2W^2\delta^t_{\mu}\delta^t_{\nu}+WK^2
(\delta^t_{\mu}\delta^\phi_{\nu}+\delta^\phi_{\mu}\delta^t_{\nu})
+L^{-2}\delta^r_{\mu}\delta^r_{\nu}
+K^2\delta^\phi_{\mu}\delta^\phi_{\nu},\nonumber\\
&{}&h_{ij}=
L^{-2}\delta^r_{i}\delta^r_{j}+K^2\delta^\phi_{i}\delta^\phi_{j},
\,\det(h_{ij})=K^2/L^2,\nonumber\\
&{}&h^{\mu\nu}=L^2\delta^{\mu}_{r}\delta^{\nu}_{r}+
\delta^{\mu}_{\phi}\delta^{\nu}_{\phi}/K^2,
\, {h^{\mu}}_{\nu}=W\delta^{\mu}_{t}\delta^{\phi}_{\nu}
+\delta^{\mu}_{r}\delta^{r}_{\nu}
+\delta^{\mu}_{\phi}\delta^{\phi}_{\nu},\nonumber\\
&{}&{h_{\mu}}^\nu=W\delta^t_{\mu}\delta^{\nu}_{\phi}
+\delta^r_{\mu}\delta^{\nu}_{r}
+\delta^\phi_{\mu}\delta^{\nu}_{\phi},{h_{i}}^j=\delta^j_{i}.
\end{eqnarray}
The components of the projection tensor $\gamma$ amount to
\begin{eqnarray}
&{}&\gamma_{\mu\nu}=g_{\mu\nu}-n_{\mu}n_{\nu},\nonumber\\
&{}&\gamma_{\mu\nu}=-(N^2-K^2W^2)\delta^t_{\mu}\delta^t_{\nu}+WK^2
(\delta^t_{\mu}\delta^\phi_{\nu}+\delta^\phi_{\mu}\delta^t_{\nu})
+K^{2}\delta^\phi_{\mu}\delta^\phi_{\nu},\nonumber\\
&{}&\gamma^{\mu\nu}=-\frac{1}{N^2}\delta^{\mu}_{t}\delta^{\nu}_{t}+\frac{W}{N^2}
(\delta^{\mu}_{t}\delta^{\nu}_{\phi}+\delta^{\mu}_{\phi}\delta^{\nu}_{t})
+\frac{N^2-K^2W^2}{K^2N^2}\delta^{\mu}_{\phi}\delta^{\nu}_{\phi},
\nonumber\\
&{}&{\gamma_{\mu}}^\nu=\delta^t_{\mu}\delta^{\nu}_{t}+
\delta^\phi_{\mu}\delta^{\nu}_{\phi},\det(\gamma_{AB})=-K^2\,N^2.
\end{eqnarray}
Notice that the indices $\mu, \nu$ associated to the
three-dimensional spacetime can be replaced by indices $A,B$
running $0\sim t,3\sim \phi$. The components of the projection
tensor  $\sigma$ amount to
\begin{eqnarray}
&{}&\sigma_{\mu\nu}=g_{\mu\nu}+u_{\mu}u_{\nu}-n_{\mu}n_{\nu},\nonumber\\
&{}&\sigma_{\mu\nu}={K^2W^2}\delta^t_{\mu}\delta^t_{\nu}
+{W}{K^2}(\delta^t_{\mu}\delta^\phi_{\nu}+\delta^\phi_{\mu}\delta^t_{\nu})+
K^2\delta^\phi_{\mu}\delta^\phi_{\nu},\,
\sigma_{\imath\jmath}=K^2\delta^\phi_{\imath}\delta^\phi_{\jmath},
\nonumber\\
&{}&\sigma^{\mu\nu}=\delta_\phi^{\mu}\delta_\phi^{\nu}/K^2,\,
\sigma^{\imath\jmath}=\delta_\phi^{\imath}\delta_\phi^{\jmath}/K^2,\,
\det(\sigma_{\imath\jmath})=K^2,
\end{eqnarray}
where $\imath,\,\jmath$ run only $3\sim\phi$.

To evaluate extrinsic curvatures one needs the expressions of the
symmetric Christoffel symbols, which amount to
\begin{eqnarray}
&{}&{\Gamma^t}_{tr}=\frac{1}{2N^2}\left(
2NN_{,r}-K^2WW_{,\,r}\right),\,{\Gamma^t}_{r\phi}=-\frac{1}{2N^2}K^2W_{,\,r},
\nonumber\\
&{}&{\Gamma^r}_{tt}={L^2}\left(
NN_{,r}-KW^2K_{,\,r}-K^2WW_{,\,r}\right),\nonumber\\
&{}&{\Gamma^r}_{t\phi}=-\frac{1}{2}L^2K \left(
2WK_{,r}+KW_{,\,r}\right),
\nonumber\\
&{}&{\Gamma^r}_{rr}=-\frac{1}{L}L_{,r},
\,{\Gamma^r}_{\phi\phi}=-{L^2}KK_{,r},\nonumber\\
&{}&{\Gamma^\phi}_{tr}=\frac{1}{2KN^2}\left(
-2WKNN_{,r}+W^2K^3W_{,\,r}+2N^2WK_{,\,r}+N^2KW_{,\,r}\right),\nonumber\\
&{}&{\Gamma^\phi}_{r\phi}=\frac{1}{2KN^2}\left(
WK^3W_{,r}+2N^2K_{,\,r}\right),
\end{eqnarray}
while all other components vanish.

The extrinsic curvature to the hypersurface $\Sigma:t=\rm const.$ is
given by the spatial tensor $K_{\mu\nu}$, namely
\begin{eqnarray}
&{}&K_{\mu\nu}=-h_{\mu}^{\alpha}\nabla_{\alpha}u_{\nu}
=-h_{\mu\beta}g^{\beta\alpha}u_{\nu;\alpha}\nonumber\\
&{}&K_{\mu\nu}=\frac{1}{2N}{K^2WW_{,\,r}}
(\delta^t_{\mu}\delta^r_{\nu}+\delta^r_{\mu}\delta^t_{\nu})
+\frac{1}{2N}{K^2W_{,\,r}}(\delta^\phi_{\mu}\delta^r_{\nu}
+\delta^r_{\mu}\delta^\phi_{\nu}),\nonumber\\
&{}&K^{\mu\nu}=\frac{1}{2N}L^2W_{,\,r}
(\delta_\phi^{\mu}\delta_r^{\nu}+\delta_r^{\mu}\delta_\phi^{\nu}),\nonumber\\
&{}&{K_{\mu}}^{\nu}=\frac{1}{2N}L^2K^2WW_{,\,r}
\delta^{t}_{\mu}\delta^{\nu}_{r}
+\frac{1}{2N}L^2K^2W_{,\,r}\delta^{\phi}_{\mu}\delta^{\nu}_{r}
+\frac{1}{2N}W_{,\,r}\delta^{r}_{\mu}\delta^{\nu}_{\phi},
\end{eqnarray}
thus the trace of $K_{\mu\nu}$ is zero, $K^{\mu}_{\mu}=0$.

The momentum tensor
$P^{\mu\nu}=\frac{1}{2\,\kappa}\sqrt{\det(h_{ij})}\left[
K^{\alpha}_{\alpha}\,h^{\mu\nu}-K^{\mu\nu}\right]$ for the
hypersurface $\Sigma$ becomes
\begin{eqnarray}\label{tensorP}
&{}&P^{\mu\nu}=-\frac{1}{4\kappa\,N}LKW_{,\,r}
(\delta^\mu_{r}\delta^\mu_{\phi}+\delta^\mu_{\phi}\delta^\mu_{r}),
\end{eqnarray}
while the surface momentum density vector
$j_\mu=-2\sigma_{\mu\nu}P^{\nu\alpha}n_{\alpha}/\sqrt{\det{h_{ij}}}$
amounts to
\begin{eqnarray}\label{vectorj}
&{}&j_{\mu}=\frac{1}{2\kappa}\frac{L}{N}K^2WW_{,\,r}\delta^{t}_{\mu}
+\frac{1}{2\kappa}\frac{L}{N}K^2W_{,\,r}\delta^{\phi}_{\mu}.
\end{eqnarray}
Consequently the surface momentum density $j_{a}$ reduces in the
studied case to
\begin{eqnarray}\label{compjphi}
j_{\phi}=-2\frac{1}{K}\sigma_{\phi\phi}P^{\phi\,r}
=\frac{1}{2\kappa}\frac{L}{N}\,K^2\,W_{, \,r}.\nonumber\\
\end{eqnarray}
modulo the additive constant related to the reference spacetime.

The energy density is evaluated by using the tensor
$$k_{\mu\nu}=-\sigma_{\mu}^{\alpha}h_{\alpha}^{\beta}n_{\lambda;
\beta}h^{\lambda}_{\nu}=h_{\mu\alpha}\Theta^{\alpha\beta}h_{\beta\nu},\,
\Theta_{\mu\nu}=-\gamma^{\beta}_{\mu}n_{\nu;\,\beta},$$ which
amounts to
\begin{eqnarray}
k_{\mu\nu}=- L
KWK_{,\,r}[W\delta^{t}_{\mu}\delta^{t}_{\nu}+2\delta^{\phi}_{(\mu}\delta^{t}_{\nu)}]
- L K K_{,\,r}\delta^{\phi}_{\mu}\delta^{\phi}_{\nu},
\end{eqnarray}
while
\begin{eqnarray}
{k^{\mu}}_{\nu}=-\frac{1}{K}L
WK_{,\,r}\delta^{\mu}_{\phi}\delta^{t}_{\nu} - \frac{1}{K}
L\,K_{,\,r}\delta^{\mu}_{\phi}\delta^{\phi}_{\nu}.
\end{eqnarray}
Rising with $\sigma_{\phi\phi}=K^2=\sigma^{\phi\phi}$ one of the
indexes of the component $k_{\phi\phi}=-L K K_{,\,r}$ of the
extrinsic curvature $k$ associated to the metric of $\stackrel{2}{B}$, one
arrives at
\begin{eqnarray}\label{tracek}
k:=k_{\imath}^{\imath}=\sigma^{\phi\phi}k_{\phi\phi}=-\frac{1}{K}LK_{,\,r}.
\end{eqnarray}
Therefore the energy density $\epsilon$ becomes
\begin{eqnarray}\label{energsurf}
\epsilon=\frac{1}{\kappa}k|^{cl}_{0}=-\frac{1}{\kappa\,K}LK_{,\,r}|_{R}-\epsilon_{0}.
\end{eqnarray}
As far as to the integral characteristics is concerned, the total
quasilocal energy
$E=\int_{B}dx\sqrt{\sigma}\epsilon=2\pi\,K\,\epsilon$ is given by
\begin{eqnarray}\label{globalE}
E=-2\frac{\pi}{\kappa}LK_{,\,r}|_{R}-2\pi\,K(R)\epsilon_{0},
\end{eqnarray}
while the mass related to the timelike Killing vector
$\xi^\mu=(\frac{\partial}{\partial t})^\mu=\delta^{\mu}_{t}$,
$M(\frac{\partial}{\partial\,t})
=-\int_{B}dx\sqrt{\sigma}(\epsilon\,u_{\mu}+j_{\mu})\xi^{\mu}$
amounts to
\begin{eqnarray}\label{globalM1}
M({\partial}/{\partial
t})=-2\frac{\pi}{\kappa}NLK_{,\,r}|_{R}
-\frac{\pi}{\kappa}\frac{L}{N}K^3WW_{,\,r}|_{R}-2\pi\,N
K|_{R}\epsilon_{0}.
\end{eqnarray}
Finally the total momentum
$J(\frac{\partial}{\partial\,\phi})=\int_{B}dx\sqrt{\sigma}j_{\mu}\zeta^{\mu}$
associated to the Killing vector
 $\zeta^{\mu}=(\frac{\partial}{\partial\,\phi})^\mu=\delta^{\mu}_{\phi}
$, is given by
\begin{eqnarray}\label{globalJ}
J({\partial}/{\partial \phi})=\int^{2\pi}_{0}d\phi\,j_{\phi}\,K=
\frac{\pi}{\kappa}\frac{L}{N}K^3\,W_{,\,r}|_{R},\,j_{\phi}
=\frac{1}{2\pi}\frac{J(R)}{K(R)}.
\end{eqnarray}
Incidentally, other representations of the mass and momentum density
are:
\begin{eqnarray}\label{globalM}
M({\partial}/{\partial t})=N(R)E(R)-W(R)J(R).
\end{eqnarray}
The extrinsic curvature
$\Theta_{\mu\nu}=-\gamma^{\alpha}_{\mu}\nabla_{\alpha}n_{\nu}
=-n_{\nu;\alpha}{\gamma^{\alpha}}_{\mu}$
of the surface boundary $\stackrel{2}{B}$ reduces to
\begin{eqnarray}\label{tensorTH}
\Theta_{\mu\nu}&=&-L(KW^2K_{,\,r}+K^2WW_{,\,r}-
NN_{,\,r})\delta^{t}_{\mu}\delta^{t}_{\nu}\nonumber\\
&{}&-L
K(2WK_{,\,r}+KW_{,\,r})\delta^{t}_{(\mu}\delta^{\phi}_{\nu)}
-LKK_{,\,r}\delta^{\phi}_{\mu}\delta^{\phi}_{\nu},
\end{eqnarray}
with trace $\Theta$ equals to
\begin{eqnarray}\label{traceTH1}
\Theta&=&-\frac{L}{N K}(KN_{,\,r}+NK_{,\,r}),
\end{eqnarray}
which, used in the definition of the boundary momentum
$$\pi^{\mu\nu}
 =-\frac{1}{2\kappa}\sqrt{-\det{\gamma_{AB}}}
 (\Theta\,\gamma^{\mu\nu}-\Theta^{\mu\nu}),$$
taking into account that $\det{\gamma_{AB}}=-K^2\,N^2$, gives
\begin{eqnarray}\label{traceTH}
\pi^{\mu\nu}&=&-\frac{L}{2\kappa\,N}K_{,\,r}\delta^{\mu}_{t}\delta^{\nu}_{t}
+\frac{L}{2\kappa\,N}(2WK_{,\,r}
+KW_{,\,r})\delta^{(\mu}_{t}\delta^{\nu)}_{\phi}\nonumber\\
&{}&+\frac{L}{2\kappa\,N\,K}(NN_{,\,r}-W^2KK_{,\,r}
-K^2WW_{,\,r})\delta^{\mu}_{\phi}\delta^{\nu}_{\phi}.
\end{eqnarray}
This tensor is used in the construction of the stress tensor
\begin{eqnarray}\label{tensorstress}
s^{\alpha\beta}&=&\frac{2}{\sqrt{\sigma}\,N}\sigma^{\alpha}_{\mu}
\,\pi^{\mu\nu}\sigma^{\beta}_{\nu}-{s_{0}}^{\alpha\beta}.
\end{eqnarray}

\section{ Ba\~nados--Teitelboim--Zanelli black hole solution}\label{BTZenergy}

To get an insight of how efficiently the definitions of energy
and momentum densities work for spacetimes with non--flat
asymptotic at spatial infinity let us consider the asymptotically
anti--de Sitter $(2+1)$--dimensional stationary black hole
solution--the BTZ--metric--
which is given by
\begin{eqnarray}\label{BTZmetric}
&&ds^2=-N(\rho)^2dt^2+\frac{1}{L(\rho)^{2}}\,d\rho^2
+\rho^2[d\phi+W(\rho)dt]^2,\nonumber\\
&&N^2(\rho)=L^{2}(\rho)=-M+\frac{\rho^2}{l^2}
+\frac{J^2}{4\,\rho^2},\,K(\rho)=\rho,
\,W(\rho)=-\frac{J}{2\,\rho^2}.
\end{eqnarray}
For the choice
\begin{eqnarray}
u_{\mu}&=&-N\delta^{t}_{\mu},\,u^{\mu}=\frac{1}{N}\delta^{\mu}_{t}
-\frac{W}{N}\delta^{\mu}_{\phi},\,
n_{\mu}=\frac{1}{L}\delta^{\rho}_{\mu},\,
n^{\mu}=L\delta_{\rho}^{\mu},
\end{eqnarray}
one has
\begin{eqnarray}
ds^2|_{\Sigma,t=\rm const.}&=&L^{-2}\,d\rho^2
+\rho^2\,d\phi^2=h_{ij}dx^i\,dx^j,
\nonumber\\
ds^2|_{\stackrel{2}{B},\rho=R=\rm const.}&=&-N^2\,dt^2
+R^2(d\phi+Wdt)^2=\gamma_{AB}dx^A\,dx^B,
\nonumber\\
ds^2|_{B,t=\rm const.,\rho=R=\rm const.}&=&
R^2\,d\phi^2=\sigma_{\imath\jmath}dx^\imath\,dx^\jmath.
\end{eqnarray}
In what follows $\kappa$ is choosing as $\kappa=\pi$. The surface
tensors amount to
\begin{eqnarray}
&{}&P^{\mu\nu}=-\frac{1}{4\pi}\frac{J}{\rho^2}
(\delta^\mu_{\rho}\delta^\mu_{\phi}+\delta^\mu_{\phi}\delta^\mu_{\rho}),
\end{eqnarray}
\begin{eqnarray}
&{}&j_{\mu}=-\frac{1}{4\pi}\frac{J^2}{\rho^3}\delta^{t}_{\mu}
+\frac{1}{2\pi}\frac{J}{\rho}\delta^{\phi}_{\mu},
\end{eqnarray}
\begin{eqnarray}
k=-\frac{1}{\rho}\sqrt{-M+\frac{\rho^2}{l^2}+\frac{J^2}{4\,\rho^2}}.
\end{eqnarray}
\subsection{ Energy, mass and momentum for the BTZ black hole}\label{BTZenergyhole}
The corresponding surface energy and momentum densities are given by
\begin{eqnarray}\label{BTZj}
&&\epsilon(R,\epsilon_{0})=-\frac{1}{\pi\,R}\sqrt{-M+\frac{R^2}{l^2}
+\frac{J^2}{4\,R^2}}-\epsilon_{0},
\nonumber\\
&&j_{\phi}(R)=\frac{1}{2\pi}\frac{J}{R}.
\end{eqnarray}
Consequently the total momentum, energy, and mass are
\begin{eqnarray}\label{BTZenergyJ}
&&
J({\partial}/{\partial \phi})=J,
\nonumber\\&&
E(R,\epsilon_{0})=-2\sqrt{-M+\frac{R^2}{l^2}+\frac{J^2}{4\,R^2}}-2\,\pi\,\epsilon_{0},
\nonumber\\&&
M({\partial}/{\partial
t})=N(R)\,E(R,\epsilon_{0})+\frac{J^2}{2\,R^2}=2M-2\frac{R^2}{l^2}-2\,\pi
\,\epsilon_{0}\sqrt{-M+\frac{R^2}{l^2}+\frac{J^2}{4\,R^2}}.
\end{eqnarray}
These  expressions for surface densities and global quantities are
in full agreement with the corresponding ones reported in
Ref.~\cite{BrownCM-prd94}, section IV.

Notice that the energy and mass independent of $\epsilon_{0}$ behave
at infinity $R$, which will be denoted from now on by
the same coordinate Greek letter $\rho$ accompanied
by $\rightarrow\infty$ and the
approximation $\approx$ sign , as
\begin{eqnarray}\label{enerBTZapproxEpsBTZ}
\epsilon(\rho\rightarrow\infty,\epsilon_{0}=0)&\approx&
-\frac{1}{\pi\,l}+\frac{l\,M}{2\pi\,\rho^2}
,\nonumber\\
E(\rho\rightarrow\infty,\epsilon_{0}
=0)&\approx&-\frac{2\rho}{l}+\frac{l\,M}{\rho}
,\nonumber\\
M(\rho\rightarrow\infty,\epsilon_{0}
=0)&\approx&2M-2\frac{\rho^2}{l^2}.
\end{eqnarray}
Although the expression of $M(\rho,\epsilon_{0}=0)$ holds in the
whole spacetime and not only in the boundary at spatial infinity,
the approximation $\approx$ sign is used  instead of the $=$ equal to be consistent
with the point under consideration.

The reference energy density to be used in this work is
the one corresponding to the anti--de Sitter metric with parameter $M_{0}$,
$\epsilon_{0}(M_{0})=
-\frac{1}{\pi\rho}\sqrt{\frac{\rho^2}{l^2}
-M_{0}},\,\epsilon_{0\mid\infty}(M_{0})\approx
-\frac{1}{\pi\,l}+\frac{l\,M_{0}}{2\pi\,\rho^2},$
then the expansions of the physical characteristics at
spatial infinity, $\rho \rightarrow \infty$,
are given  as
\begin{eqnarray}\label{enerBTZapproxEps1}
\epsilon(\rho\rightarrow\infty, \epsilon_{0\mid\infty}(M_{0}))
&&\approx\frac{l}{2\pi \,\rho^2}(M-M_{0}),\nonumber\\
E(\rho\rightarrow\infty, \epsilon_{0\mid\infty}(M_{0}))
&&\approx\,l\,\frac{(M-M_{0})}{\rho},\nonumber\\
M(\rho\rightarrow\infty, \epsilon_{0\mid\infty}(M_{0}))&&\approx\,M-M_{0}.
\end{eqnarray}

Another reference energy density
$\epsilon_{0}$ of common use is the one corresponding to the proper anti--de Sitter
space with $M_{0}=-1$, namely
$\epsilon_{0}=-\frac{1}{\pi\rho}\sqrt{1
+\frac{\rho^2}{l^2}},\,\epsilon_{AdS\mid\infty}\approx
-\frac{1}{\pi\,l}-\frac{l}{2\pi\,\rho^2},$ then the expansions
of the functions (\ref{BTZenergy}) at
spatial infinity, $\rho \rightarrow \infty$,
are given  as
\begin{eqnarray}\label{enerBTZapproxEps2}
\epsilon(\rho\rightarrow\infty, \epsilon_{AdS\mid\infty})
&&\approx\frac{l}{2\pi \,\rho^2}(1+M),\nonumber\\
E(\rho\rightarrow\infty, \epsilon_{AdS\mid\infty})
&&\approx\,l\,\frac{(1+ M)}{\rho},\nonumber\\
M(\rho\rightarrow\infty, \epsilon_{AdS\mid\infty})&&\approx\,1+M.
\end{eqnarray}

These results, concerning approximations of energies and masses at
spatial infinity for the BTZ solution,
Eq.~(\ref{enerBTZapproxEpsBTZ}), Eq.~(\ref{enerBTZapproxEps1}),
and Eq.~(\ref{enerBTZapproxEps2}), will be used for comparison with the
corresponding expressions related to other solutions to be treated
in what follows.\\

In the forthcoming sections, in some cases, the order
in the approximations could appear high compared with
the order needed but the adopted expansions will be done
to establish to what extend the
solutions are comparable or similar at spatial infinity.

\subsection{Mass, energy and momentum of the BTZ
solution counterpart}\label{BTZcounter}
As it has been pointed by Peldan~\cite{Peldan93},
the~sentence after Eq.(84), one should
be not able to integrate the magnetic branch of solutions
if the Schwarzschild gauge, i.e. $g_{\phi\phi}=\rho^2$,
were been adopted. For a $\rho$--gauge different of the
Schwarzschild one, as the reference metric it is
more adequate to consider the stationary or
static BTZ solution counterpart and its mass,
energy, and momentum characteristics at spatial infinity.
In the representation
\begin{eqnarray}\label{metricSBTZcount}
{g}&&=-\rho^2\left({dt}+\frac{J}{2\rho^2}{d
\phi}\right)^2+\frac{{d\rho}^2}{f(\rho)}+{f(\rho)}{d\phi}^2\nonumber\\
&&=-\rho^2\,\frac{f(\rho)}{\rho^2/l^2+M}{{dt}}^2+\frac{{d\rho}^2}{f(\rho)}
+(\rho^2/l^2+M)[{d\phi}-\frac{J_0}{2(\rho^2/l^2+M)}\,{dt}]^2,\nonumber\\
f(\rho)&&=\frac{\rho^2}{l^2}+M+\frac{J^2}{4\rho^2},
\end{eqnarray}
of the BTZ metric the coordinate $\phi$ loses its
interpretation of
circular angular coordinate; the same fact takes
place in the case of magnetic solutions
to be studied below. \\
It becomes apparent that this metric form is just another real cut
of the metric~(\ref{BTZmetric}) when subjecting it to the complex
transformations $t\rightarrow{i\,\phi}$ and $\phi\rightarrow{i\,t}$.\\
Moreover, by accomplishing in the above metric the transformation of
the radial coordinate
\begin{eqnarray}\label{TransBTZCtoBTZ}
t\rightarrow{t/l},
\,\rho\rightarrow{l\,\sqrt{\rho^2/l^2-M}},\,\phi\rightarrow{l\,\phi}
\end{eqnarray}
one gets the standard BTZ metric~(\ref{BTZmetric}).\\
As far as to the energy-momentum characteristics are concerned, for
the second form of the metric~(\ref{metricSBTZcount}), one has
\begin{subequations}\label{metric_rho_AdScount_energyx}
\begin{eqnarray}\label{metric_rho_AdScount_energyx1}
j_{\phi}(\rho)&&=\frac{J}{2\,\pi\,l^2\sqrt{\rho^2/\,l^2+M}},\,
j({\rho\rightarrow\infty})\approx \frac{J}{2\,\pi\,l\rho},\nonumber\\
J_{\phi}(\rho)&&=\frac{J}{l^2},\, J({\rho\rightarrow\infty})
\approx\frac{J}{l^2},
\end{eqnarray}
\begin{eqnarray}\label{metric_rho_AdScount_energyx2}
\epsilon(\rho,\epsilon_{0})&&=-\frac{\rho\,}{l^2\pi
}\frac{\sqrt{f(\rho,M,J)}}{\rho^2/\,l^2+M}-\epsilon_{0},\nonumber\\
\epsilon({\rho\rightarrow\infty},0)&&\approx-\frac{1}{\pi
l}+\frac{l}{2\pi}\frac{M}{\rho^2},\nonumber\\
\epsilon({\rho\rightarrow\infty},\epsilon_{0\mid\infty}(M_{0}))
&&\approx\frac{l}{2\pi}\frac{M-M_0}{\rho^2},
\end{eqnarray}
\begin{eqnarray}\label{metric_rho_AdScount_energyx3}
E(\rho,\epsilon_{0})&&=-\frac{2}{l^2}\,
\rho\frac{\sqrt{f(\rho,M,J)}}{\sqrt{\rho^2/\,l^2+M}}-2\pi
\,\epsilon_{0}{\sqrt{\rho^2/\,l^2+M}},\nonumber\\
E({\rho\rightarrow\infty},0)&&\approx-2\frac{\rho}{l^2},\nonumber\\
E({\rho\rightarrow\infty},\epsilon_{0\mid\infty}(M_{0}))
&&\approx\frac{1}{\rho}(M-M_0),
\end{eqnarray}
\begin{eqnarray}\label{metric_rho_AdScount_energyx4}
M(\rho,\epsilon_{0})&&=-\frac{2}{l^2}\rho^2
-2\pi\,\rho\,\epsilon_{0}\,\sqrt{f(\rho,M,J)},
\nonumber\\
M(\rho,0)&&=-2\frac{\rho^2}{l^2},\nonumber\\
M({\rho\rightarrow\infty},\epsilon_{0\mid\infty}(M_{0}))
&&\approx{M-M_0},
\end{eqnarray}
\begin{eqnarray}\label{metric_rho_AdScount_energyx5}
f(\rho,M,J)&&=\frac{\rho^2}{l^2}+M+\frac{J^2}{4\rho^2},
\nonumber\\
\epsilon_0(\rho,\,M_0)&&=
-\frac{\rho}{\pi\,l^2\,\sqrt{\rho^2/\,l^2+M_0}},\nonumber\\
\,\epsilon_{0\mid\infty}(M_{0})
&&\approx-\frac{1}{\pi\,l}+\frac{l\,M_0}{2\pi\,\rho^2}.
\end{eqnarray}
\end{subequations}
It is clear that the BTZ solution
counterpart, for $J=0$, gives rise to the
AdS metric in a slightly modified
representation--which one may call ``the AdS metric counterpart''.
The evaluated energy and mass can be
considered as the reference energy and mass at spatial infinity for
magnetic solutions. The point is that
for this class of magnet--static solutions or
stationary solutions generated from them via $SL(2,R)$
there is no room for a Schwarzschild radial
$\rho$ coordinate such that $g_{\phi\phi}=\rho^2$.

\subsection{Symmetries of the stationary and
static cyclic symmetric BTZ solutions}

Although it is known that the BTZ solution possesses two Killing
vectors--the timelike symmetry
along the time coordinate and the spacelike symmetry along
the orbits of the periodic angular variable--in my opinion, some comments on
this respect can be added to clarify
how the number of six  Killing vectors solutions
for the BTZ metric structure reduces to the quoted two.
In this framework, the six symmetries
of the anti--de Sitter
space with parameter $M_{0}$, $AdS(M_{0})$,
are derived; the $AdS(M_{0})$
allowing for $time + polar$ coordinates
possesses a timelike and one $2\pi$--periodic
circular symmetries. For the anti--de Sitter
space with parameter $M_{0}=-1$, denoted simply by $AdS$, there are
six symmetries: time, circular, and four boots symmetries.

The study of the symmetries of the stationary and static cyclic
symmetric BTZ families and AdS classes of solutions starts  with
the stationary metric for the standard BTZ solution
\begin{eqnarray}\label{BrownCAG}
{g}&=&-F(r)^2\,{d\,t}^2+\frac{{dr}^2}{F(r)^2}
+{r}^{2}\left[{d\phi}-\frac{J}{2\,r^2}{d\,t}\right]^2,
{F(r)}^{2}=\frac{r^2}{l^2}-M+\frac{J^2}{4\,r^2}.\nonumber\\
\end{eqnarray}

The covariant Killing vectors are derived in
Appendix~\ref{appendixSym}, the contravariant
vectors' components ${k_{i}}^{\mu}$, $\bm{\partial}_{k_{i}}=
{k_{i}}^{\mu}\frac{\partial}{{\partial{x^{\mu}}}}=C_{i}{V}_{i}^{\mu}$,
$i=1,...6$ are given below. The reason to include the integration
constants in the definitions of the Killing vectors ${k_{i}}^{\mu}$ is related
with the domain of definition of the spatial coordinates; in the case of the
existence of a periodic coordinate some Killing vectors
vanish, which can be easily achieved by setting certain structural
constant equal to zero. Explicitly these Killing vectors are:
\begin{subequations}
\begin{eqnarray}
C_1,\,\bm{\partial}_{k_{1}};\,{k_{1}}^{\mu}
={\it C_{1}}{{\exp}\left({\frac {\sqrt {Ml-J} }{{l}^{3/2}
}}\left( l\phi+t \right)\right )}
&&\left[\frac{1}{4}\, \frac{ Jl-2\,{r}^{2}}
{\sqrt {Ml-J}} \frac {1}{r\,F(r)}
\frac{1}{l^{3/2}},\frac{1}{2}\frac{F(r)}{l},
\right.\nonumber\\&&\left.-\frac{1}{4}\,
 \frac{ Jl+2\,{r}^{2}-2\,M{l}^{2}}{\sqrt{Ml-J}} \frac {1}{
{l}^{5/2}}\frac {1}{{r}\,F(r)}\right],\nonumber\\
\end{eqnarray}
\begin{eqnarray}
C_2,\,\bm{\partial}_{k_{2}};\,{k_{2}}^{\mu}
={\it C_{2}}{{\exp}\left({\frac {\sqrt {J+Ml} \left( l\phi-t \right)
}{{l}^{3/2}}}\right)}&&\left[\frac{1}{4}\,
 \frac{ l\,J+2\,{r}^{2} }{\sqrt{J+Ml }}
 \,\frac {1}{ l^{3/2}}{\frac {1}{{{r}\,F(r)}}} ,
\frac{1}{2}\,\frac{F(r)}{l}
,\right.\nonumber\\&&\left.
\frac{1}{4}\, \frac{\left( l\,J+2\,M{l}^{2}
-2\,{r}^{2} \right)}{\sqrt{J+Ml }}\,
\frac{1}{{l}^{5/2}}\frac{1}{{{r}\,F(r)}}\right],\nonumber\\
\end{eqnarray}
\begin{eqnarray}
C_3,\,\bm{\partial}_{k_{3}};\,{k_{3}}^{\mu}={\it C_{3}}{
{\exp}\left(-{\frac {\sqrt {J+Ml}  }{{l}^{3/2}}}( l\phi-t )\right)}
&&\left[-\frac{1}{4}\, \frac{ l\,J+2\,{r}^{2} }{\sqrt{J+Ml }}
\,\frac {1}{ l^{3/2}}{\frac {1}{{{r}\,F(r)}}} ,
\frac{1}{2}\,\frac{F(r)}{l},\right.\nonumber\\&&\left.
-\frac{1}{4}\, \frac{\left( l\,J+2\,M{l}^{2}-2\,{r}^{2} \right)}
{\sqrt{J+Ml }}\, \frac{1}{{l}^{5/2}}\frac{1}{{{r}\,F(r)}}\right],\nonumber\\
\end{eqnarray}
\begin{eqnarray}
C_4,\,\bm{\partial}_{k_{4}};\,{k_{4}}^{\mu}
={\it C_{4}}{{\exp}\left(-{\frac {\sqrt {M\,l-J} \left( l\phi+t \right) }{{l}^{3/
2}}}\right)} &&\left[-\frac{1}{4}\,\frac{ J\,l-2\,{r}^{2} }
{\sqrt {M\,l-J}} \frac {1}{l^{3/2}}{\frac {1
}{{r}\,F(r)}}
,\frac{1}{2}\,\frac{F(r)}{l},
\right.\nonumber\\&&\left.
\frac{1}{4}\,\frac{ J\,l+2\,{r}^{2}-2\,M{l}^{2} }
{\sqrt {M\,l-J}} \frac{1}{{l}^{5/2}}\frac {1}{{r}
\,F(r)}\right],\nonumber\\
\end{eqnarray}
\begin{eqnarray}
\bm{\partial}_{k_{5}};\,{k_{5}}^{\mu}={\it C_{5}}[0,0,1/2],
\end{eqnarray}
\begin{eqnarray}
\bm{\partial}_{k_{6}};\,{k_{6}}^{\mu}={\it C_{6}}[-2,0,0].
\end{eqnarray}
\end{subequations}
For completeness, the list of the independent commutators is given
\begin{subequations}\label{KillingBTZ}
\begin{eqnarray}
&&{\bf\partial}_{[k_{6}}{\bf\partial}_{k_{1}]}
=-2{\it C_{6}}\,{\frac {\sqrt {Ml-J}}{{l}^{3/2}}}{\bf\partial}_{k_{1}},
{\bf\partial}_{[k_{6}}{\bf\partial}_{k_{3}]}
=-2{\it C_{6}}\,
{\frac {\sqrt {Ml+J}}{{l}^{3/2}}}{\bf\partial}_{k_{3}},
\nonumber\\&&
{\bf\partial}_{[k_{6}}{\bf\partial}_{k_{4}]}
=2{\it C_{6}}\,{\frac {\sqrt {Ml-J}}{{l}^{3/2}}}{\bf\partial}_{k_{4}},
{\bf\partial}_{[k_{6}}{\bf\partial}_{k_{2}]}
=2{\it C_{6}}\,{\frac {\sqrt {Ml+J}}{{l}^{3/2}}}{\bf\partial}_{k_{2}},
\end{eqnarray}
\begin{eqnarray}
&&{\bf\partial}_{[k_{5}}{\bf\partial}_{k_{1}]}
={\it C_{5}}\frac{1}{2}\,{\frac {\sqrt {Ml-J}}{{l}^{1/2}}}{\bf\partial}_{k_{1}},
{\bf\partial}_{[k_{5}}{\bf\partial}_{k_{3}]}
=-{\it C_{5}}\frac{1}{2}\,{\frac {\sqrt {Ml+J}}{{l}^{1/2}}}{\bf\partial}_{k_{3}},
\nonumber\\&&
{\bf\partial}_{[k_{5}}{\bf\partial}_{k_{4}]}
=-{\it C_{5}}\frac{1}{2}\,{\frac {\sqrt {Ml-J}}{{l}^{1/2}}}{\bf\partial}_{k_{4}},
{\bf\partial}_{[k_{5}}{\bf\partial}_{k_{2}]}
={\it C_{5}}\frac{1}{2}\,\frac {\sqrt {Ml+J}}{ {l}^{1/2}}{\bf\partial}_{k_{2}},
\end{eqnarray}
\begin{eqnarray}
{\bf\partial}_{[k_{1}}{\bf\partial}_{k_{4}]}
=-\frac{1}{2}\,\frac{{\it C_{1}}{\it C_{4}}}{{\it C_{6}}}
\,\frac {1}{{l}^{5/2}\sqrt {Ml-J}}{\bf\partial}_{k_{6}}+
{2}\,\frac{{\it C_{1}}{\it C_{4}}}{{\it C_{5}}}
\,\frac {1}{{l}^{7/2}\sqrt {Ml-J}}{\bf\partial}_{k_{5}},
\end{eqnarray}
\begin{eqnarray}
{\bf\partial}_{[k_{3}}{\bf\partial}_{k_{2}]}
=-\frac{1}{2}\,\frac{{\it C_{3}}{\it C_{2}}}{{\it C_{6}}}
\,\frac {1}{{l}^{5/2}\sqrt {Ml+J}}{\bf\partial}_{k_{6}}-
{2}\,\frac{{\it C_{3}}{\it C_{2}}}{{\it C_{5}}}
\,\frac {1}{{l}^{7/2}\sqrt {Ml+J}}{\bf\partial}_{k_{5}},
\end{eqnarray}
\end{subequations}

All  anti--de Sitter metrics for coordinates
$\{t,\rho,\phi\}$--merely names--ranging
$-\infty\leq t\leq\infty$, $-\infty\leq \rho\leq\infty$,
$-\infty\leq \phi\leq\infty$ allows for six symmetries, i.e., six Killing vectors.
All these spaces in these coordinates are maximally symmetric spaces.\\

Another is the situation if
the spatial coordinates are constrained to range\\
$$ 0\leq \rho\leq\infty,\,0\leq \phi\leq\,2\pi,$$
in such case $\rho$ and $\phi$
are polar coordinates with $\phi$ being the angular
coordinate with  period  $2\pi$. Since
the expressions of four of the Killing
vector fields depending on $\phi$ the do not exhibit
the angular symmetry in $2\pi$, invariance under the
change $\phi\rightarrow\phi+2\pi$,
therefore there is no room for the corresponding
symmetries and the integration constants associated with those
vectors ought to be zero. Consequently
the metric with positive $M$ allowing for
polar angular coordinate, and only that, possesses
only two Killing vectors  $\partial_{t}$
and  $\partial_{\phi}$ (two symmetries: the time translation
and the $2\pi$--periodic angular rotation).
This spacetime is known
as the stationary BTZ black hole.
On this respect see also~\cite{Ayon04}.

A similar situation takes place in the case
of the static anti--de Sitter metric. By setting the
rotation parameter equal to zero, $J=0$, the
above--expressions~(\ref{KillingBTZ}) give
the Killing vectors for the static anti--de Sitter space--time.
Again, in the case of the coordinates
restricted to ranges $ 0\leq \rho\leq\infty,\,0\leq \phi\leq\,2\pi,$
the static anti--de Sitter metric allows
only for two Killing vectors: $\partial_{t},
\partial_{\phi}$, i.e., the time translation
and the $2\pi$--periodic angular rotation,
otherwise, when there are six constants, the space is maximally symmetric.

\subsection{\label{sec:symmBTZ}Symmetries
of the anti--de Sitter metric for negative $M$, $M=-\alpha^2$}\label{SymmBTZaneg}

If $M$ is negative, one equates it to $-\alpha^2$. Moreover,
it results better to use trigonometric sine and cosine
functions instead of complex exponential functions.
Thus, one can give the  Killing vector components as
\begin{subequations}
\begin{eqnarray}
\bm{\partial}_{k_{1}};\,{k_{1}}^{\mu}
= &&{\it C_{1}}\,[\frac{r}{l{\alpha}{\sqrt {{\alpha}^{2}{l}^{2}
+{r}^{2}}}}\sin \left(
\alpha\,\phi \right) \cos \left( {\frac {\alpha\,t}{l}} \right),
\frac{\sqrt{{\alpha}^{2}{l}^{2}
+{r}^{2}}}{{l}^{2}}\sin \left( \alpha\,\phi \right) \sin
 \left( {\frac {\alpha\,t}{l}} \right) ,\nonumber\\&&
 \frac{\sqrt {{\alpha}^{2}{l}
^{2}+{r}^{2}}}{{l}^{2}\alpha\,r}
\cos \left( \alpha\,\phi \right) \sin \left( {\frac {
\alpha\,t}{l}} \right) ]
\end{eqnarray}
\begin{eqnarray}
\bm{\partial}_{k_{2}};\,{k_{2}}^{\mu}= &&{\it C_{2}}
\,[-\frac{r}{l{\alpha}{\sqrt {{\alpha}^{2}{l}^{2}+{r}^{2}}}}\sin \left(
\alpha\,\phi \right) \sin \left( {\frac {\alpha\,t}{l}} \right),
\frac{\sqrt{{\alpha}^{2}{l}^{2}+{r}^{2}}}{{l}^{2}}
\sin\left( \alpha\,\phi \right) \cos
 \left( {\frac {\alpha\,t}{l}} \right) ,\nonumber\\&&
 \frac{\sqrt {{\alpha}^{2}{l}
^{2}+{r}^{2}}}{{l}^{2}\alpha\,r}
\cos \left( \alpha\,\phi \right) \cos \left( {\frac {
\alpha\,t}{l}} \right) ]
\end{eqnarray}
\begin{eqnarray}
\bm{\partial}_{k_{3}};\,{k_{3}}^{\mu}
= &&{\it C_{3}}\,
[\frac{r}{l{\alpha}{\sqrt {{\alpha}^{2}{l}^{2}+{r}^{2}}}}\cos\left(
\alpha\,\phi \right) \cos \left( {\frac {\alpha\,t}{l}} \right),
\frac{\sqrt{{\alpha}^{2}{l}^{2}+{r}^{2}}}{{l}^{2}}
\cos\left( \alpha\,\phi \right) \sin
 \left( {\frac {\alpha\,t}{l}} \right)
 ,\nonumber\\&&
 -\frac{\sqrt {{\alpha}^{2}{l}
^{2}+{r}^{2}}}{{l}^{2}\alpha\,r}
\sin \left( \alpha\,\phi \right) \sin \left( {\frac {
\alpha\,t}{l}} \right) ]
\end{eqnarray}
\begin{eqnarray}
\bm{\partial}_{k_{4}};\,{k_{4}}^{\mu}=
&&{\it C_{4}}\,[-\frac{r}{l{\alpha}{\sqrt {{\alpha}^{2}{l}^{2}
+{r}^{2}}}}\cos \left(
\alpha\,\phi \right) \sin \left( {\frac {\alpha\,t}{l}} \right),
\frac{\sqrt{{\alpha}^{2}{l}^{2}
+{r}^{2}}}{{l}^{2}}\cos \left( \alpha\,\phi \right) \cos
 \left( {\frac {\alpha\,t}{l}} \right) ,\nonumber\\&&
- \frac{\sqrt {{\alpha}^{2}{l}
^{2}+{r}^{2}}}{{l}^{2}\alpha\,r}
\sin \left( \alpha\,\phi \right) \cos \left( {\frac {
\alpha\,t}{l}} \right) ]
\end{eqnarray}
\begin{eqnarray}
\bm{\partial}_{k_{5}};\,{k_{5}}^{\mu}={\it C_{5}}\,\mbox(0,0,1)
\end{eqnarray}
\begin{eqnarray}
\bm{\partial}_{k_{6}};\,{k_{6}}^{\mu}
={\it C_{6}}\,\mbox (-{l}^{2},0,0).
\end{eqnarray}
\end{subequations}

For completeness, the commutators are given explicitly as
\begin{subequations}
\begin{eqnarray}
&&{\bf\partial}_{[k_{6}}{\bf\partial}_{k_{1}]}
=\alpha\frac{{\it C_{6}}{\it C_{1}}}{{\it C_{3}}}
\, {\bf\partial}_{k_{3}},
{\bf\partial}_{[k_{6}}{\bf\partial}_{k_{2}]}
=\alpha\frac{{\it C_{6}}{\it C_{2}}}{{\it C_{4}}}
\, {\bf\partial}_{k_{4}},
\nonumber\\&&
{\bf\partial}_{[k_{6}}{\bf\partial}_{k_{3}]}
=-\alpha\frac{{\it C_{6}}{\it C_{3}}}{{\it C_{1}}}\, {\bf\partial}_{k_{1}},
{\bf\partial}_{[k_{6}}{\bf\partial}_{k_{4}]}
=-\alpha\frac{{\it C_{6}}{\it C_{4}}}{{\it C_{2}}}\, {\bf\partial}_{k_{2}},
\end{eqnarray}
\begin{eqnarray}
&&
{\bf\partial}_{[k_{5}}{\bf\partial}_{k_{1}]}
=-l\,\alpha\frac{{\it C_{5}}{\it C_{1}}}{{\it C_{2}}}\, {\bf\partial}_{k_{2}},
{\bf\partial}_{[k_{5}}{\bf\partial}_{k_{2}]}
=l\,\alpha\frac{{\it C_{5}}{\it C_{2}}}{{\it C_{1}}}
\, {\bf\partial}_{k_{1}},\nonumber\\
&&
{\bf\partial}_{[k_{5}}{\bf\partial}_{k_{3}]}
=-l\,\alpha\frac{{\it C_{5}}{\it C_{3}}}{{\it C_{4}}}\, {\bf\partial}_{k_{4}},
{\bf\partial}_{[k_{5}}{\bf\partial}_{k_{4}]}
=l\,\alpha\frac{{\it C_{5}}{\it C_{4}}}{{\it C_{3}}}\, {\bf\partial}_{k_{3}},
\end{eqnarray}
\begin{eqnarray}
&&{\bf\partial}_{[k_{4}}{\bf\partial}_{k_{3}]}
=-\frac{1}{\alpha\,l^5}
\,\frac{{\it C_{4}}{\it C_{3}}}{{\it C_{5}}}\, {\bf\partial}_{k_{5}},
{\bf\partial}_{[k_{4}}{\bf\partial}_{k_{2}]}
=\frac{1}{\alpha\,l^4}
\,\frac{{\it C_{4}}{\it C_{2}}}{{\it C_{6}}}\, {\bf\partial}_{k_{6}},
\end{eqnarray}
\begin{eqnarray}
&&{\bf\partial}_{[k_{3}}{\bf\partial}_{k_{1}]}
=\frac{1}{\alpha\,l^4}
\,\frac{{\it C_{3}}{\it C_{1}}}{{\it C_{6}}}\, {\bf\partial}_{k_{6}},
{\bf\partial}_{[k_{2}}{\bf\partial}_{k_{1}]}
=\frac{1}{\alpha\,l^5}
\,\frac{{\it C_{2}}{\it C_{1}}}{{\it C_{5}}}\, {\bf\partial}_{k_{5}}.
\end{eqnarray}
\end{subequations}
This anti--de Sitter metric,
(cosmological constant negative--$\lambda=-1/l^2$), for the coordinates
$\{t,\rho,\phi\}$--merely names--ranging
$-\infty\leq t\leq\infty$, $-\infty\leq \rho\leq\infty$;
$-\infty\leq \phi\leq\infty$ allows
for six symmetries, i.e., six Killing vectors.
For these coordinate ranges
the space is maximally symmetric.\\
Moreover, if the spatial coordinates are restricted to range\\
$$ 0\leq \rho\leq\infty,\,0\leq \phi\leq\,2\pi, $$
and $\alpha$ is equated to unity,
$\alpha=1=-M$, then in such case $\rho$ and $\phi$
become polar coordinates with $\phi$ being the angular
coordinate with  period  $2\pi$.
This spacetime--the (proper)
anti--de Sitter space (with $M=-1)$--allows for six symmetries, and as such it is maximally
symmetric.

\section{Peldan electrostatic solution}\label{staticPeldanM}

It seems that static Einstein--Maxwell solutions with cosmological constant were first
derived in the Peldan's work~\cite{Peldan93}; in that publication Peldan mentioned his
failure in finding any work done on explicit solutions to Einstein--Maxwell solutions
with cosmological constant, although there was pointed the existence of
static and rotation symmetric solutions with vanishing cosmological constant, namely
the Deser--Mazur~\cite{DeserM85}, and Melvin~\cite{Melvin86} solutions.

The electrostatic Peldan
solution, see also~\cite{GarciaAnnals09}~Eq.~(4.15), is given by the metric
\begin{eqnarray}
ds^2&&=-N(\rho)^2dt^2+\frac{1}{L(\rho)^{2}}\,d\rho^2
+K(\rho)^2[d\phi+W(\rho)dt]^2,\nonumber\\
L(\rho)=N(\rho)&&=\sqrt{\frac{\rho^2}{l^2}
-2b^2\ln{\rho}-M},\,K(\rho)=\rho,\, W(\rho)=0.
\end{eqnarray}

\subsection{ Mass, energy and momentum for the
electrostatic Peldan solution}\label{staticPeldan}

The surface energy density $\epsilon$ occurs to be
\begin{eqnarray}
\epsilon(\rho,\epsilon_{0})=-\frac{1}{\pi\rho}N(\rho)
- \epsilon_{0}.
\end{eqnarray}
Consequently the global energy and mass are given by
\begin{eqnarray}\label{den-catga}
E(\rho,\epsilon_{0})&=& -2N(\rho)
-2\pi\,\rho\,\epsilon_{0},\nonumber\\
M(\rho,\epsilon_{0})&=&-2\frac{\rho^2}{l^2}
+2m+4b^2\ln{\rho}-2\pi\rho\,N(\rho)\epsilon_{0}.
\end{eqnarray}
Thus, for the natural choice of a vanishing reference energy density
$\epsilon_{0}=0$, one has at the spatial infinity
$\rho\rightarrow\infty$ that
\begin{eqnarray}
\epsilon(\rho\rightarrow\infty,\epsilon_{0}=0)&\approx&-\frac{1}{\pi\,l}
+\frac{l\,M}{2\pi\,\rho^2}+\frac{l\,b^2}{\pi\,\rho^2}\ln{\rho},\nonumber\\
E(\rho\rightarrow\infty,\epsilon_{0}=0)&\approx&-2\frac{\rho}{l}
+\frac{M\,l}{\rho}+2\frac{l\,b^2}{\rho}\ln{\rho},\nonumber\\
M(\rho\rightarrow\infty,\epsilon_{0}=0)&\approx&-2\frac{\rho^2}{l^2}+2M+4b^2\ln{\rho},
\end{eqnarray}
while if the reference energy is the one corresponding to the
anti--de Sitter spacetime  $AdS(M_{0})$,
$\epsilon_{0}=-\frac{1}{\pi\,\rho}\sqrt{\frac{\rho^2}{l^2}-M_{0}}$,
$\epsilon_{0\mid\infty}(M_{0})\approx
-\frac{1}{\pi\,l}+\frac{lM_{0}}{2\pi\,\rho^2}$,
then the energies and mass at spatial infinity are expressed as
\begin{eqnarray}
\epsilon(\rho\rightarrow\infty, \epsilon_{0\mid\infty}(M_{0}))&\approx&
\,l\,\frac{M-M_{0}}{2\pi\,\rho^2}
+\frac{l\,b^2}{\pi\,\rho^2}\ln{\rho},\nonumber\\
E(\rho\rightarrow\infty,
\epsilon_{0\mid\infty}(M_{0}))
&\approx&\,l\frac{M-M_{0}}{\rho}+2\frac{lb^2}{\rho}\ln{\rho},\nonumber\\
M(\rho\rightarrow\infty,
\epsilon_{0\mid\infty}(M_{0}))&\approx&\, M-M_{0}+ 2\,b^2\ln{\rho}.
\end{eqnarray}
Comparing with the static BTZ one recognize $M$ as the BTZ $M$.
Notice that the energy and mass include an amount of energy due to
the electric field through the logarithmical term; because of this
dependence, these quantities diverge at infinity logarithmically.

\subsection{Field, energy--momentum,
and Cotton tensors for the electrostatic Peldan solution}

The electromagnetic tensor field associated with the Peldan
solution is given by
\begin{eqnarray}
({F^\alpha}_{\beta})= \left[
\begin {array}{ccc} 0&{\frac {b}{{ L^2}\,\rho}}&0
\\\noalign{\medskip}{\frac {b\,{ L^2}}{{\rho}
}}&0&0\\\noalign{\medskip}0&0&0\end {array} \right] ,
\end{eqnarray}
Searching for its eigenvectors, one arrives at
\begin{eqnarray}
\lambda_{1}&&=0;{\bf V}1=[{V^{1}}=0,{\it V2}=0,{V^{3}}={V^{3}}],
 V_{\mu}V^\mu ={\rho}^{2}{V^{3}}^2,\,{\bf V}1={\bf S}1,\nonumber\\
\lambda_2&&=\frac {b}{\rho};
{\bf V}2=[{V^{1}}=\frac {V^{2}}{
L^2},{V^{2}}={V^{2}},{V^{3}}=0],
\,V^{\mu}V_{\mu}=0,\,{\bf V}2={\bf N}2,\nonumber\\
\lambda_{3}&&=-\frac {b}{\rho};
{\bf V}3=[{V^{1}}=-\frac {V^{2}}{
L^2},{V^{2}}={V^{2}},{V^{3}}=0],\,
V^{\mu}V_{\mu}=0,\,{\bf V}3={\bf N}3,\nonumber\\
&&{\rm Type:}\{S,N,N\}.
\end{eqnarray}
thus its type is
$${\rm Type:}\{S,N,N\}$$
As far as to the electromagnetic
energy momentum tensor is concerned, its matrix
amounts to
\begin{eqnarray}
({T^\alpha}_{\beta}) =\left[ \begin {array}{ccc} -\frac{1}{8}\,{
\frac {{b}^{2}}{\pi \,{\rho}^{2}}}&0&0\\\noalign{\medskip}0&-\frac{1}{8}\,{
\frac {{b}^{2}}{\pi \,{\rho}^{2}}}&0\\\noalign{\medskip}0&0&\frac{1}{8}\,{
\frac {{b}^{2}}{\pi \,{\rho}^{2}}}\end {array} \right],
\end{eqnarray}
with the following eigenvalues and their
corresponding eigenvectors
\begin{eqnarray}
\lambda_1&&=-\frac{1}{8\pi}\,\frac {{b}^{2}}{{\rho}^{2} };
{\bf V}1=[V^{1},V^{2},0],
\,V1_{\mu}=V^{1}g_{\mu t}+V^{2}g_{\mu \rho},
\,V1^{\mu}V1_{\mu}=(V^{1})^2g_{t \,t}+(V^{2})^2g_{\rho\rho},\nonumber\\
{\bf V}1&&=\{{\bf T}1, {\bf S}1,{\bf N}1\},\nonumber\\
\lambda_2&&=-\frac{1}{8\pi}\,\frac {{b}^{2}}{{\rho}^{2} };
{\bf V}2=[\tilde V^{1},\tilde V^{2},0],
\,V2_{\mu}=\tilde V^{1}g_{\mu t}+\tilde V^{2}g_{\mu \rho},
\,V2^{\mu}V2_{\mu}=(\tilde V^{1})^2g_{t \,t}+(\tilde V^{2})^2g_{\rho\rho},\nonumber\\
{\bf V}2&&=\{{\bf T}2, {\bf S}2,{\bf N}2\},\nonumber\\
\lambda_3&&=\frac{1}{8\pi}\,\frac {{b}^{2}}{{\rho}^{2} };
{\bf V}3=[0,0,V^{3}],\,V_{\mu}=  V^{3}g_{\mu\phi},\,V^{\mu}V_{\mu}
=({ V}^{3})^2g_{\phi \,\phi},\,\nonumber\\\tilde
{\bf V}3&&={\bf S}3.
\end{eqnarray}
For ${\bf V}1$ and ${\bf V}2$, the character of these vectors depends on
the sing of their magnitudes; for instance, choosing
\begin{eqnarray*}
{ V}^{1}&&=s\,\sqrt{g_{\rho \,\rho}}/\sqrt{|g_{t \,t}|}\,{ V}^{2},
\,s={\rm constant},\,{{\bf V}1}^{\mu}{{\bf V}1}_{\mu}=(1-s^2)g_{\rho
\,\rho}\,(V^{2})^2; \nonumber\\&& s>1\rightarrow{{\bf V}1={\bf
T}},\, s=\pm 1\,\rightarrow{{\bf V}1={\bf N}},\,s<1\rightarrow{{\bf
V}1={\bf S}}.
\end{eqnarray*}
The space--like vector ${\bf V}3$ is
aligned along the circular Killing
direction $\bf {\partial}_{\phi}$. Thus
one may have the space--time arraignment $\{{\bf T}1, {\bf S}2,
{\bf S}3\}$, or  $\{{\bf N}1, {\bf N}2,
{\bf S}3\}$, and so on.

The Cotton tensor for electrostatic cyclic symmetric gravitational field is
given by
\begin{eqnarray}
({C^\alpha}_{\beta})=\left[
\begin {array}{ccc} 0&0&\,{\frac {{b}^{2}
}{2\rho^{2}}}\\\noalign{\medskip}0&0&0\\
\noalign{\medskip}-\frac{1}{2}\,{\frac {{b}^{2} \left(
{l}^{2}M+{\rho}^{2 }-2\,{b}^{2}{l}^{2}\ln  \left( \rho \right)
\right) }{{l}^{2}{\rho}^{ 4}}}&0&0\end {array} \right] =\left[
\begin {array}{ccc} 0&0&\,{\frac {{b}^{2}
}{2\rho^{2}}}\\\noalign{\medskip}0&0&0\\
\noalign{\medskip}-\,{\frac {{b}^{2}
}{2{\rho}^{4}}}L^2&0&0\end {array} \right].
\end{eqnarray}
The search for its eigenvectors yields
\begin{eqnarray}
\lambda_1=0;{\bf V}1&&=[0,V^{2},0],\,V1_{\mu}=
V^{2}g_{\rho\rho}\delta^\rho_{\mu},\,V1^{\mu}V1_{\mu}=
(V^{2})^2g_{\rho\rho},\,{\bf V}1={\bf S}1,\nonumber\\
\lambda_{2}=\frac{i}{2}\,{\frac {L\,b^{2}}{{\rho}^{3}}}
;{\bf V}2&&=[V^{1},0,V^{3}=\frac{i\,L}{\rho}V^{1}],\,V2_{\mu}=
V^{1}g_{\mu t}+V^{3}g_{\mu \phi},\,{\bf V}2={\bf Z},\nonumber\\
\lambda_{3}=-\frac{i}{2}\,{\frac {L\,b^{2}}{{\rho}^{3}}};
{\bf V}3&&=[ V^{1},0, V^{3}=-\frac{i\,L}{\rho}{ V^{1}}],\,V3_{\mu}=
{V}^{1}g_{\mu t}+{ V}^{3}g_{\mu \phi},\,
{\bf V}3={\bf {\bar Z}},
\end{eqnarray}
therefore the corresponding tensor type is
$${\rm Type:}\{S,Z,\bar Z\}.$$

The eigenvectors ${\bf V}2$ and ${\bf V}3$ are complex
conjugated, or, if one wishes, one may
consider the component ${V}^{1}$ different
for each of the complex vectors. For the zero
eigenvalue $\lambda_1$, the vector ${\bf V}1$
is a spacelike vector, it points along the
$\rho$--coordinate direction.

It is worthwhile to point out that the
field and Cotton tensors of the
solutions ge\-ne\-rated via coordinate transformations,
in particular  $SL(2,R)$ transformations,
applied onto this electrostatic cyclic
symmetric Peldan solution
will shear the  eigenvalues $\lambda_i$ of the
corresponding field and
Cotton tensors of the Peldan solution;
recall that eigenvalues are invariant characteristics of tensors, although
the components of the eigenvectors, in general, look different in
different coordinate systems--this remark also
applies to the (eigenvalues) eigenvectors of the
seed and  resulting solutions.

\subsection{Field, energy--momentum, and Cotton tensors for a
modified electrostatic Peldan solution}

The solution to  be studied is a slight
modification of the Peldan electrostatic
solution used in the previous
paragraph, namely the one with metric
\begin{eqnarray}\label{HWmetric}
ds^2&&=-\frac {{\it L^2}\,{\rho}^{2}}{{\rho}^{2}+{
\it Mg}}dt^2+\frac{1}{L^{2}}\,d\rho^2
+({\rho}^{2}+{\it Mg})d\phi^2,\nonumber\\
{\it L^2}&&:={\frac { \left[ K0+{\rho}^{2}+{\it Mg}-{b}^{2}{l}^{2}\ln
 \left( {\rho}^{2}+{\it Mg} \right)  \right]  \left( {\rho}^{2}+{\it
Mg} \right) }{{l}^{2}{\rho}^{2}}},\nonumber\\
\end{eqnarray}
and electromagnetic Maxwell field tensor
\begin{eqnarray}\label{HWelecTensor}
({F^\alpha}_{\beta})= \left[
\begin {array}{ccc} 0&{\frac {b}{{\it L^2}\,\rho}}&0
\\\noalign{\medskip}{\frac {b\rho\,{\it L^2}}{{\rho}^{2}+{\it Mg}
}}&0&0\\\noalign{\medskip}0&0&0\end {array} \right].
\end{eqnarray}
Searching for its eigenvectors, one arrives at
\begin{eqnarray}
\lambda_{1}&&=0;{\bf V}1=[{V^{1}}=0,{V^{2}}=0,{V^{3}}={V^{3}}],
 V_{\mu}V^\mu =\left( {\rho}^{2}+{\it Mg} \right)
 {V^{3}}^2,\,{\bf V}1={\bf S}1,\nonumber\\
\lambda_2&&={\frac {b}{\sqrt {{\rho}^{2}+{\it Mg}}}};
{\bf V}2=[{V^{1}}={\frac {{V^{2}}\,\sqrt {{\rho}^{2}+{\it Mg}}}{{\it
L^2}\,\rho}},{V^{2}}={V^{2}},{V^{3}}=0],\nonumber\\&&
\,V^{\mu}V_{\mu}=0,\,{\bf V}2={\bf N}2,\nonumber\\
\lambda_{3}&&=-{\frac {b}{\sqrt {{\rho}^{2}+{\it Mg}}}};
{\bf V}3=[{V^{1}}=-{\frac {{V^{2}}\,\sqrt {{\rho}^{2}+{\it Mg}}}{{\it
L^2}\,\rho}},{V^{2}}={V^{2}},{V^{3}}=0],\,\nonumber\\&&
V^{\mu}V_{\mu}=0,\,{\bf V}3={\bf N}3,
\end{eqnarray}
thus its type is
$$\{S,N,N\}.$$
For the energy--momentum tensor
\begin{eqnarray}
({T^\alpha}_{\beta})= \left[
\begin {array}{ccc} -\frac{1}{8}\,{\frac {{b}^{2}}
{ \left( {\rho}^{2}+{\it Mg}
 \right) \pi }}&0&0\\\noalign{\medskip}0&-\frac{1}{8}\,{\frac {{b}^{2}}{
 \left( {\rho}^{2}+{\it Mg} \right) \pi }}&0\\
 \noalign{\medskip}0&0&\frac{1}{8}\,{\frac {{b}^{2}}
{ \left( {\rho}^{2}+{\it Mg} \right) \pi }}
\end {array} \right],
\end{eqnarray}
one has the following set of eigenvectors
\begin{eqnarray}
\lambda_{1,2}&&=-\frac{1}{8}\,\frac {{b}^{2}}
{ \left( \rho^{2}+{\it Mg}\right) \pi };
{\bf V}1,2=[{V^{1}}={V^{1}},{V^{2}}={V^{2}},{V^{3}}=0],
\nonumber\\&&
V_{\mu}V^\mu =-\frac{{\rho}^{2}{\it L^2}}{{\rho}^{2}+
{\it Mg} }{V^{1}}^2+\frac{1}{\it L^2}{V^{2}}^2\nonumber\\
&&\,{\bf V}1={\bf T}1,\,{\bf N }1,\,{\bf S}1,
\,\,{\bf V}2={\bf T}2,{\bf N}2,\,{\bf S}2,\nonumber\\
\lambda_3&&=\frac{1}{8}\,\frac {{b}^{2}}
{ \left( \rho^{2}+{\it Mg}\right) \pi };
{\bf V}3=[{V^{1}}=0,{V^{2}}=0,{V^{3}}={V^{3}}],
\nonumber\\&&
V_{\mu}V^\mu ={{V^{3}}^{2}}{ \left( \rho^{2}+{\it Mg}\right) },{\bf V}3={\bf S}3,
\end{eqnarray}
hence its type present several possibilities
$${\rm Type:}\{2T,S\},\,\{2N,S\},\,\{2S,S\}.$$

The Cotton tensor
\begin{eqnarray}
({C^\alpha}_{\beta}) = \left[
\begin {array}{ccc} 0&0&\frac{1}{2}\,
{\frac {{b}^{2}}{{\rho}^{2}+{\it Mg}}}
\\\noalign{\medskip}0&0&0\\\noalign{\medskip}
-\frac{1}{2}\,{\frac {{b}^{2}{
\it L^2}\,{\rho}^{2}}{ \left( {\rho}^{2}
+{\it Mg} \right) ^{3}}}&0
&0\end{array}\right],
\end{eqnarray}
allows for the eigenvectors
\begin{eqnarray}
\lambda_{1}&&=0;
{\bf V}1=[{V^{1}}=0,{V^{2}}={V^{2}},{V^{3}}=0],
\,V_{\mu}V^\mu ={V^{2}}^2/{
\it L^2},\,{\bf V}1={\bf S}1,
\nonumber\\
\lambda_{2}&&=\frac{1}{2}\,
{\frac {\sqrt {-{\it L^2}}{b}^{2}\rho}{ \left( {\rho}^{2}
+{\it Mg} \right) ^{2}}};\nonumber\\&&
{\bf V}2=[{V^{1}}={V^{1}},{V^{2}}=0,{V^{3}}=
{\frac {\sqrt {-{\it L^2}}\rho\,
{V^{1}}}{{\rho}^{2}+{\it Mg}}}],
{\bf V}2={\bf Z},
\nonumber\\
\lambda_{3}&&=-\frac{1}{2}\,
{\frac {\sqrt {-{\it L^2}}{b}^{2}\rho}
{ \left( {\rho}^{2}+{\it Mg
} \right) ^{2}}};\nonumber\\&&
{\bf V}3=[{V^{1}}={V^{1}},{V^{2}}=0,
{V^{3}}=
-{\frac {\sqrt {-{\it L^2}}\rho\,
{V^{1}}}{{\rho}^{2}+{\it Mg}}}],
{\bf V}3={\bf {\bar Z}},
\end{eqnarray}
hence the type of this Cotton tensor is
$${\rm Type}:\,\{S,Z,\bar Z\}.$$

\subsection{Field, energy--momentum,
and Cotton tensors for the generalized--via $SL(2,R)$
transformations--Peldan solution}

Many solutions in 2+1 gravity are generated via $SL(2,R)$
transformations of the form
\begin{eqnarray}
t&&=\alpha\,T+\beta\,\Phi,\,\,\alpha\delta-\beta\gamma=1,\nonumber\\
\phi&&=\gamma\,T+\delta\,\Phi,
\end{eqnarray}
applied, in this particular case,
onto the seed electrostatic solution (\ref{HWmetric})
and (\ref{HWelecTensor}), yielding
to the stationary metric
\begin{eqnarray}\label{rotelecPeldan}
g&&= \left[ \begin {array}{ccc} {\frac {-{\alpha}^{2}{
\it L^2}\,{\rho}^{2}+{\gamma}^{2} \left( {\rho}^{2}+{\it Mg}
 \right) ^{2}}{{\rho}^{2}+{\it Mg}}}&0&{\frac {-\alpha\,\beta\,{\it
L^2}\,{\rho}^{2}+\gamma\,\delta\, \left( {\rho}^{2}+{\it Mg}
 \right) ^{2}}{{\rho}^{2}+{\it Mg}}}\\\noalign{\medskip}0&1/{{\it
L^2}}&0\\\noalign{\medskip}{\frac {-\alpha\,\beta\,{\it
L^2}\,{\rho}^{2}+\gamma\,\delta\, \left( {\rho}^{2}+{\it Mg}
 \right) ^{2}}{{\rho}^{2}+{\it Mg}}}&0&{\frac {-{\beta}^{2}{\it
L^2}\,{\rho}^{2}+{\delta}^{2} \left( {\rho}^{2}+{\it Mg} \right)
^{2}}{{\rho}^{2}+{\it Mg}}}\end {array} \right] ,\nonumber\\
{\it L^2}&&:={\frac { \left[ K0+{\rho}^{2}+{\it Mg}-{b}^{2}{l}^{2}\ln
 \left( {\rho}^{2}+{\it Mg} \right)  \right]  \left( {\rho}^{2}+{\it
Mg} \right) }{{l}^{2}{\rho}^{2}}},
\end{eqnarray}
accompanied by the electromagnetic field tensor
\begin{eqnarray}
({F^\alpha}_{\beta})=\left[
\begin {array}{ccc} 0&{\frac {\delta\,b}{ {\it L^2}
\,\rho}}&0\\\noalign{\medskip}{
\frac {\alpha\,{\it L^2}\,b\rho}{{\rho}^{2}+{\it Mg}}}&0&{\frac {
\beta\,{\it L^2}\,b\rho}{{\rho}^{2}+{\it Mg}}}
\\\noalign{\medskip}0&-{\frac {b\gamma}
{ {\it L^2}\,\rho}}&0\end {array} \right],
\end{eqnarray}
with eigenvalues
\begin{eqnarray}
\lambda_{1}&&=0;{\bf V}1=[{V^{1}}=-\frac{
\beta}{\alpha}{V^{3}},{V^{2}}=0,{V^{3}}={V^{3}}],
 V_{\mu}V^\mu =({\rho}^{2}+{\it Mg}){V^{3}}^2\,{\bf V}1
 ={\bf S}1,\nonumber\\
\lambda_2&&={\frac {b}{\sqrt {{\rho}^{2}+{\it Mg}}}};
\nonumber\\&&
{\bf V}2=[{V^{1}}={V^{1}},{V^{2}}={\frac {{\it L^2}
\,\rho\, {V^{1}}}{\delta\,\sqrt {{\rho}^{
2}+{\it Mg}}}},{V^{3}}=-{\frac {\gamma}{\delta}\,{V^{1}}}]
,\,V^{\mu}V_{\mu}=0,\,{\bf V}2={\bf N}2,\nonumber\\
\lambda_{3}&&=-{\frac {b}{\sqrt {{\rho}^{2}+{\it Mg}}}};
\nonumber\\&&
{\bf V}3=[{V^{1}}={V^{1}},{V^{2}}=-{\frac {{\it L^2}
\,\rho\, {V^{1}}}{\delta\,\sqrt {{\rho}^{
2}+{\it Mg}}}},{V^{3}}=-{\frac {\gamma}{\delta}\,{V^{1}}}],
\,V^{\mu}V_{\mu}=0,\,{\bf V}3={\bf N}3
\end{eqnarray}
$$\{S,N,N\},$$
The evaluation of the electromagnetic energy--momentum tensor brings
\begin{eqnarray}
({T^\alpha}_{\beta})= \left[
\begin {array}{ccc} -\frac{{b}^{2}}{8\,\pi}\,
{\frac { \left( \alpha\,\delta+\beta\,
\gamma \right) }{
 \left( {\rho}^{2}+{\it Mg} \right)}}&0&
 -\frac{{b}^{2}}{4\,\pi}\,{\frac {\beta\,
\delta}{
 \left( {\rho}^{2}+{\it Mg} \right)}}
 \\\noalign{\medskip}0&-\frac{{b}^{2}}{8\,\pi}\,
\frac {1}{ \left( {\rho}^{2}+{\it Mg} \right) }&0
\\\noalign{\medskip}\frac{{b}^{2}}{4\,\pi}
\,{\frac {\alpha\,\gamma}{ \left( {\rho}^{2}+{\it Mg}
 \right)}}&0&\frac{{b}^{2}}{8\,\pi}\,{\frac {\left( \alpha\,\delta+\beta\,
\gamma \right) }{
 \left( {\rho}^{2}+{\it Mg} \right) }}\end {array} \right]
\end{eqnarray}
characterized by the following eigenvectors
\begin{eqnarray}
\lambda_{1,2}&&=-\frac{1}{8\pi}
{\frac {{b}^{2}}{{\rho}^{2} +{\it Mg}}};
\nonumber\\
{\bf V}1,2&&=[{V^{1}}={V^{1}},{V^{2}}={V^{2}},{V^{3}}
=-{\frac {\gamma}{\delta}\,{V^{1}}}],\,V^{\mu}V_{\mu}=
-\frac{{V^{1}}^2}{\delta^2}\frac{{\rho}^{2}{\it L^2}}
{\left( {\rho}^{2}+{\it Mg} \right)}+\frac{{V^{2}}^2}{{\it L^2}},\nonumber\\
&&{\bf V}1={\bf T}1,\,{\bf N }1,\,{\bf S}1,\,
\,{\bf V}2={\bf T}2,{\bf N}2,\,{\bf S}2,\nonumber\\
\lambda_3&&=\frac{1}{8\pi}{\frac {{b}^{2}}
{{\rho}^{2} +{\it Mg}}};\nonumber\\
{\bf V}3&&=[{V^{1}}=-\frac{
\beta}{\alpha}{V^{3}},{V^{2}}=0,{V^{3}}
={V^{3}}],
 V_{\mu}V^\mu =\frac{{V^{3}}^2}{\alpha^2}
 \left( {\rho}^{2}+{\it Mg} \right),\,{\bf V}3={\bf S}3,
\end{eqnarray}
thus its type could be, among other variants:
$$\{2T,S\},\,\{2N,S\},\,\{2S,S\}.$$

The transformed Cotton tensor is given by
\begin{eqnarray}
({C^\alpha}_{\beta})= \left[
\begin {array}{ccc} \frac{1}{2}\,\frac {b^{2}  \alpha\,\beta\,{\it
L^2}\,\rho^{2}}{ \left( {\rho}^{2}+{\it Mg} \right) ^{3} }
+\frac{1}{2}\frac {b^{2}\gamma\,\delta }{  \left( {\rho}^{2}+{\it Mg}
 \right) }&0&\frac{1}{2}\,\frac {b^{2} \beta^2\,{\it
L^2}\,\rho^{2}}{ \left( {\rho}^{2}+{\it Mg} \right) ^{3} }
+\frac{1}{2}\frac {b^{2}\delta^2 }{  \left( {\rho}^{2}+{\it Mg}
 \right) }
\\\noalign{\medskip}0&0&0\\\noalign{\medskip}-\frac{1}{2}\,\frac {b^{2}  \alpha^2\,{\it
L^2}\,\rho^{2}}{ \left( {\rho}^{2}+{\it Mg} \right) ^{3} }
-\frac{1}{2}\frac {b^{2}\gamma^2}{  \left( {\rho}^{2}+{\it Mg}
 \right) }&0&-\frac{1}{2}\,\frac {b^{2}  \alpha\,\beta\,{\it
L^2}\,\rho^{2}}{ \left( {\rho}^{2}+{\it Mg} \right) ^{3} }
-\frac{1}{2}\frac {b^{2}\gamma\,\delta}{  \left( {\rho}^{2}+{\it Mg}
 \right) }\end {array} \right],
\end{eqnarray}
\begin{eqnarray}
\lambda_{1}&&=0;{\bf V}1=[{V^{1}}=0,{V^{2}}={V^{2}},{V^{3}}=0],
 V_{\mu}V^\mu ={V^{2}}^2/{\it
L^2},\,{\bf V}1={\bf S}1,\nonumber\\
\lambda_{2}&&=i\frac{1}{2}\,
{\frac {{{\it L}}{b}^{2}\rho}{ \left( {\rho}^{2}+{\it Mg} \right) ^{2}}};
\nonumber\\&&
{\bf V}2=[ {V^{1}}=-{\frac {{V^{3}}\, \left( {\beta}^{2}{\it L^2}\,{\rho}^
{2}+{\delta}^{2} \left( {\rho}^{2}+{\it Mg} \right) ^{2} \right) }{
\gamma\,\delta\, \left( {\rho}^{2}+{\it Mg} \right) ^{2}
+\alpha\,\beta\,{\it L^2}\,{\rho}^{2}-i{{\it
 L}}{\rho}({\rho}^{2}+Mg)  }},{V^{2}}=0,{V^{3}}={V^{3}} ]
,
\nonumber\\&&\,\,{\bf V}2={\bf Z},\nonumber\\
\lambda_{3}&&=-i\frac{1}{2}\,
{\frac { {{\it L}}{b}^{2}\rho}{ \left( {\rho}^{2}
+{\it Mg} \right) ^{2}}};
\nonumber\\&&
{\bf V}3=[{V^{1}}=-{\frac {{V^{3}}\, \left( {\beta}^{2}{\it L^2}\,{\rho}^{2}+{
\delta}^{2} \left( {\rho}^{2}+{\it Mg} \right) ^{2} \right) }{\gamma\,
\delta\, \left( {\rho}^{2}+{\it Mg} \right) ^{2}+\alpha\,\beta\,{\it L^2}\,{\rho}^{2}+i{{\it  L
}}\rho\,  \left( {\rho}^{2
}+{\it Mg} \right)} },{V^{2}}=0,{V^{3}}={V^{3}}].
\nonumber\\&&\,{\bf V}3={\bf \bar Z}.
\end{eqnarray}
The type of this generalized Cotton tensor is
$$\{S,Z,\bar Z\}.$$

\section{Martinez--Teitelboim--Zanelli solution}

Martinez--Teitelboim--Zanelli solution ~\cite{MartinezTZ00} is defined by the metric
\begin{eqnarray}
ds^2&&=-N(\rho)^2dt^2+\frac{1}{L(\rho)^{2}}\,d\rho^2
+K(\rho)^2[d\phi+W(\rho)dt]^2,\nonumber\\
H(\rho)&:=&\rho^2+\frac{l^2\,\omega^2}{l^2-\omega^2}(M+\frac{Q^2}{4}\ln{\rho^2}),\nonumber\\
L(\rho)&=&\sqrt{\frac{\rho^2}{l^2}-M-\frac{Q^2}{4}\ln{\rho^2}},\nonumber\\
K(\rho)&=&\sqrt{H(\rho)},\,
N(\rho)=\rho\frac{L(\rho)}{\sqrt{H(\rho)}},\nonumber\\
W(\rho)&=&-\frac{\omega\,l^2}{l^2-\omega^2}\frac{1}{H(\rho)}(M+\frac{Q^2}{4}\ln{\rho^2}).
\end{eqnarray}
This solution can be also derived from the rotating--under $SL(2,R)$
transformations--Peldan solution (\ref{rotelecPeldan}) with
\begin{eqnarray}
\alpha={\frac {1}{\sqrt {1-{\frac {{\omega}^{2}}{{l}^{2}}}}}},\beta=-
{\frac {\omega}{\sqrt {1-{\frac {{\omega}^{2}}{{l}^{2}}}}}},\gamma=-
{\frac {\omega}{{l}^{2}\sqrt {1-{\frac {{\omega}^{2}}{{l}^{2}}}}}},
\delta={\frac {1}{\sqrt {1-{\frac {{\omega}^{2}}{{l}^{2}}}}}},
\end{eqnarray}
see also \cite{GarciaAnnals09}~ Eq.~(11.33).

\subsection{ Mass, energy and momentum for the MTZ solution}\label{MTZsol}

The evaluation of the surface energy and momentum densities yield
\begin{eqnarray}\label{den-MTZ1}
\epsilon(\rho,\epsilon_{0})&=&-\frac{1}{2\pi}\,\frac{L}{K^2}
\,\left(2\rho+\frac{l^2\,\omega^2}{l^2
-\omega^2}\frac{Q^2}{2\rho}\right)-\epsilon_{0},\nonumber\\
j(\rho)&=&\frac{\rho}{\pi}\frac{l^2\omega}{l^2-\omega^2}\frac{L}{N\,K^2}
\left(M+\frac{1}{2}Q^2\ln{\rho}-\frac{Q^2}{4}\right),
\end{eqnarray}
while the integral quantities can be evaluated from the generic
expressions
\begin{eqnarray}\label{den-MTZ2}
&&J(\rho)=2\,\pi\,K(\rho)\,j(\rho),\nonumber\\
&&E(\rho)=2\,\pi\,K(\rho)\,\epsilon(\rho),\nonumber\\
&&M(\rho)=N(\rho)\,E(\rho)-W(\rho)\,J(\rho).
\end{eqnarray}

The evaluation of the corresponding functions with $\epsilon_{0}=0$
behave at infinity according to
\begin{eqnarray}
\epsilon({\rho\rightarrow\infty},\epsilon_{0}=0)&\approx&-\frac{1}{\pi\,l}
+\frac{[2\,M\,(l^2+\omega^2)-\omega^2\,Q^2]}{4\,\pi(l^2-\omega^2)\rho^2}
+\frac{l\,Q^2}{4\,\pi\rho^2}\frac{l^2+\omega^2}{l^2-\omega^2}
\,\ln{\rho},\nonumber\\
j({\rho\rightarrow\infty})&\approx&\frac{\omega\,l^2}{4\pi\rho}
\frac{4\,M-Q^2}{l^2-\omega^2}
+\frac{\omega\,Q^2\,l^2}{2\pi\rho(l^2-\omega^2)}\,\ln{\rho}
,\nonumber\\
J({\rho\rightarrow\infty})&\approx&\frac{\omega\,l^2}{2}
\frac{4\,M-Q^2}{l^2-\omega^2}
+\frac{\omega\,Q^2\,l^2}{l^2-\omega^2}\,\ln{\rho}\nonumber\\
&=&{J}_{\textrm{(MTZ~\cite{MartinezTZ00},Eq.82)}}(\omega \rightarrow \omega/l)
+\frac{\omega\,Q^2\,l^2}{l^2-\omega^2}\,\ln{\rho}
,\nonumber\\
E({\rho\rightarrow\infty},\epsilon_{0}=0)&\approx&-\frac{2\rho}{l}
+\frac{l}{2\rho}\frac{2 M
l^2-\omega^2\,Q^2}{l^2-\omega^2}+\frac{l^3\,Q^2}{2\rho(l^2-\omega^2)}\ln{\rho}
,\nonumber\\
M({\rho\rightarrow\infty},\epsilon_{0}=0)&
\approx&\,-\frac{2\rho^2}{l^2}+\frac{4\,M
l^2-\omega^2\,Q^2}{2(l^2-\omega^2)}
+\frac{l^2\,Q^2}{l^2-\omega^2}\ln{\rho}.
\end{eqnarray}

Using in the expressions (\ref{den-MTZ1}) and (\ref{den-MTZ2})
as reference energy
density the quantity
$\epsilon_{0}=-\frac{1}{\pi\rho}\sqrt{-M_{0}+\frac{\rho^2}{l^2}}$,
which at the spatial infinity behaves as $\epsilon_{0\mid\infty}(M_{0})\approx
-\frac{1}{\pi\,l}+\frac{lM_{0}}{2\pi\,\rho^2}$, the series
expansions of the corresponding quantities at $\rho=$ infinity
result in
\begin{eqnarray}
\epsilon({\rho\rightarrow\infty},\epsilon_{0\mid\infty}(M_{0}))&\approx&
\,\frac{l}{2\pi\,\rho^2}\,(M-M_{0})
+\frac{l\,\omega^2}{4\,\pi(l^2-\omega^2)\rho^2}(4\,M-Q^2)
+\frac{l\,Q^2}{4\,\pi\rho^2}
\frac{l^2+\omega^2}{l^2-\omega^2}\,\ln{\rho},\nonumber\\
E({\rho\rightarrow\infty},
\epsilon_{0\mid\infty}(M_{0}))
&\approx&\,\frac{l}{\rho}
\,(M-M_{0})+\frac{l\omega^2}{2\rho\,(l^2-\omega^2)}(4\,M-Q^2)
+\frac{l\,Q^2}{2\,\rho}
\frac{l^2+\omega^2}{l^2-\omega^2}\,\ln{\rho},\nonumber\\
M({\rho\rightarrow\infty},
\epsilon_{0\mid\infty}(M_{0}))&\approx&\,M-M_{0}
+\frac{\omega^2}{2(l^2-\omega^2)}(4\,M-Q^2)
+\frac{Q^2}{2}\frac{l^2+\omega^2}{l^2-\omega^2}\,\ln{\rho}\nonumber\\
&=&-M_{0}+{M}_{\textrm{(MTZ~\cite{MartinezTZ00},Eq.81)}}(\omega\rightarrow\omega/l)
+\frac{Q^2}{2}\frac{l^2+\omega^2}{l^2-\omega^2}\,\ln{\rho}.
\end{eqnarray}

Notice that the charges $Q$ used above differs
from ${Q}_{\textrm{(MTZ~\cite{MartinezTZ00},Eq.83)}}$,
\begin{eqnarray}
{Q}_{\textrm{(MTZ~\cite{MartinezTZ00},Eq.83)}}=\frac{l}{\sqrt{l^2-\omega^2}}Q
\end{eqnarray}
Therefore, comparing with the energy characteristics of the BTZ
solution, one concludes that the parameter $M$ can be considered as the BTZ
mass, and the energy and mass functions logarithmically diverges at
spatial infinity.

\subsection{Field, energy--momentum, and Cotton tensors for the MTZ solution}

The Maxwell field tensor for this MTZ solution is given by
\begin{eqnarray}
({F^\alpha}_\beta)&&= \left[
\begin {array}{ccc} 0&\frac{1}{2}\,{\frac {Ql}{\rho\,\sqrt {{l}^{2}-{\omega}^{2}}{
\it L^2}}}&0\\\noalign{\medskip}\frac{1}{2}
\,{\frac {Ql{\it L^2}}{\rho\,\sqrt {{l}^{2}
-{\omega}^{2}}}}&0&-\frac{1}{2}\,{\frac {\omega\,
l{\it L^2}\,Q}{\rho\,\sqrt {{l}^{2}-{\omega}^{2}}}}
\\\noalign{\medskip}0&\frac{1}{2}\,
{\frac {\omega\,Q}{l\rho\,\sqrt {{l}^{2}-{\omega}
^{2}}{\it L^2}}}&0\end {array} \right],\nonumber\\&&
{\it L^2}= \frac{\rho^{2}}{{l}^{2}}
-m-\frac{1}{4}\,{Q}^{2}\ln \left( {\rho}^{2}
 \right) ,
\end{eqnarray}
while its eigenvalues and the corresponding eigenvectors amount to
\begin{eqnarray}
\lambda_{1}&&=0;
{\bf V}1=[{V^{1}}={V^{3}}\,\omega,{V^{2}}=0,{V^{3}}={V^{3}}],
\,\nonumber\\&&
V_{\mu}V^\mu =\frac{l^2-\omega^2}{l^2}\rho^2{V^{3}}^2,
\,{\bf V}1=\,{\bf S}1,\nonumber\\
\lambda_2&&=-\frac{1}{2}\,{\frac {Q}{\rho}};{\bf V}2=[{V^{1}}={V^{1}},{V^{2}}=
{\frac {{V^{1}}\,\sqrt {{l}^{2}-
{\omega}^{2}}}{l}}{\it L^2},\,{V^{3}}=\,{V^{1}}{\frac {\omega}{{l
}^{2}}}],
\nonumber\\&&
V_{\mu}V^\mu =0,{\bf V}2={\bf N}2\nonumber\\
\lambda_3&&=\frac{1}{2}\,{\frac {Q}{\rho}};
{\bf V}3=[{V^{1}}={V^{1}},{V^{2}}=-{\frac {{
V^{1}}\,\sqrt {{l}^{2}-{\omega}^{2}}}{l}}{\it L^2},{V^{3}}
={V^{1}}\frac{\omega}{l^2}],
\nonumber\\&&
V_{\mu}V^\mu =0,{\bf V}3={\bf N}3,\nonumber\\
{\rm Type:}&&\{S,N,N\}
\end{eqnarray}
The energy--momentum tensor
\begin{eqnarray}
({T^\alpha}_\beta)= \left[
\begin {array}{ccc} -\frac{1}{32\pi}\,{\frac {{Q}^{2} \left( {l}^{2}+{\omega}^{2}
 \right) }{{\rho}^{2} \, \left( {l}^{2}-{\omega}^{2} \right) }}
 &0&\frac{1}{16\pi}\,{\frac {{l}^{2}\omega\,{Q}^{2}}{{\rho}^{2}
 \, \left( {l}^{2}-{
\omega}^{2} \right) }}\\\noalign{\medskip}0&
-\frac{1}{32\pi}\,{\frac {{Q}^{2}}{{
\rho}^{2}}}&0\\\noalign{\medskip}-\frac{1}{16\pi }\,{\frac {\omega\,{Q}^{2}}{{
\rho}^{2} \, \left( {l}^{2}-{\omega}^{2} \right) }}
&0&\frac{1}{32\pi}\,{\frac
{{Q}^{2} \left( {l}^{2}+{\omega}^{2} \right) }{{\rho}^{2} \,
 \left( {l}^{2}-{\omega}^{2} \right) }}\end {array} \right],
\end{eqnarray}

possesses the following eigenvalues and eigenvectors
\begin{eqnarray}
\lambda_{1}&&=\frac{1}{32\pi}\,{\frac {{Q}^{2}}{{\rho}^{2} }};
{\bf V}1=[{V^{1}}={V^{3}}\,\omega,{V^{2}}=0,{V^{3}}={V^{3}}],
\,\nonumber\\&&
V_{\mu}V^\mu ={\frac {{{V^{3}}
}^{2}{\rho}^{2} \left( {l}^{2}-{\omega}^{2} \right) }{{l}^{2}}},
\,{\bf V}1={\bf S}1,\nonumber\\
\lambda_{2,3}&&=-\frac{1}{32\pi}\,{\frac {{Q}^{2}}{{\rho}^{2} }};
{\bf V}2=[{V^{1}}={\frac {{l}^{2}{V^{3}}}{\omega}},
{V^{2}}={V^{2}},{V^{3}}={V^{3}}],
\nonumber\\&&
V_{\mu}V^\mu =-{\frac {{l}^{2}\, \left( {l}^{2}
-{\omega}^{2} \right) {
{V^{3}}}^{2}}{{\omega}^{2}}}{\it L^2}
+{\frac {{{V^{2}}}^{2}}{{\it L^2}}},\nonumber\\
&&
{\bf V}2=\{{\bf T}2,\,{\bf N}2, {\bf S}2\},\,
{\bf V}3=\{{\bf T}3,\,{\bf N}3,\,{\bf S}3\},\nonumber\\
{\rm Type:}&&\{S,2N\}
\end{eqnarray}
$${\rm Type:}\{S,2N\}$$
For the Cotton tensor
\begin{eqnarray}
({C^\alpha}_\beta)= \left[
\begin {array}{ccc} -\frac{1}{8}\,{\frac { \left({\rho}
^{2}+ {l}^{2}{\it L^2} \right)
\omega\,{Q}^{2}}{{\rho}^{4} \left( {l}^{2}-{\omega}^{2}
 \right) }}&0&\frac{1}{8}\,{\frac { \left( {\rho}^{
2} +{\omega}^{2}{\it L^2}\right) {l}^{2}{Q}^{2}}{{\rho}^{4}
\left( {l}^{2}-{\omega}^{2}
 \right) }}\\\noalign{\medskip}0&0&
 0\\\noalign{\medskip}-\frac{1}{8}\,{\frac {
 \left({\omega}^{2}{\rho}^{2}+ {l}^{4}{\it L^2} \right) {Q}^{2}}{
{\rho}^{4}{l}^{2} \left( {l}^{2}-{\omega}^{2} \right) }}
&0&\frac{1}{8}\,{
\frac { \left( {\rho}^{2}+{l}^{2}{\it L^2} \right) \omega\,{Q}^{2
}}{{\rho}^{4} \left( {l}^{2}-{\omega}^{2} \right) }}\end {array}
 \right],
\end{eqnarray}
the eigenvalues and the corresponding eigenvectors are
\begin{eqnarray}
\lambda_{1}&&=0;
{\bf V}1=[{V^{1}}=0,{V^{2}}={V^{2}},{V^{3}}=0],\,
V_{\mu}V^\mu =\frac{{V^{2}}^2}{\it L^2}
\,{\bf V}1={\bf S}1,\nonumber\\
\lambda_2&&=\frac{1}{8}\,{\frac {{Q}^{2}
\sqrt {-{\it L^2}}}{{\rho}^{3}}};{\bf V}2
=[{V^{1}}={V^{1}},{V^{2}}=0,{V^{3}}
={\frac {{V^{1}}\, \left( \omega\,\rho+
\sqrt {-{\it L^2}}{l}^{2} \right) }{{l}^{2}
 \left( \rho+\sqrt {-{\it L^2}}\omega \right) }}],\,{\bf V}2={\bf Z},
\nonumber\\
\lambda_3&&=-\frac{1}{8}\,{\frac {{Q}^{2}
\sqrt {-{\it L^2}}}{{\rho}^{3}}};{\bf V}3
=[{V^{1}}={V^{1}},{V^{2}}=0,{V^{3}}
=\frac {{V^{1}}\,
\left( \omega\,\rho-\sqrt {-{\it L^2}}{l}^{2} \right) }{{l}^{2}
 \left(\rho -\sqrt {-{\it L^2}}\omega\right) }],\,
{\bf V}3={\bf \bar Z},\nonumber\\
{\rm Type:}&&\{S,Z,\bar Z\}.
\end{eqnarray}

\section{Clement spinning solution}

Clement rotating charged solution~\cite{Clement96} is defined by the metric
functions
\begin{eqnarray}
ds^2&&=-N(\rho)^2dt^2+\frac{1}{L(\rho)^{2}}\,d\rho^2
+K(\rho)^2[d\phi+W(\rho)dt]^2,\nonumber\\
H(\rho)&:=&\rho^2+4\pi\,G\,\omega^2\,{Q^2}
\ln{(\frac{\rho^2}{\rho_{0}^2})},\nonumber\\
F(\rho)&:=&\frac{\rho^2}{l^2}
-\frac{4\pi\,G\,(l^2-\omega^2)\,{Q^2}}{l^2}
\ln{(\frac{\rho^2}{\rho_{0}^2})},\nonumber\\
L(\rho)&=&\sqrt{F(\rho)},
K(\rho)=\sqrt{H(\rho)},
N(\rho)=\rho\,\sqrt{\frac{F(\rho)}{H(\rho)}},\nonumber\\
W(\rho)&=&-\omega\frac{4\pi\,G\,Q^2}
{H(\rho)}\ln{(\frac{\rho^2}{\rho_{0}^2})},
\end{eqnarray}
see also \cite{GarciaAnnals09}~ Eq.~(11.24).\\

\subsection{Mass, energy and momentum for the Clement
spinning solution}\label{ClementSpin}

The evaluation of the surface energy and momentum densities yield
\begin{eqnarray}\label{den-Cl1}
\epsilon(\rho, \epsilon_{0})&=&-\frac{1}{\pi\rho}
\,\frac{\sqrt{F(\rho)}}{H(\rho)}
\,\left(\rho^2+4\pi\,G\,\omega^2\,{Q^2}\right)
-\epsilon_{0},\nonumber\\
j(\rho)&=&-4\,G\,\omega\,{Q^2}
\rho[1-\ln{(\frac{\rho^2}{\rho_{0}^2})}]\frac{1}{\sqrt{H(\rho)}},
\end{eqnarray}
while the integral quantities can be evaluated from the generic
expressions
\begin{eqnarray*}\label{den-Cl2}
J(\rho)=2\,\pi\,K(\rho)\,j(\rho),
\,E(\rho)=2\,\pi\,K(\rho)\,\epsilon(\rho),\,
M(\rho)=N(\rho)\,E(\rho)-W(\rho)\,J(\rho).
\end{eqnarray*}

In this  manner one arrives at
\begin{eqnarray}\label{den-Cl3}
&&J(\rho)=-8\pi\,G\,\omega\,{Q^2}
(1-\ln{(\frac{\rho^2}{\rho_{0}^2})}),\nonumber\\
&&E(\rho,\epsilon_{0})=-2\frac{L}{\rho\,K}
\,\left(\rho^2+4\pi\,G\,\omega^2\,{Q^2}\right)
-2\,\pi\,K(\rho)\,\epsilon_{0},\nonumber\\
&&M(\rho,\epsilon_{0})=-2\frac{\rho^2}{l^2}
-8\frac{\pi\,G\omega^2\,Q^2}{l^2}
+8\pi\,G
\,Q^2\ln{(\frac{\rho^2}{\rho_{0}^2})}
-2\,\pi\,K(\rho)\,N(\rho)\,\epsilon_{0}.
\end{eqnarray}
The evaluation of the corresponding functions for the base energy
$\epsilon_{0}=0$ yield at spatial infinity  $\rho\rightarrow{\infty}$
\begin{eqnarray}
j({\rho\rightarrow\infty})&\approx&-4\frac{G\,\omega\,{Q^2}}{\rho}
+8\frac{G\,\omega{Q^2}}{\rho}\ln{(\frac{\rho}{\rho_{0}})}
,\nonumber\\
J({\rho\rightarrow\infty})&\approx&
-8\pi\,G\,\omega\,{Q^2}(1-2\ln{(\frac{\rho}{\rho_{0}})})
,\nonumber\\
\epsilon({\rho\rightarrow\infty},
\epsilon_{0}=0)&\approx&
-\frac{1}{\pi\,l}
-4\frac{G\,\omega^2\,{Q^2}}{l\rho^2}
+4\frac{G\,{Q^2}}{l\rho^2}
(l^2+\omega^2)\ln{(\frac{\rho}{\rho_{0}})},\nonumber\\
E({\rho\rightarrow\infty},\epsilon_{0}=0)
&\approx&-\frac{2\rho}{l}
-8\frac{\pi\,G\,\omega^2\,{Q^2}}{l\rho}
+8\frac{\pi\,G\,{Q^2}l}{\rho}\ln{(\frac{\rho}{\rho_{0}})}
,\nonumber\\
M({\rho\rightarrow\infty},\epsilon_{0}=0)& \approx&\,-2\frac{\rho^2}{l^2}
-8\frac{\pi\,G\omega^2\,Q^2}{l^2}
+16\pi\,G\,Q^2\ln{(\frac{\rho}{\rho_{0}})}.
\end{eqnarray}
Using in the expressions (\ref{den-Cl1}) and (\ref{den-Cl3}) as the reference energy
density the quantity
$\epsilon_{0}=-\frac{1}{\pi\rho}\sqrt{-M_{0}+\frac{\rho^2}{l^2}}$,
which at the spatial infinity behaves as $\epsilon_{0\mid\infty}(M_{0})\approx
-\frac{1}{\pi\,l}+\frac{lM_{0}}{2\pi\,\rho^2}$, the series
expansions of the corresponding quantities at $\rho\rightarrow{\infty}$
result in
\begin{eqnarray}
\epsilon({\rho\rightarrow\infty},\epsilon_{0\mid\infty}(M_{0}))&\approx&\,
-\frac{l\,M_{0}}{2\pi\,\rho^2}-4\frac{G\,\omega^2\,{Q^2}}{l\rho^2}
+4\frac{G\,{Q^2}}{l\rho^2}(l^2+\omega^2)\ln{(\frac{\rho}{\rho_{0}})}
,\nonumber\\
E({\rho\rightarrow\infty},\epsilon_{0\mid\infty}(M_{0}))&\approx&\,
-\frac{l\,M_{0}}{\rho}-8\frac{\pi\,G\,\omega^2\,{Q^2}}{l\rho}
+8\frac{\pi\,G\,{Q^2}}{l\rho}(l^2+\omega^2)\ln{(\frac{\rho}{\rho_{0}})},\nonumber\\
M({\rho\rightarrow\infty},\epsilon_{0\mid\infty}(M_{0}))&\approx&
\,-M_{0}-8\frac{\pi\,G\omega^2\,Q^2}{l^2}
+8\pi\,G\,Q^2\frac{l^2+\omega^2}{l^2}\ln{(\frac{\rho}{\rho_{0}})}.
\end{eqnarray}
Comparing with the energy characteristics of the BTZ
solution, one concludes that a mass parameter $M$ similar to the BTZ mass
is absent, instead a term in the mass function due to the product of the
rotation $\omega$ and the charge $Q$ is present. Notice that
$E(\rho)$ and $M(\rho)$ logarithmically diverges at
spatial infinity. The momentum parameter is due to the
product of $\omega\,Q$, and hence is not a free parameter.

\subsection{Cotton tensor for the Clement spinning solution}

The Cotton characterization of this solution is given by
\begin{eqnarray}
Cotton= \left[ \begin {array}{ccc} \frac{1}{8}\,{
\frac {\omega\, \left( F \left( \rho \right) {l}^{2}+{\rho}^{2}
 \right) }{\rho\,
 \left( {l}^{2}-{\omega}^{2} \right) }}
 {\frac {d^{3}}{d{\rho}^{3}}}F \left( \rho \right)&0&-\frac{1}{8}\,
 {\frac { \left( F\left( \rho \right) {\omega}^{2}
 +{\rho}^{2} \right) {l}^{2} }{\rho\, \left( {l}^{2}
 -{\omega}^{2} \right) }}{\frac {d^
{3}}{d{\rho}^{3}}}F \left( \rho \right)\\\noalign{\medskip}0&0&0
\\\noalign{\medskip}\frac{1}{8}
\,{\frac { \left( F \left( \rho \right) {l}^{4}+{\omega}^{2}{\rho}^{2}
 \right) }{{l}^{2}
\rho\, \left( {l}^{2}-{\omega}^{2} \right) }}
{\frac {d^{3}}{d{\rho}^{3}}}F \left( \rho \right)
&0&-\frac{1}{8}\,{\frac {\omega\,
 \left( F \left( \rho \right) {l}^{2}
 +{\rho}^{2} \right)  }{\rho\, \left( {l}^{2}-{\omega}^
{2} \right) }}{\frac {d^{3}
}{d{\rho}^{3}}}F \left( \rho \right)\end {array} \right] ,
\end{eqnarray}
where
\begin{eqnarray}
F(\rho)={\frac {{\rho}^{2}}{{l}^{2}}}+ {4\,{Q}^{2}\pi\,{\it gr}\,
\ln  \left( { \rho_{0}^{2}} \right) -4\,{Q}^{2}\pi\,{\it gr}\,\ln
 \left( {\rho}^{2} \right) },\,
\frac {d^{3}
}{d{\rho}^{3}}F \left( \rho \right)=
-16\,{\frac {{Q}^{2}\pi\,{\it gr}}{{\rho}^{3}}}
\end{eqnarray}

\begin{eqnarray}
\lambda_{1}&&=0;\nonumber\\&&
{\bf V}1=[{V^{1}}=0,{V^{2}}={V^{2}},{V^{3}}=0],
\,V_{\mu}V^\mu ={{V^{2}}}^{2},\,{\bf V}1={\bf S}1,\nonumber\\
\lambda_2&&=2\,{\frac {\sqrt {-F
 \left( \rho \right) }{Q}^{2}\pi\,{\it gr}}{{\rho}^{3}}};
\nonumber\\&&
{\bf V}2=[{V^{1}}={V^{1}},{V^{2}}=0,{V^{3}}
={\frac {{V^{1}}\, \left(
\omega\,\rho+\sqrt {-F \left( \rho \right) }{l}^{2} \right) }{{l}^{2}
 \left( \rho+\sqrt {-F \left( \rho \right) }\omega \right) }}]
 ,\,{\bf V}2={\bf { Z}},\nonumber\\
\lambda_{3}&&=
-2\,{\frac {\sqrt {-F \left( \rho \right) }{Q}^{2}\pi\,{\it gr}}{{\rho}^{3}}};
\nonumber\\&&
{\bf V}3=[{V^{1}}={V^{1}},{V^{2}}=0,{V^{3}}
={\frac {{V^{1}}\, \left(
\omega\,\rho-\sqrt {-F \left( \rho \right) }{l}^{2} \right) }{{l}^{2}
 \left( \rho-\sqrt {-F \left( \rho \right) }\omega \right) }}]
 ,\,{\bf V}3={\bf {\bar Z}}.
\end{eqnarray}

\section{Garcia solution}

The Garcia solution \cite{GarciaAnnals09}~ Eq.~(10.9),
is defined by the metric functions
\begin{eqnarray}
ds^2&&=-N(\rho)^2dt^2+\frac{1}{L(\rho)^{2}}\,d\rho^2
+K(\rho)^2[d\phi+W(\rho)dt]^2,\nonumber\\
H(\rho)&:=&\frac{H_{n}}{H_{d}};\nonumber\\
H_{n}&=&4\,\rho^2\,({\rho^2-M\,\l^2})(M^2\,l^2-J^2)-{J^2Q^4}\,{l^6}\,
{R_{-}^2}\,(\ln{\mid Z(\rho)\mid})^2\nonumber\\
&&
-2{Q^2}\,{l^3}\sqrt{M^2\,l^2-J^2}\,
[M\,J^2\,l^2-2\,\rho^2R_{-}\sqrt{M^2\,l^2-J^2}]
\,\ln{\mid Z(\rho)\mid},\nonumber\\
H_{d}&=&4\,(M^2\,l^2-J^2)
(\rho^2-M\,\l^2)-2\,l^3Q^2\,R_{-}^2\sqrt{M^2\,l^2-J^2}
\,\ln{\mid Z(\rho)\mid},\nonumber\\
L(\rho)^2&=&\frac{\rho^2}{l^2}-M
+\frac{J^2}{4\,\rho^2}+\frac{l\,Q^2}{2\,\rho^2}(2\rho^2\,R_{-}-l\,J^2)
\ln{\mid Z(\rho)\mid},\nonumber\\
K(\rho)^2&=&{H(\rho)},\nonumber\\
N(\rho)^2&=&\rho^2\frac{L(\rho)^2}{{H(\rho)}},\nonumber\\
W(\rho)\,H_{n}&=&J\,Q^4\,l^5\,R_{-}^3\,(\ln{\mid Z(\rho)\mid})^2
+Q^2\,l^2\,J\,\sqrt{M^2\,l^2-J^2}
\,[J^2\,l+2\,l\,R_{-}^2-2\rho^2\,R_{-}]
\ln{\mid Z(\rho)\mid}\nonumber\\
&&-2\,J\,({M^2\,l^2-J^2})(\rho^2-M\,\l^2),\nonumber\\
Z(\rho)&:=&\rho^2-\frac{l}{2}(M\,l-\sqrt{M^2\,l^2-J^2})
=\,\rho^2-\frac{l\,R_{-}}{2},\nonumber\\
R_{\pm}&:=&\,M\,l\pm\sqrt{M^2\,l^2-J^2},
\end{eqnarray}
with electromagnetic vector potential
\begin{eqnarray}
\bm{ A}:=-\frac{\sqrt{l}\,Q}{\sqrt{2}}
\frac {\ln{\mid\rho\mid}}{\sqrt [4]{{l}^{2}{M}^{2}-{J}^{2}}}
\left(R_{-}\bm{ dt}-\frac{J}{l}\bm{ d\phi}\right),
\end{eqnarray}
therefore the non--vanishing covariant
components of the electromagnetic field tensor $F_{\mu\nu}$ are
\begin{eqnarray}
F_{t\rho}=\frac{\sqrt{l}\,Q}{\sqrt{2}}
\frac {R_{-}}{\sqrt [4]{{l}^{2}{M}^{2}
-{J}^{2}}}\frac{1}{\rho},\,F_{\phi\rho}
=\frac{Q}{\sqrt{2}}\frac{{J}}{\sqrt{l}}
\frac {1}{\sqrt [4]{{l}^{2}{M}^{2}-{J}^{2}}}\frac{1}{\rho}.
\end{eqnarray}

Notice that the above gravitational--electromagnetic field,
as it was pointed previously out, when the electromagnetic
field is switched off, $Q=0$, becomes the rotating BTZ solution,
while for vanishing rotation, $J=0$, the corresponding
solution is represented by the static BTZ metric, i.e.,
the AdS metric with $M$--parameter.\\

\subsection{ Mass, energy and momentum for the
Garcia black hole}\label{Garciahole}

The evaluation of the surface energy density
at spatial infinity $\rho\rightarrow\infty$  for $\epsilon_{0}=0$
yields
\begin{eqnarray}
\epsilon({\rho\rightarrow\infty},\epsilon_{0}=0)
&\approx&-\frac{1}{\pi\,l}
+\,\frac{l\,M}{2\pi\rho^2}
-\,\frac{l^2\,J^2\,Q^2}{2\pi\rho^2\,\sqrt{M^2l^2-J^2}}
+\frac{l^3\,Q^2\,M}{\pi\,\rho^2\,\sqrt{M^2l^2-J^2}}
\,R_{-}\,\ln{(\rho)},\nonumber\\
\end{eqnarray}
while the momentum quantities amount to
\begin{eqnarray}
j({\rho\rightarrow\infty})&\approx&\frac{J}{2\pi\,\rho}
-\,\frac{l^2\,{J}Q^2}{2\pi\rho\,\sqrt{M^2l^2-J^2}}\,R_{-}
+\frac{l^2{J}Q^2}{\pi\,\rho\,\sqrt{M^2l^2-J^2}}
\,R_{-}\,\ln{(\rho)}\nonumber\\
J({\rho\rightarrow\infty})&\approx&\,{J}
-\,\frac{l^2\,{J}Q^2}{\sqrt{M^2l^2-J^2}}\,R_{-}
+\frac{2\,l^2{J}Q^2}{\sqrt{M^2l^2-J^2}}\,R_{-}\,\ln{(\rho)}
.\nonumber\\
\end{eqnarray}
The integral energy and mass
characteristic at spatial infinity can
be evaluated from the generic expressions
$E(\rho)=2\,\pi\,K\,\epsilon(\rho),
\, M(\rho)=N\,E(\rho)-W\,J(\rho).
$

The evaluation of the corresponding
functions with $\epsilon_{0}=0$
behave at infinity as $\rho\rightarrow\infty$ according to
\begin{eqnarray}
E({\rho\rightarrow\infty},\epsilon_{0}=0)&
\approx&-\frac{2\rho}{l}+\frac{l\,M}{\rho}
-\frac{l^2\,J^2\,Q^2}{\rho\sqrt{M^2l^2-J^2}}
+\frac{l^2\,Q^2}{\rho\,\sqrt{M^2l^2-J^2}}\,R_{-}^2\,\ln{(\rho)}
,\nonumber\\
M({\rho\rightarrow\infty},\epsilon_{0}=0)&
\approx&-2\frac{\rho^2}{l^2}
+2\,M-\,\frac{l\,J^2\,Q^2}{\sqrt{M^2l^2-J^2}}
+\frac{2\,l\,Q^2}{\sqrt{M^2l^2-J^2}}
\,R_{-}^2\,\ln{(\rho)}.\nonumber\\
\end{eqnarray}
The series expansions of the expressions of
the energy and mass evaluated for the reference energy
density
$\epsilon_{0}=-\frac{1}{\pi\rho}\sqrt{-M_{0}+\frac{\rho^2}{l^2}}$, which
at the spatial infinity behaves as
$\epsilon_{0\mid\infty}(M_{0})\approx
-\frac{1}{\pi\,l}+\frac{l\,M_{0}}{2\pi\,\rho^2}$,
at $\rho\rightarrow\infty$ occur to be
\begin{eqnarray}
\epsilon({\rho\rightarrow\infty},
\epsilon_{0\mid\infty}(M_{0}))&
\approx&\frac{l}{2\pi\rho^2}(M-M_{0})
-\,\frac{l^2\,J^2\,Q^2}{2\pi\rho^2\,\sqrt{M^2l^2-J^2}}
+\frac{l^3Q^2\,M\,R_{-}}{\pi
\,\rho^2\,\sqrt{M^2l^2-J^2}}\,\ln{(\rho)},
\nonumber\\
E({\rho\rightarrow\infty},
\epsilon_{0\mid\infty}(M_{0}))&\approx&\,\frac{l(M-M_{0})}{\rho}
-\,\frac{l^2\,J^2\,Q^2}{\rho\sqrt{M^2l^2-J^2}}+\frac{2\,l^3Q^2\,M}{
\rho\,\sqrt{M^2l^2-J^2}}\,R_{-}\,\ln{(\rho)},\nonumber\\
M({\rho\rightarrow\infty},\epsilon_{0\mid\infty}(M_{0}))&\approx&\,M-M_{0}
-\,\frac{l\,J^2\,Q^2}{\sqrt{M^2l^2-J^2}}
+\frac{2\,l^2\,M\,Q^2}{\sqrt{M^2l^2-J^2}}\,R_{-}\,\ln{(\rho)}.
\end{eqnarray}\\

For vanishing electromagnetic field charge $Q$, which gives rise to
the rotating BTZ black hole, the mass, and the energy--momentum
quantities become just the mass--energy--momentum characteristics of
the BTZ solution Eqs.~(\ref{BTZj})--(\ref{enerBTZapproxEps1}), hence
one concludes that the parameters $M$ and $J$ are related with the
mass and momentum respectively. Moreover in the electromagnetic
case, the momentum, mass, energy functions logarithmically diverges
at spatial infinity.

\subsection{Field, energy--momentum, and
Cotton tensors for the Garcia solution}

To determine the eigenvector structure of
the Garcia solution it is more convenient to use
another of its representation in the
coordinates $\{\tau,r,\sigma\}$, namely
\begin{eqnarray}
g= \left[
\begin {array}{ccc} -{F }/{H }+H^{2}&0&H W \\
\noalign{\medskip}0&  1/F&0\\
\noalign{\medskip}H W &0&H \end {array}
 \right] ,
\end{eqnarray}
where the metric functions are
\begin{eqnarray}
&&F(r)=4\,{\frac {{r}^{2}}{{l}^{2}}}+2\,\frac{r}{l}
\,{ { \left( l{\it w_{1}}+{
\sqrt {{l}^{2}{{\it w_{1}}}^{2}-4}} \right)
\left( {\it w_{0}}+{\it W0}\,\ln  \left( r \right)
 \right) }},\nonumber\\
 &&H(r)=-{\it w_{0}}-{\it W0}\,
 \ln  \left( r \right) +{\frac {r}{l}}
 \sqrt {{l}^{2}{{\it w_{1}}}^{2}-4},\nonumber\\
 &&
W(r)= \frac{\left( {\it w_{0}}+{\it W0}\,\ln  \left( r \right) +{\it
w_{1}}\,r \right)}{\left( -{\it w_{0}}-{\it W0}\,\ln \left( r \right)
+{\frac {r}{l}} \sqrt {{l}^{2}{{\it w_{1}}}^{2}-4}\,\right)},\nonumber\\
 &&
 {\it W0}=-\frac{{l}^{2}{\alpha}^{2}}{2}
 \left( {l}^{2}{{\it w_{1}}}^{2}-2-
l{\it w_{1}}\,\sqrt {{l}^{2}{{\it w_{1}}}^{2}-4}\right)
=-\frac{{l}^{2}{\alpha}^{2}}{4}
\left( l{\it w_{1}}-\sqrt {{l}^{2}{{\it w_{1}}}^{2}-4} \right)^2.
\end{eqnarray}
To achieve the metric structure studied in the
previous paragraph, one subject the above metric to the
coordinate transformation
$$\tau=\frac{1}{\sqrt {2\,{J}{l}}}
\frac{\left( J\,t-{l}^{2}M\,\phi \right)}
{\sqrt [4]{{l}^{2}{M}^{2}-{J}^{2}}} ,r
=-{\rho}^{2}+\frac{{l}^{2}M}{2}\,
+\frac{l}{2}\,\sqrt {{l}^{2}{M}^{2}-{J}^{2}},\sigma
=\frac{\sqrt{l}}{\sqrt {2}\sqrt{J}}
\sqrt [4]{{l}^{2}{M}^{2}-{J}^{2}}\,\phi,$$
together with
$${\it w_{0}}
=-\frac{\sqrt{{l}^{2}{M}^{2}-{J}^{2}}}{J}R_{-},
\,{\it w_{1}}=2\,{\frac {M}{J}},\,{\it W0}=-\frac{{l}^{2}{
\alpha}^{2}}{J^2}R_{-}^2,\,R_{-}
=\left( lM-\sqrt {{l}^{2}{M}^{2}-{J}^{2}} \right).
$$
followed by the change of the charge $\alpha\rightarrow J^{1/2}Q$.

In these coordinates, the  Maxwell electromagnetic field tensor is given by
\begin{eqnarray}
\left({F^\alpha}_{\beta}\right)= \left[
\begin {array}{ccc} 0&{\alpha}/{F }&0
\\\noalign{\medskip}\alpha\, \left[ F -
 H^2  W(W -1) \right] /{H }&0&-H \alpha
\, \left( W -1 \right) \\\noalign{\medskip}0&-
{\alpha}/{F }&0\end {array} \right],
\end{eqnarray}
and it is characterized by the following set of
eigenvectors
\begin{eqnarray}
\lambda_{1}&&=0;{\bf V}1=[{V^{1}}={\frac {
 H^2 \left( W -1\right) }{F -  H^{2} W(W-1)}}
 {V^{3}},0,{V^{3}}],\,\nonumber\\&&
 V_{\mu}V^\mu ={\frac {H F  \left(F- H^2 \left( W -1
 \right) ^{2}  \right) }{ \left[F - H^2W
 \left( W -1 \right)  \right] ^{2}}}
 {{V^{3}}}^{2},\,{\bf V}1={\bf T}1,\,{\bf S}1,\nonumber\\
\lambda_2&&={\frac {\sqrt {F - H^{2} (W-1)^{2} }
}{\sqrt{H F }}}\alpha;
\nonumber\\&&
{\bf V}2=[{V^{1}}={\frac {\alpha\,{V^{2}}}{{\it \lambda_2}
\,F \left( r
 \right) }},{V^{2}},{V^{3}}
 =-{\frac {\alpha\,{V^{2}}}{{\it
\lambda_2}\,F }}],\,V^{\mu}V_{\mu}=0,\,{\bf V}2
={\bf N}2,{\bf { Z}},\nonumber\\
\lambda_{3}&&=-{\frac {\sqrt {F - H^{2} (W-1)^{2} }
}{\sqrt{H F }}}\alpha;
\nonumber\\&&
{\bf V}3=[{V^{1}}={\frac {\alpha\,{V^{2}}}{{\it \lambda_3}\,F }}
,{V^{2}},{V^{3}}=-{\frac {\alpha\,{V^{2}}}{{\it
\lambda_3}\,F }}],\,V^{\mu}V_{\mu}=0,\,{\bf V}3
={\bf N}3,{\bf {\bar Z}}.
\end{eqnarray}
For the Maxwell energy--momentum tensor
\begin{eqnarray}
\left({T^\alpha}_{\beta}\right)= \left[
\begin {array}{ccc} -\frac{1}{8\pi}\,
{\frac {{\alpha}^{2} \left[ F - H^2 \left( W^2  -1 \right)\right] }{F
   \,H }}&0&\frac{1}{4\pi}\,
   {\frac {H  {\alpha}^{2} \left( W -1 \right) }{F
  }}\\\noalign{\medskip}0&-\frac{1}{8\pi}\,{\frac {{\alpha}^{2
} \left[ F - H^2
 \left( W -1 \right) ^{2} \right] }
 {F  \,H }}&0\\\noalign{\medskip}\frac{1}{4\pi}\,{
\frac {{\alpha}^{2} \left[ F - H ^{2}W  \left( W -1
 \right)  \right] }{F \,H }}&0&
\frac{1}{8\pi}\,{\frac {{\alpha}^{2}
\left[ F -  H^{2} \left( W -1 \right)
 \left( W +1 \right)  \right] }{F
 \,H }}\end {array} \right] ,
\end{eqnarray}
one has the following eigenvalues and eigenvectors
\begin{eqnarray}
\lambda_{1,2}&&=-\frac{1}{8\pi }\,{\frac {{\alpha}^{2}
\left[ F - H^{2} \left( W -1 \right)^{2} \right] }{F \,H }};
{\bf V}1,2=[{V^{1}}=-{V^{3}},{V^{2}}={V^{2}},{V^{3}}={V^{3}}],
\,\nonumber\\&&
V_{\mu}V^\mu ={\frac {{{V^{2}}}^{2}}{F }}
-{\frac {{{V^{3}}}^{2}[F-H^2 \left( W -1 \right) ^{2}]}{H }},
\,{\bf V}1={\bf T}1,\,{\bf S}1,{\bf V}2={\bf T}2,\,{\bf S}2,\nonumber\\
\lambda_3&&=\frac{1}{8\pi }\,{\frac {{\alpha}^{2}
\left[ F - H^{2} \left( W -1 \right)^{2} \right] }{F \,H }};{\bf V}3=
[{V^{1}}={\frac { H^2{V^{3}}\,
 \left( W -1 \right) }{F -  H^{2} W(W-1)}},{V^2}=0,{V^{3}}={V^{3}}],
\nonumber\\&&
V_{\mu}V^\mu =-{\frac {{{V^{3}}}^{2}H F  \left[
F - H^2 \left( W -1 \right) ^{2} \right] }{ \left[ F - H^2W
 \left( W -1 \right)  \right] ^{2}}},{\bf V}3={\bf T}3,\,{\bf S}3
\end{eqnarray}

Finally, the Cotton tensor
\begin{eqnarray}
\left({C^\alpha}_{\beta}\right)= \left[
\begin {array}{ccc} {{C^1}_{1}}&0&{{C^1}_{3} }\\\noalign{\medskip}0&{
{C^2}_{2}}&0\\\noalign{\medskip}{{C^3}_{1} }&0&-{{C^1}_{1}}-{{C^2}_{2}}\end {array}
 \right],
\end{eqnarray}
\begin{eqnarray}
{C^1}_{1}&&=-\frac{1}{32\,\pi }\,{\frac {H W {\alpha}^{2}
 \left( W -1 \right) ^{2} \left(  HF_{,r} -F
H_{,r}\right) }{ F^{2}}}+\frac{1}{32\,\pi }\,{\frac {{\alpha}^{2}
\left( W -1 \right) F_{,r} }{F }}\nonumber\\&&
 -\frac{1}{32\,\pi }\,{\frac {{\alpha}^{2} \left( 3\,W-2 \right) H_{,r} }{
\,H }}-\frac{1}{16\,\pi }\,{\alpha}^{2}W_{,r},
\end{eqnarray}
\begin{eqnarray}
{C^1}_{3}=-\frac{1}{32\,\pi }\,{\frac {{\alpha}^{2}H
\left( W  -1 \right)^{2} \left(  H \,F_{,r}  -F\,H_{,r}  \right) }
{ F ^{2}}}-\frac{1}{32\,\pi }\,{\frac {{\alpha}^{2}H_{,r} }
{\,H }},
\end{eqnarray}
\begin{eqnarray}
{C^2}_{2}=-\frac{1}{16\,\pi }\,{\frac {{\alpha}^{2} \left( W -1 \right)
 \left(   F_{,r} H -2\,F H_{,r}
 \right) }{F\,H }}+\frac{1}{16\,\pi }\,{
\frac {{\alpha}^{2} \left(  H^2
 \left( W -1 \right) ^{2}+F
 \right) W_{,r} }{F }},
\end{eqnarray}
\begin{eqnarray}
{C^3}_{1}&&=\frac{1}{32\,\pi }\,{\frac {{\alpha}^{2} \left( W -1 \right)
 \left( -F  \left( 1+W  \right) +
 W^2  H^{2} \left( W -1 \right)  \right)
F_{,r} }{ F^{2}}}\nonumber\\&&
-\frac{1}{32\,\pi }\,{\frac { \left( F +  H ^{2} W^2
 \right) {\alpha}^{2} \left( -F +  H ^{2} \left( W -1 \right) ^{2}
 \right) H_{,r} }{ H ^{3}F}},
 \nonumber\\&&
 -\frac{1}{16\,\pi }\,{\frac {W
 {\alpha}^{2} \left( -F + H^{2} \left( W -1 \right) ^{2} \right) W_{,r} }{F}},
\end{eqnarray}
\begin{eqnarray}
{C^3}_{3} &&=\frac{1}{32\,\pi }
\,{\frac { H^2W {\alpha}^{2} \left( W -1 \right) ^{2}F_{,r} }{ F^{2}}}
-\frac{1}{32\,\pi }\,{\frac {{\alpha}^{2} \left( H ^{2}W  \left( W -1\right) ^{2}
+F  \left( W -2\right)  \right) H_{,r} }{F\,H }}
 \nonumber\\&&
 -\frac{1}{16\,\pi }\,{\frac {{\alpha}^{2}
 H^2 \left( W -1\right) ^{2}W_{,r} }{F}},
 \end{eqnarray}
possesses, in general, three different eigenvalues, with
the possibility of complex conjugated roots, namely
\begin{eqnarray}
&&\lambda_{1}={{C^2}_{2}},\nonumber\\&&
\lambda_{2}=-1/2\,{{C^2}_{2}}+1/2\,\sqrt {({C^1}_{1}
+{C^2}_{2})^2+4\,{{C^1}_{3}}\,{{C^3}_{1}}},\nonumber\\&&
\lambda_{3}=-1/2\,{{C^2}_{2}}-1/2\,\sqrt {({C^1}_{1}
+{C^2}_{2})^2+4\,{{C^1}_{3}}\,{{C^3}_{1}}}].
\end{eqnarray}
The set of eigenvector equations reduces to
\begin{eqnarray}
&&{V^{1}}\,({C^1}_{1}-\lambda)+{C^1}_{3}\,{V^{3}}=0,
\nonumber\\&&
 \left( {C^2}_{2}-\lambda
 \right) {V^{2}}=0,
\nonumber\\&&
 {C^3}_{1}\,{V^{1}}-{V^{3}}({C^1}_{1}+\,{C^2}_{2}+\,\lambda)=0.
\end{eqnarray}
with solutions
\begin{eqnarray}
\lambda_{1}&&={C^2}_{2};
{\bf V}1=[{V^{1}}=0,{V^{2}}={V^{2}},{V^{3}}=0],
\,V_{\mu}V^\mu ={V^{2}}^2/F,\,{\bf V}1={\bf S}1,
\nonumber\\
\lambda_{2}&&=-1/2\,{{C^2}_{2}}+1/2\,\sqrt {({C^1}_{1}+{C^2}_{2})^2
+4\,{{C^1}_{3}}\,{{C^3}_{1}}};\nonumber\\&&
{\bf V}2=[{V^{1}}=-{\frac {{{C^1}_{3}}\,{V^{3}}}{{C^1}_{1}-{\lambda_{2}}}},{V^{2}}=0
,{V^{3}}={V^{3}}],\nonumber\\&&
V^{\mu}V_{\mu}={\frac {
 H^2 \left( {{C^1}_{1}}-W {
{C^1}_{3}} \right) ^{2}+  H^{2}{\lambda_{2}}\,
\left( {\lambda_{2}}-2\,{C^1}_{1}+2\,W {{C^1}_{3}} \right)
-({C^1}_{3})^{2}F }
{ \left( {C^1}_{1}-{\lambda_{2}} \right) ^{2}H }}{V^{3}}^{2},
\nonumber\\&&
 {\bf V}2={\bf S}2,{\bf N}2,{\bf { Z}},
 \nonumber\\
\lambda_{3}&&=-1/2\,{{C^2}_{2}}-1/2\,\sqrt {({C^1}_{1}+{C^2}_{2})^2
+4\,{{C^1}_{3}}\,{{C^3}_{1}}};\nonumber\\&&
{\bf V}2=[{V^{1}}=-{\frac {{{C^1}_{3}}\,{V^{3}}}{{C^1}_{1}
-{\lambda_{3}}}},{V^{2}}=0
,{V^{3}}={V^{3}}],\nonumber\\&&
V^{\mu}V_{\mu}={\frac {
 H^2 \left( {{C^1}_{1}}-W {
{C^1}_{3}} \right) ^{2}+  H^{2}{\lambda_{3}}\,
\left( {\lambda_{3}}-2\,{C^1}_{1}+2\,W {{C^1}_{3}} \right)
-({C^1}_{3})^{2}F }
{ \left( {C^1}_{1}-{\lambda_{3}} \right) ^{2}H }}{V^{3}}^{2},
\nonumber\\&&
 {\bf V}3={\bf S}3,{\bf N}3,{\bf {\bar Z}},
\end{eqnarray}

\section{Spinning electro--magnetic solution}

By accomplishing in~\cite{GarciaAnnals09}~(9.13),
the coordinate transformations
$$t\rightarrow C_{1}/2\,t+ j_{0}\phi,\,
\phi\rightarrow\phi,\,r\rightarrow\rho^2+M_{p}-C_{0},$$
one arrives at a stationary electromagnetic solution
\begin{eqnarray}
ds^2&&=-N(\rho)^2dt^2+\frac{1}{L(\rho)^{2}}\,d\rho^2
+K(\rho)^2[d\phi+W(\rho)dt]^2,\nonumber\\
h(\rho)&:=&\frac{\rho^2}{l^2}+\frac{M_{p}-M\,l^2}{l^2}
-b^2\ln{(\rho^2+M_{p})},\nonumber\\
H(\rho)&:=&\rho^2+M_{p}-j_{0}^2\,h(\rho),\nonumber\\
L(\rho)&=&\frac{1}{\rho}\sqrt{(\rho^2+M_{p})h(\rho)},\nonumber\\
K(\rho)&=&\sqrt{H(\rho)},\nonumber\\
N(\rho)&=&\sqrt{\rho^2+M_{p}}\sqrt{\frac{h(\rho)}{H(\rho)}},
\nonumber\\
W(\rho)&=&-j_{0}\frac{h(\rho)}{H(\rho)},
\end{eqnarray}
To establish  the angular deficit
of this metric at the neighborhood of the
rotation point $\rho=0$ one looks at
the behavior of $g_{\phi\phi}=H(\rho)$
as $\rho\rightarrow 0$, which gives rise to the series
$$ {\it Mp} - \left( {\frac {{\it Mp}}{{l}^{2}}}
{-M}-{b}^{2}\ln  \left( {\it Mp} \right) \right) {j_{0}}^{2}
\,+\left(1 -\frac{1}{{l}^{2}}{j_{0}}^{2}
+{\frac {{b}^{2}}{{\it Mp}}}{j_{0}}^{2}\right){\rho}^{2}+...,
$$
since the order zero in $\rho$ has to vanish,
then one gets the condition
\begin{eqnarray}\label{EuclCOND}
{\it Mp} - \left( {\frac {{\it Mp}}{{l}^{2}}}{-M}
-{b}^{2}\ln  \left( {\it Mp} \right) \right) {j_{0}}^{2}=0,
\end{eqnarray}
which solved for ${\it Mp}$ yields
\begin{eqnarray}
{\it Mp}=\exp[-\frac{M}{{b}^{2}}
-{\it LambertW}(\frac{l^2
-j_{0}^{2}}{j_{0}^{2}{b}^{2}{l}^{2}}{\exp}(-\frac {M}{{b}^
{2}}))]
\end{eqnarray}
Using (\ref{EuclCOND}) one brings $H$ to the form
$$
K(\rho)^2=H(\rho)={b}^{2}{j_{0}}^{2}\ln \left( 1
+{\frac {{\rho}^{2}}{{\it Mp}}} \right) +{\frac {{\rho}^{2}
 \left({l}^{2} -{j_{0}}^{2} \right) }{{l}^{2}}},$$
which behaves at $\rho\rightarrow 0$ as it should be as:
$H(\rho\rightarrow 0)=\left(1 -\frac{1}{{l}^{2}}{j_{0}}^{2}
+{\frac {{b}^{2}}{{\it Mp}}}{j_{0}}^{2}\right){\rho}^{2}
+O\left( {\rho}^{4} \right).$
On the other hand $L(\rho)^2$ is given explicitly by
$$L(\rho)^2=\left( {\rho}^{2}+{\it Mp} \right)
\left[ {\frac {{\rho}^{2}}{{l}^{2}}}+{\frac {{\it Mp}}{{j_{0}}^{2}
}}-{b}^{2}\ln  \left( {\frac {{\rho}^{2}+{\it Mp}}{{
\it Mp}}} \right)  \right] {\rho}^{-2}.
$$
and
$\rho\rightarrow 0$ behaves as a polynomial in $\rho$.
Thus the factor $1 -\frac{1}{{l}^{2}}{j_{0}}^{2}
+{\frac {{b}^{2}}{{\it Mp}}}{j_{0}}^{2}$
determines the deficit in the angle $\phi$,
$0\leq \phi\leq2\pi/\sqrt{1 -\frac{1}{{l}^{2}}{j_{0}}^{2}
+{\frac {{b}^{2}}{{\it Mp}}}{j_{0}}^{2}}$,
$\Delta \phi=2\pi(1-1/\sqrt{1 -\frac{1}{{l}^{2}}{j_{0}}^{2}
+{\frac {{b}^{2}}{{\it Mp}}}{j_{0}}^{2}}).$

Moreover this solution  allows for the existence
of a black hole; the vanishing of the
function $h(\rho)$ determines the horizons
\begin{eqnarray}
\rho_{\pm}^2=-{\it Mp}-{b}^{2}{l}^{2}{\it LambertW}
\left( -{\frac {{\it Mp}}{{b}^{2}{l}^{2}}} \right).
\end{eqnarray}

This metric can be considered as a modified Clement
rotating electro--magnetic solution~\cite{Clement93}~(Cl.24).

\subsection{ Mass, energy and momentum for the spinning
electro--magnetic black hole}\label{spinninghole}

The evaluation of the surface energy and momentum
densities yield
\begin{eqnarray}\label{den-CLS1}
\epsilon(\rho)&=&-\frac{1}
{\pi\,l^2}\,\frac{\sqrt{h(\rho)}}{H(\rho)}\frac{
[(l^2-j_{0}^2)(\rho^2+M_{p})+j_{0}^2b^2]}{(\rho^2+M_{p})}
-\epsilon_{0},\nonumber\\
j(\rho)&=&-\frac{j_{0}}{\pi}\,\frac{1}{\sqrt{H(\rho)}}
[M-b^2+b^2\,\ln{(\rho^2+M_{p})}],
\end{eqnarray}
while the integral quantities can be evaluated from the generic
expressions
\begin{eqnarray}\label{den-CLS2}
&&J(\rho)=2\,\pi\,K(\rho)\,j(\rho)=-2{j_{0}}\,
[M-b^2+b^2\ln{(\rho^2+M_{p})}],\nonumber\\
&&E(\rho)=2\,\pi\,K(\rho)\,\epsilon(\rho),\nonumber\\
&&M(\rho)=N(\rho)\,E(\rho)-W(\rho)\,J(\rho)
=-\frac{2\rho^2}{l^2}+2M
-\frac{2}{l^2}M_{p}+2{b^2}{\ln{(\rho^2+M_{p})}}
-2\pi\,NK\epsilon_{0}.\nonumber\\
\end{eqnarray}
The momentum--energy functions, evaluated for $\epsilon_{0}=0$,
behave at spatial infinity $\rho\rightarrow{\infty}$
according to
\begin{eqnarray}
j({\rho\rightarrow\infty})&\approx&-\frac{j_{0}l}{\pi\,\rho\sqrt{l^2-j_{0}^2}}\,
[M-b^2+2b^2\ln{\rho}]
,\nonumber\\
J({\rho\rightarrow\infty})&\approx&-2{j_{0}}\,
[M-b^2+2b^2\ln{\rho}]
,\nonumber\\
\epsilon({\rho\rightarrow\infty},\epsilon_{0}=0)
&\approx&-\frac{1}{\pi\,l}
+\frac{b^2\,l}{\pi\,\rho^2}\frac{l^2+j_{0}^2}{l^2-j_{0}^2}\ln{(\rho)}
+\frac{M\,l}{2\pi\rho^2}\frac{l^2+j_{0}^2}{l^2-j_{0}^2}
-\frac{j_{0}^2}{\pi\,\,\rho^2}\frac{b^2\,l}{l^2-j_{0}^2},
\nonumber\\
E({\rho\rightarrow\infty},\epsilon_{0}=0)
&\approx&-\frac{2\rho}{l^2}\sqrt{l^2-j_{0}^2}
+\frac{Ml^2}{\rho\,\sqrt{l^2-j_{0}^2}}+2\frac{b^2l^2}
{\rho\,\sqrt{l^2-j_{0}^2}}\ln{(\rho)}\nonumber\\
&-&\frac{\sqrt{l^2-j_{0}^2}}{l^2\rho}M_{p}
-2\frac{b^2j_{0}^2}{\rho\sqrt{l^2-j_{0}^2}},
\nonumber\\
M({\rho\rightarrow\infty},\epsilon_{0}=0)&
\approx&\,-\frac{2\rho^2}{l^2}+2M
-\frac{2}{l^2}M_{p}+4{b^2}{\ln{(\rho)}}.
\end{eqnarray}
Using in the expressions (\ref{den-CLS1})
and (\ref{den-CLS2}) as reference energy
density the quantity
$\epsilon_{0}=-\frac{1}{\pi\rho}\sqrt{-M_{0}+\frac{\rho^2}{l^2}}$,
which at the spatial infinity behaves as
$\epsilon_{0\mid\infty}(M_{0})\approx
-\frac{1}{\pi\,l}+\frac{lM_{0}}{2\pi\,\rho^2}$, their series
expansions at
$\rho\rightarrow{\infty}$ result in
\begin{eqnarray}
\epsilon({\rho\rightarrow\infty},
\epsilon_{0\mid\infty}(M_{0}))&
\approx&\,-\frac{lM_{0}}{2\pi\rho^2}
+\frac{b^2\,l}{\pi\,\rho^2}\frac{l^2+j_{0}^2}{l^2-j_{0}^2}\ln{(\rho)}
+\frac{Ml}{2\pi\rho^2}\frac{l^2+j_{0}^2}{l^2-j_{0}^2}
-\frac{j_{0}^2}{\pi\,\rho^2}\frac{b^2\,l}{l^2-j_{0}^2}
,\nonumber\\
E({\rho\rightarrow\infty},\epsilon_{0\mid\infty}(M_{0}))
&\approx&\,-\frac{M_{0}}{\rho}\sqrt{l^2-j_{0}^2}
+\frac{M}{\rho}\frac{l^2+j_{0}^2}{\sqrt{l^2-j_{0}^2}}-
2\frac{b^2j_{0}^2}{\rho\sqrt{l^2-{-M}+^2}}
\nonumber\\
&&+2b^2\frac{l^2+j_{0}^2}{\rho\sqrt{l^2-j_{0}^2}}\ln{(\rho)},\nonumber\\
M({\rho\rightarrow\infty},\epsilon_{0\mid\infty}(M_{0}))&\approx&\,M-{M_{0}}
+2{b^2}{\ln{(\rho)}}.
\end{eqnarray}
Therefore, comparing with the energy characteristics of the BTZ
solution, one concludes that the mass parameter
is equal to $M$. The momentum parameter
is proportional to $j_{0}$. The contribution of
the charge $b$ in the mass--energy--momentum
functions at infinity is through logarithmical
terms and hence they diverges at
spatial infinity. For vanishing rotation
parameter $j_{0}$ one arrives at the
expressions of the mass--energy--momentum
functions of the Peldan electrostatic solution, Section~\ref{staticPeldan}.

\section{Kamata-Koikawa solution}

The Kamata--Koikawa solution~\cite{KamataK95},
see also~\cite{GarciaAnnals09}~Eq.~(7.7), is
defined by the metric and the structural functions
\begin{eqnarray}\label{GenKK}
ds^2&&=-N(\rho)^2dt^2+\frac{1}{L(\rho)^{2}}\,d\rho^2
+K(\rho)^2[d\phi+W(\rho)dt]^2,\nonumber\\
L(\rho)&=&\frac{\sqrt{\Lambda}}{\rho}({\rho^2-\rho_{0}^2}),
\,\sqrt{\Lambda}=1/l,\nonumber\\
K(\rho)&=&\sqrt{\rho^2+\frac{Q^2}{\Lambda}
\ln{(\frac{\rho^2}{\rho_{0}^2}-1})},\nonumber\\
N(\rho)&=&\rho\,{L}/\,{K}
=\sqrt{\Lambda}({\rho^2-\rho_{0}^2})/\sqrt{\rho^2
+\frac{Q^2}{\Lambda}\ln{(\frac{\rho^2}{\rho_{0}^2}-1})},
\nonumber\\
W(\rho)&=&\frac{({\rho^2-\rho_{0}^2})\sqrt{\Lambda}}
{[\rho^2+\frac{Q^2}{\Lambda}\ln{(\frac{\rho^2}{\rho_{0}^2}-1)}]}
-\sqrt{\Lambda}.
\end{eqnarray}

\subsection{ Mass, energy and momentum for
the KK solution}\label{KKspinning}

The surface energy and momentum densities
are respectively given by
\begin{eqnarray}
\epsilon(\rho,\epsilon_{0})&=&-\frac{1}{\pi\sqrt{\Lambda}}\,
\frac{Q^2-\Lambda\rho_{0}^2+\Lambda\rho^2}
{[\rho^2+\frac{Q^2}{\Lambda}
\ln{(\frac{\rho^2}{\rho_{0}^2}-1})]}
-\epsilon_{0},\nonumber\\
j(\rho)&=&\frac{1}{\pi\sqrt{\Lambda}}\,
\frac{\Lambda\rho_{0}^2-Q^2+{Q^2}
\ln{(\frac{\rho^2}{\rho_{0}^2}-1})}
{\sqrt{\rho^2+\frac{Q^2}{\Lambda}
\ln{(\frac{\rho^2}{\rho_{0}^2}-1})}},
\end{eqnarray}
while the integral quantities amount to
\begin{eqnarray}\label{den-KK}
J(\rho)&=&2\frac{\Lambda \rho_{0}^2-Q^2}{\sqrt{\Lambda}}
 +2\frac{Q^2}{\sqrt{\Lambda}}
 \ln{(\frac{\rho^2}{\rho_{0}^2}-1)},\nonumber\\
E(\rho,\epsilon_{0})&=&-\frac{2}{\sqrt{\Lambda}}\,
\frac{Q^2-\Lambda\rho_{0}^2+\Lambda\rho^2}
{\sqrt{\rho^2+\frac{Q^2}{\Lambda}
\ln{(\frac{\rho^2}{\rho_{0}^2}-1})}}
-2\pi\,K\epsilon_{0},\nonumber\\
M(\rho,\epsilon_{0})&=&-2\Lambda\,\rho^2+2(2\Lambda
\rho_{0}^2-Q^2)+2Q^2\ln{(\frac{\rho^2}{\rho_{0}^2}-1)}
-2\pi\,\sqrt{\Lambda}\,(\rho^2-\rho_{0}^2)\epsilon_{0}.
\end{eqnarray}
The evaluation of the functions above  for vanishing $\epsilon_{0}$,
i.e. $\epsilon_{0}=0$, behave at $\rho\rightarrow\infty$ according to
\begin{eqnarray}\label{enerKAM}
\epsilon(\rho\rightarrow\infty,\epsilon_{0}=0)&
\approx&-\frac{\sqrt{\Lambda}}{\pi}
+\frac{\Lambda \rho_{0}^2-Q^2}{\pi\sqrt{\Lambda}\,\rho^2}
 +2\frac{Q^2}{\pi\sqrt{\Lambda}\,\rho^2}
 \ln{(\frac{\rho}{\rho_{0}})},\nonumber\\
j(\rho\rightarrow\infty)&\approx&
\frac{\Lambda \rho_{0}^2-Q^2}{\pi\sqrt{\Lambda}\,\rho}
 +2\frac{Q^2}{\pi\sqrt{\Lambda}\,\rho}
 \ln{(\frac{\rho}{\rho_{0}})},\nonumber\\
J(\rho\rightarrow\infty)&
\approx&2\frac{\Lambda \rho_{0}^2-Q^2}{\sqrt{\Lambda}}
 +4\frac{Q^2}{\sqrt{\Lambda}}\ln{(\frac{\rho}{\rho_{0}})},
 \nonumber\\
E(\rho\rightarrow\infty,\epsilon_{0}=0)&
\approx&\,-2{\sqrt{\Lambda}}{\rho}
+2\frac{\Lambda \rho_{0}^2-Q^2}{\sqrt{\Lambda}\,\rho}
+2\frac{Q^2}{\sqrt{\Lambda}\,\rho}\ln{(\frac{\rho}{\rho_{0}})}
,\nonumber\\
M(\rho\rightarrow\infty,\epsilon_{0}=0)&\approx&
-2\Lambda\,\rho^2+2(2\Lambda \rho_{0}^2-Q^2)
+4Q^2\ln{(\frac{\rho}{\rho_{0}})}.
\end{eqnarray}
Using in the expressions~(\ref{den-KK}) as reference energy
density the quantity
$\epsilon_{0}=-\frac{1}{\pi\rho}\sqrt{-M_{0}+\frac{\rho^2}{l^2}}$,
which at the spatial infinity behaves as
$\epsilon_{0\mid\infty}(M_{0})\approx
-\frac{1}{\pi\,l}+\frac{lM_{0}}{2\pi\,\rho^2}
=-\frac{\sqrt{\Lambda}}{\pi}
+\frac{M_{0}}{2\pi\,\sqrt{\Lambda}\,\rho^2},$  the series
expansions of the corresponding
quantities at $\rho\rightarrow\infty$
result in
\begin{eqnarray}
\epsilon({\rho\rightarrow\infty},
\epsilon_{0\mid\infty}(M_{0}))&\approx&
\,-\frac{M_{0}}{2\pi\sqrt{\Lambda}
\,\rho^2}+\frac{\Lambda
\rho_{0}^2-Q^2}{\pi\sqrt{\Lambda}\,\rho^2}
 +2\frac{Q^2}{\pi\sqrt{\Lambda}\,\rho^2}
 \ln{(\frac{\rho}{\rho_{0}})},\nonumber\\
E({\rho\rightarrow\infty},\epsilon_{0\mid\infty}(M_{0}))
&\approx&-\frac{M_{0}}{\sqrt{\Lambda} \,\rho}+2\frac{\Lambda
\rho_{0}^2-Q^2}{\sqrt{\Lambda}\,\rho}
+4\frac{Q^2}{\sqrt{\Lambda}\,\rho}
\ln{(\frac{\rho}{\rho_{0}})},\nonumber\\
M({\rho\rightarrow\infty},\epsilon_{0\mid\infty}(M_{0}))
&\approx& -M_{0}+2(\Lambda
\rho_{0}^2-Q^2)+4Q^2\,\ln{(\frac{\rho}{\rho_{0}})}.
\end{eqnarray}
Therefore, comparing with the energy characteristics of the BTZ
solution, one arrives to the conclusion that there
is no a mass parameter of the kind $M$ present in the
BTZ solution. All characteristic functions logarithmically diverges at
spatial infinity.\\

\subsection{ Field, energy--momentum and Cotton
tensors for the Kamata-Koikawa solution}

The electromagnetic field  tensor
\begin{eqnarray}
({F^\alpha}_{\beta})== \left[
\begin {array}{ccc} 0&
{-\frac {Q\rho\,q}{ \Lambda\left( {\rho}^{2}-{{\rho}_{0}}
^{2} \right) ^{2}}}&0
\\\noalign{\medskip}{\frac {qQ\Lambda \left( {
\rho}^{2}-{{\rho}_{0}}^{2} \right) }{\rho}}&
0&{-\frac {qQ\sqrt {\Lambda} \left(
{\rho}^{2}-{{\rho}_{0}}^{2} \right) }{\rho}}
\\\noalign{\medskip}0&
{-\frac {Q\rho\,q}{\sqrt {\Lambda} \left( {\rho}
^{2}-{{\rho}_{0}}^{2} \right) ^{2}}}&0\end {array} \right],
\end{eqnarray}
allows for a triple zero eigenvalue and the
following set of eigenvectors
\begin{eqnarray}
&&\lambda_{1,2,3}=0;\,{\bf V}=(V^{1},V^{2},V^{3}=
\sqrt{\Lambda}\,V^{1}),\,V^{\mu}\,V_{\mu}=0,\,{\bf V}=
{\bf N},\nonumber\\
&&
{\rm Type}:\{3S\}.
\end{eqnarray}
The electromagnetic energy momentum tensor with
vanishing invariants occurs to be
\begin{eqnarray}
({T^\alpha}_{\beta})=\frac{1}{4\,\pi }\,\frac {{
Q}^{2}}{\sqrt{\Lambda}\,
\left( {\rho}^{2}-{\rho_0}^{2} \right) }
\left[ \begin {array}{ccc} -\sqrt{\Lambda}&0&-\Lambda
\\\noalign{\medskip}0&0&0\\\noalign{\medskip}
1&0&\sqrt{\Lambda}\end {array} \right],
\end{eqnarray}
while the Cotton tensor for this electromagnetic--gravitational
stationary cyclic symmetric field is
given by
\begin{eqnarray}
({C^\alpha}_{\beta})=\frac {{Q}^{2}}
{\left( {\rho}^{2}-{\rho_0}^{2} \right)}
\left[ \begin {array}{ccc} -\sqrt{\Lambda}&0&1
\\\noalign{\medskip}0&0&0\\\noalign{\medskip}-\Lambda
&0&\sqrt{\Lambda}\end {array} \right].
\end{eqnarray}
It is clear that both the Cotton and Maxwell tensors
possess the same eigenvalues, namely
the triple zero eigenvalue $\lambda=0$.
Searching for the eigenvectors of these tensors, one arrives at
\begin{eqnarray}
&&\lambda_{1,2,3}=0;\,{\bf V}=(V^{1},V^{2},V^{3}=
\sqrt{\Lambda}\,V^{1}),\,V^{\mu}\,V_{\mu}=
{\frac {({V^{2}})^{2}{\rho}^{2}}
{\Lambda \left(\rho - {\rho_0}\right) ^{2}}},\,{\bf V}=
{\bf S},\,{\bf V}(V^{2}=0)={\bf N},\nonumber\\
&&
{\rm Type}:\{3S\},\{2S,N\},\{S,2N\},\{3N\}.
\end{eqnarray}
The eigenvectors are spacelike or null vectors depending on
the non-vanishing or vanishing value of the component $V^{2}$ .
One may consider them different one to another, having
different $ V^1$ and $V^2$ components.
The most degenerate case are $\{3{ S}\}$ and $\{{ S},2{ N}\}$.

\subsection{ Proper Kamata-Koikawa
solution, $\rho_{0}=\pm Q/\sqrt{\Lambda}$}

The proper Kamata--Koikawa solution is
defined by the metric and structural
functions of (\ref{GenKK}) for $\Lambda \rho_{0}^2-Q^2=0 $, i.e.,
$\rho_{0}=\pm Q/\sqrt{\Lambda}$, namely
\begin{eqnarray}
ds^2&&=-N(\rho)^2dt^2+\frac{1}{L(\rho)^{2}}\,d\rho^2
+K(\rho)^2[d\phi+W(\rho)dt]^2,\nonumber\\
L(\rho)&=
&\frac{\sqrt{\Lambda}}{\rho}({\rho^2-\rho_{0}^2}),
\nonumber\\
K(\rho)&=&\sqrt{\rho^2+\frac{Q^2}{\Lambda}
\ln{(\frac{\rho^2}{\rho_{0}^2}-1})},\nonumber\\
N(\rho)&=&\rho\,{L}/\,{K}
=\sqrt{\Lambda}({\rho^2-\rho_{0}^2})/\sqrt{\rho^2
+\frac{Q^2}{\Lambda}\ln{(\frac{\rho^2}{\rho_{0}^2}-1})},
\nonumber\\
W(\rho)&=&\frac{({\rho^2-\rho_{0}^2})\sqrt{\Lambda}}
{[\rho^2+\frac{Q^2}{\Lambda}\ln{(\frac{\rho^2}{\rho_{0}^2}-1)}]}
-\sqrt{\Lambda},
\end{eqnarray}
The surface energy and momentum densities are respectively given by
\begin{eqnarray}
\epsilon(\rho,\epsilon_{0})&=&-\frac{\sqrt{\Lambda}}{\pi}
\, \frac{\rho^2}
{[\rho^2+\frac{Q^2}{\Lambda}
\ln{(\frac{\rho^2}{\rho_{0}^2}-1})]},\nonumber\\
j(\rho)&=&\frac{1}{\pi\sqrt{\Lambda}}\,
\frac{{Q^2}\ln{(\frac{\rho^2}{\rho_{0}^2}-1})}
{\sqrt{\rho^2+\frac{Q^2}{\Lambda}
\ln{(\frac{\rho^2}{\rho_{0}^2}-1})}},
\end{eqnarray}
while the integral quantities amount to
\begin{eqnarray}\label{den-PKK}
J(\rho)&=&2\,\frac{Q^2}{\sqrt{\Lambda}}
\ln{(\frac{\rho^2}{\rho_{0}^2}-1)},\nonumber\\
E(\rho,\epsilon_{0})&=&-{2\sqrt{\Lambda}}\,
\frac{\rho^2}
{\sqrt{\rho^2+\frac{Q^2}{\Lambda}
\ln{(\frac{\rho^2}{\rho_{0}^2}-1})}}
-2\pi\,K\epsilon_{0},\nonumber\\
M(\rho,\epsilon_{0})&=&-2\Lambda\,\rho^2+2\Lambda
\rho_{0}^2+2Q^2\ln{(\frac{\rho^2}{\rho_{0}^2}-1)}
-2\pi\,\sqrt{\Lambda}\,(\rho_{0}^2-\rho_{0}^2)\epsilon_{0}.
\end{eqnarray}
The evaluation of the functions above  for
vanishing $\epsilon_{0}$,
i.e. $\epsilon_{0}=0$, behave at spatial
infinity according to
\begin{eqnarray}\label{enerKAMprop}
\epsilon(\rho\rightarrow\infty,\epsilon_{0}=0)&
\approx&-\frac{\sqrt{\Lambda}}{\pi}
+2\frac{Q^2}{\pi\sqrt{\Lambda}\,\rho^2}
\ln{(\frac{\rho}{\rho_{0}})},\nonumber\\
j(\rho\rightarrow\infty)&\approx& 2\frac{Q^2}{\pi\sqrt{\Lambda}
\,\rho}
\ln{(\frac{\rho}{\rho_{0}})},\nonumber\\
J(\rho\rightarrow\infty)&\approx
&4\frac{Q^2}{\sqrt{\Lambda}}
\ln{(\frac{\rho}{\rho_{0}})},\nonumber\\
E(\rho\rightarrow\infty,\epsilon_{0}=0)&
\approx&\,-2{\sqrt{\Lambda}}{\rho}
+2\frac{Q^2}{\sqrt{\Lambda}\,\rho}
\ln{(\frac{\rho}{\rho_{0}})}
,\nonumber\\
M(\rho\rightarrow\infty,\epsilon_{0}=0)&
\approx& -2\Lambda\,\rho^2+2\Lambda
\rho_{0}^2+4Q^2\ln{(\frac{\rho}{\rho_{0}})}.
\end{eqnarray}
Using in the expressions (\ref{den-PKK}) as reference energy
density the quantity
$\epsilon_{0}=-\frac{1}{\pi\rho}
\sqrt{-M_{0}+\frac{\rho^2}{l^2}}$,
which at the spatial infinity behaves as
$\epsilon_{0\mid\infty}(M_{0})\approx
-\frac{\sqrt{\Lambda}}{\pi}
+\frac{M_{0}}{2\pi\sqrt{\Lambda}\,\rho^2},$  the series
expansions of the corresponding quantities at $\rho=$ infinity
result in
\begin{eqnarray}
\epsilon({\rho\rightarrow\infty},
\epsilon_{0\mid\infty}(M_{0}))
&\approx&\,-\frac{M_{0}}{2\pi\sqrt{\Lambda}
\,\rho^2}+2\frac{Q^2}{\pi\sqrt{\Lambda}\,\rho^2}
\ln{(\frac{\rho}{\rho_{0}})},\nonumber\\
E({\rho\rightarrow\infty},\epsilon_{0\mid\infty}(M_{0}))
&\approx&-\frac{M_{0}}{\sqrt{\Lambda} \,\rho}
+4\frac{Q^2}{\sqrt{\Lambda}\,\rho}\ln{(\frac{\rho}{\rho_{0}})},\nonumber\\
M({\rho\rightarrow\infty},\epsilon_{0\mid\infty}(M_{0}))
&\approx& -M_{0}+4Q^2\,\ln{(\frac{\rho}{\rho_{0}})}.
\end{eqnarray}

In the work by Chan~\cite{Chan96} there are some comments
addressed to the evaluation of the global
momentum, energy and mass of the proper Kamata--Koikawa solution:
the exact and the approximated expressions of the momentum $J$ coincide
with the corresponding ones
given in (\cite{Chan96}~Eq.9) and (\cite{Chan96}~Eq.7). Moreover,
the mass $M$ at spatial infinity, (\ref{enerKAMprop}), coincides
with the $M $  (\cite{Chan96}~Eq.10.)
for a zero background energy density with the correct extra
term $-2\Lambda\,\rho^2$. From my point of view, it is recommendable to
accomplish series expansions of the quantities under consideration
to determine how fast they approach to zero or diverge at
spatial infinity. From this perspective,
the evaluation of the energy density $\epsilon$ and the
global energy $E$ yield to quantities different
from zero at spatial infinity, although they both approach
faster to zero as $\rho\rightarrow\infty$ than the momentum and mass.

Comparing with the energy characteristics of the BTZ
solution, one concludes that the mass, energy and
momentum functions logarithmically diverge at
spatial infinity.

\section{Peldan magnetostatic solution}

The magnetostatic solution~\cite{Peldan93},
see also~\cite{GarciaAnnals09}~(4.30), with a
negative cosmological constant is determined by the
metric functions
\begin{eqnarray}\label{PeldanMagstatic}
ds^2&&=-N(\rho)^2dt^2+\frac{1}{L(\rho)^{2}}\,d\rho^2
+K(\rho)^2[d\phi+W(\rho)dt]^2,\nonumber\\
L(\rho)=K(\rho)&&=\sqrt{\frac{\rho^2}{l^2}
+2\,a^2\ln{\rho}+m},\,N(\rho)=\rho,\, W(\rho)=0.
\end{eqnarray}
Notice that the metric functions $L^{2}$ and $K^{2}$ are
positive functions for values of $\rho>\rho_{\rm root}$,  where
$$\rho_{\rm root}=\exp\large[-\frac{m}{2\,a^2}
-\frac{1}{2}{\rm LambertW }(\frac{1}{l^2\,a^2}
{\rm e}^{-\frac{m}{a^2}})],
\,{\rm LW }(x)\exp({\rm LW }(x))=x,$$
where for short ${\rm LW }:={\rm LambertW }$,
$\rho_{\rm root}$ is solution of the equation
$g^{\rho\,\rho}(\rho_{\rm root})=L^{2}(\rho_{\rm root})=0$,
or explicitly, $${\rho_{\rm root}^2}/{l^2}
+2\,a^2\ln{\rho_{\rm root}}+m=0.$$ Therefore the
coordinate $\rho$ does not cover the expected range
$0\leq \rho\leq\infty$. For $\rho\leq\rho_{\rm root}$
the metric suffers an unacceptable signature change. This fact also
points out on the non-existence of a horizon $\rho=\rm const$ for
the Peldan solution. Consequently, one has to modify the choice of the $\rho$
coordinate in order to be able to rich the origin of coordinates;
with this purpose in mind a new coordinate
system is chosen in the forthcoming paragraph (\ref{ModPeldan}), see also (\ref{ModHW}).

\subsection{ Mass, energy and momentum for the Peldan solution}\label{PeldanMag}

The surface energy density $\epsilon$ is given by
\begin{eqnarray}
\epsilon(\rho)=-\frac{1}{\pi\,K}(\frac{\rho}{l^2}+\frac{a^2}{\rho})
\epsilon_{0}.
\end{eqnarray}
Consequently the global energy and mass are given by
\begin{eqnarray}\label{den-catgzh}
E(\rho,\epsilon_{0})&=& -2\frac{\rho}{l^2}
-2\frac{a^2}{\rho}-2\pi\,K\,\epsilon_{0},\nonumber\\
M(\rho,\epsilon_{0})&=&-2\frac{\rho^2}{l^2}-2{a^2}-2\pi\,\rho\,K\,\epsilon_{0}.
\end{eqnarray}
For the natural choice of a vanishing reference energy density
$\epsilon_{0}=0$, one has at the spatial infinity
$\rho\rightarrow\infty$ that
\begin{eqnarray}
\epsilon({\rho\rightarrow\infty},\epsilon_{0}=0)&
\approx&-\frac{1}{\pi\,l}+l\,\frac{m-2a^2}{2\pi\,\rho^2}
+\frac{l\,a^2}{\pi\,\rho^2}\ln{\rho},\nonumber\\
E({\rho\rightarrow\infty},\epsilon_{0}=0)&
=&-2\frac{\rho}{l^2}-2\frac{a^2}{\rho},\nonumber\\
M({\rho\rightarrow\infty},\epsilon_{0}=0)&
=&-2\frac{\rho^2}{l^2}-2{a^2},
\end{eqnarray}
while if the reference energy is the one corresponding to the
anti--de Sitter spacetime with $M_{0}$ parameter,
$\epsilon_{0}=-\frac{\rho}{\pi\,l^2}/\sqrt{\frac{\rho^2}{l^2}
+M_{0}}$, $\epsilon_{0\mid\infty}(M_{0})\approx
-\frac{\sqrt{\Lambda}}{\pi}+\frac{M_{0}}{2\pi\sqrt{\Lambda}\,\rho^2}$,
then the energies are expressed at spatial infinity as
\begin{eqnarray}
\epsilon({\rho\rightarrow\infty},\epsilon_{0\mid\infty}(M_{0}))
&\approx&\,l\,\frac{m-M_{0}-2a^2}{2\pi\,\rho^2}+
\frac{a^2}{\pi\,\rho^2}\ln{\rho},\nonumber\\
E({\rho\rightarrow\infty},
\epsilon_{0\mid\infty}(M_{0}))&\approx
&\,\frac{m-M_{0}-2a^2}{\rho}
+2\frac{a^2}{\rho}\ln{\rho},\nonumber\\
M({\rho\rightarrow\infty},\epsilon_{0\mid\infty}(M_{0}))
&\approx&\,m-M_{0}-2a^2+ 2\,a^2\ln{\rho}.
\end{eqnarray}
Comparing these quantities with the corresponding ones of the static
BTZ solution counterpart, Section~\ref{BTZcounter},
one sees a complete correspondence for vanishing
electromagnetic parameter $a$, thus one recognize $m$ as mass parameter.
Notice that the energy and mass include an amount of energy due to
the magnetic field, in a way similar to the electric one, through a
logarithmical terms; because of this dependence, these quantities
logarithmically diverge at infinity.

\subsection{Field, energy-momentum and Cotton tensors for the
magnetostatic Peldan solution}

The electromagnetic field tensor for this solution is given by
\begin{eqnarray}
({F^\alpha}_{\beta})= \left[ \begin {array}{ccc} 0&0&0
\\\noalign{\medskip}0&0&{\frac {{\it L^2}\,a}{{\rho}}}\\
\noalign{\medskip}0&-{\frac {a}{{\it L^2}\,\rho}}&0
\end {array} \right] ,
\end{eqnarray}
and is algebraically characterized by the following eigenvectors
\begin{eqnarray}
\lambda_1&&=0;
{\bf V}1=[{V^{1}}={V^{1}},{V^{2}}=0,
{V^{3}}=0 ],
\,V^{\mu}V_{\mu}=-(V^{1})^2(\rho^{2}),\,
{\bf V}1={\bf T}1,\nonumber\\
\lambda_2&&=i\frac {a}{\rho};
{\bf V}1=[{V^{1}}=0,{V^{2}}={V^{2}},{V^{3}}=
i\frac {1}{{\it L^2}}{V^{2}}],
{\bf V}1={\bf Z},\nonumber\\
\lambda_3&&=-i\frac {a}{\rho};
{\bf V}1=[{V^{1}}=0,{V^{2}}={V^{2}},
{V^{3}}=-i\frac{
1}{{\it L^2}}{V^{2}}\,],
{\bf V}1={\bf \bar Z},
\nonumber\\
{\rm Type:}&&\{T, Z, \bar Z\}.
\end{eqnarray}
As far as to the electromagnetic energy momentum tensor is concerned, its matrix
amounts to
\begin{eqnarray}
({T^\alpha}_{\beta}) \left[
\begin {array}{ccc} -\frac{1}{8\pi}\,{\frac {{a}^{2}}{{\rho}^{2} }}&0&0
\\\noalign{\medskip}0&\frac{1}{8\pi}\,{\frac {{a}^{2}}{{\rho}^{2} }}&0
\\\noalign{\medskip}0&0&\frac{1}{8\pi}\,{\frac {{a}^{2}}{{\rho}^{2} }}
\end {array} \right]
\end{eqnarray}
with the following eigenvalues and their corresponding eigenvectors
\begin{eqnarray}
\lambda_1&&=-\frac{1}{8\pi}\,\frac {{a}^{2}}{{\rho}^{2} };
{\bf V}1=(V^{1},0,0),\,V^{\mu}V_{\mu}=-\rho^2\,V^{1})^2,\,
{\bf V}1={\bf T}1,\nonumber\\
\lambda_2&&=\frac{1}{8\pi}\,\frac {{a}^{2}}{{\rho}^{2} };
{\bf V}2=(0,V^{2},V^{3}),\,
\,V^{\mu}V_{\mu}=(V^{2})^2/\,L^2+(V^{3})^2\,L^2,\,{\bf V}2
={\bf S}2,\nonumber\\
\lambda_3&&=\frac{1}{8\pi}\,\frac {{a}^{2}}{{\rho}^{2} };
{\bf V}3=(0,\tilde V^{2},\tilde V^{3}),
\,V^{\mu}V_{\mu}
=({\tilde V}^{2})^2/\,L^2+({\tilde V}^{3})^2\,L^2,
\,{\bf V}3={\bf S}3,\nonumber\\
{\rm Type:}&&\{T, 2S\}.
\end{eqnarray}
This tensor structure corresponds to that one describing a
perfect fluid energy momentum tensor, but this time for the state
equation: $\it {energy = pressure}$. Again, the solutions generated from this
metric by using coordinate transformations will possess this perfect fluid
feature because the invariance of the eigenvalues.

The Cotton tensor for electrostatic cyclic symmetric gravitational field is
given by
\begin{eqnarray}
({C^\alpha}_{\beta})=\left[
\begin {array}{ccc} 0&0&\frac{1}{2}\,{\frac {{a}^{2} \left( {\rho}^{2
}+2\,{a}^{2}{l}^{2}\ln  \left( \rho \right)  \right) }{{l}^{2}{\rho}^{
4}}}\\\noalign{\medskip}0&0&0
\\\noalign{\medskip}-\frac{1}{2}\,{\frac {{a}^{2}
}{{\rho}^{2}}}&0&0\end {array} \right]=\left[
\begin {array}{ccc} 0&0&\,{\frac {{a}^{2} L^2 }{2\rho^{
4}}}\\\noalign{\medskip}0&0&0\\\noalign{\medskip}-\,{\frac {{a}^{2}
}{2\rho^{2}}}&0&0\end {array} \right] .
\end{eqnarray}
Searching for its eigenvectors, one arrives at
\begin{eqnarray}
&&\lambda_1=0;{\bf V}1=(0,V^{2},0),
\,V^{\mu}V_{\mu}=(V^{2})^2g_{\rho\rho},
\,{\bf V}1={\bf S}1,\nonumber\\
\lambda_{2}=\frac{i}{2}\,{\frac {L\,a^{2}}{{\rho}^{3}}};
{\bf V}2&&=(V^{1}=-\frac{i\,L}{\rho}V^{3},0,V^{3}),
\,{\bf V}2={\bf Z},\nonumber\\
\lambda_{3}=-\frac{i}{2}\,{\frac {L\,a^{2}}{{\rho}^{3}}};
{\bf V}3&&=( V^{1}=\frac{i\,L}{\rho}{ V^{3}},0, V^{3}),
{\bf V}3={\bf {\bar Z}},
\nonumber\\
{\rm Type:}&&\{T, Z, \bar Z\}.
\end{eqnarray}
The eigenvectors ${\bf V}2$ and ${\bf V}3$ are complex
conjugated while the vector ${\bf V}1$, associated to the zero
eigenvalue, occurs to be spacelike--the only physically tractable
$\rho$--direction vector in this case. It is worthwhile
to point out that the solutions generated via coordinate transformations,
in particular the $SL(2,R)$ transformations,
applied onto this magneto--static cyclic symmetric metric
will shear the eigenvalues $\lambda_i$ of the Cotton tensor quoted above;
recall that eigenvalues are invariant characteristics of tensors, although
their components in different coordinate systems are different--this last also
applies to the eigenvectors of the seed and the resulting solutions.

\subsection{Field, energy-momentum and Cotton
tensors for a modified magnetostatic Peldan solution}\label{ModPeldan}

One encounters in the literature a slightly modified
Peldan solution, namely the one with metric
\begin{eqnarray}\label{MagModPeldan}
\left({g_{\alpha}}_{\beta}\right)&&= \left[ \begin {array}{ccc} -{\rho}^{2
}-{\it Mg}&0&0\\\noalign{\medskip}0&1/{{\it L^2}}&0
\\\noalign{\medskip}0&0&{\frac {{\it L^2}\,{\rho}^{2}}{{\rho}^{2}+
{\it Mg}}}\end {array} \right],\nonumber\\
{\it L^2}&&:={\frac { \left( -{l}^{2}M+{\rho}^{2}+{\it Mg}+{a}^{2}{l}^{2}\ln
 \left( {\rho}^{2}+{\it Mg} \right)  \right)  \left( {\rho}^{2}+{\it
Mg} \right) }{{l}^{2}{\rho}^{2}}}.
\end{eqnarray}
One easily establishes that in this coordinate system one may reach the
origin $\rho=0$. In fact,
$$g_{\phi\phi}=\frac {{\it L^2}\,{\rho}^{2}}{{\rho}^{2}+
{\it Mg}}=\frac{\rho^{2}}{l^2}+\frac{\it Mg}{l^2}-M+{a}^{2}\,\ln
 \left( {\rho}^{2}+{\it Mg} \right)=\frac{\rho^{2}}{l^2}+{a}^{2}\,\ln
 \left[( {\rho}^{2}+{\it Mg})\exp{(\frac{\it Mg-Ml^2}{a^2\,l^2})} \right],$$
thus, adopting the choice
$${\it Mg}\,\exp{(\frac{\it Mg-Ml^2}{a^2\,l^2})}=1\,\rightarrow{\it Mg}
={l}^{2}{\rm LW} \left( {\frac {{{\rm e}^{M}}}{{l}^{2}}} \right),\,
{\rm LW }(x)\exp({\rm LW }(x))=x, $$
where ${\rm LW }$ stand for ${\rm LambertW }$--function, one arrives
as $\rho\rightarrow 0$ at
$$g_{\phi\phi|\rho\rightarrow 0}=\rho^{2}{a}^{2}\,
\exp(\frac{\it Mg-Ml^2}{a^2\,l^2})=\rho^{2}\frac{a^{2}}{{\it Mg}},
\,g_{\rho\rho|\rho\rightarrow 0}=\frac{1}{a^{2}},$$
$$\rightarrow{{ds^2}_{|\rho\rightarrow 0}=-(d\,t\,\sqrt{\it Mg}})^2
+(d\frac{\rho}{a})^2+\left(\frac{\rho}{a}\right)^2(d\frac{a^2}
{\sqrt{\it Mg}}\phi)^2,$$
where it has been taken into  account that $1/g_{\rho\rho}={\it L^2}
=\frac{g_{\phi\phi}}{\rho^2}\,\left( {\rho}^{2}+{\it
Mg} \right).$
\noindent
Therefore, the angular coordinate exhibit an angle deficit, assuming
$0\leq\frac{a^2}{\sqrt{\it Mg}}\phi<2\pi$, then
$0\leq \phi<2\pi\frac{\sqrt{\it Mg}}{a^2}$, and
$\Delta \phi=2\pi\left(1-\frac{\sqrt{\it Mg}}{a^2}\right).$\\

The electromagnetic field tensor is given by
\begin{eqnarray}
({F^\alpha}_{\beta})= \left[ \begin {array}{ccc} 0&0&0
\\\noalign{\medskip}0&0&{\frac {{\it L^2}\,\rho\,a}{{\rho}^{2}+
{\it Mg}}}\\\noalign{\medskip}0&-{\frac {a}{{\it L^2}\,\rho}}&0
\end {array} \right] ,
\end{eqnarray}
with eigenvectors
\begin{eqnarray}
\lambda_1&&=0;
{\bf V}1=[{V^{1}}={V^{1}},{V^{2}}=0,
{V^{3}}=0 ],
\,V^{\mu}V_{\mu}=-(V^{1})^2(\rho^{2}+{\it Mg}),\,
{\bf V}1={\bf T}1,\nonumber\\
\lambda_2&&=i{\frac {a}{\sqrt {{\rho}^{2}+{\it Mg}}}};
{\bf V}1=[{V^{1}}=0,{V^{2}}={V^{2}},{V^{3}}=i{\frac {\,
\sqrt {{\rho}^{2}+{\it Mg}}}{{\it L^2}\,\rho}}{V^{2}}],
{\bf V}1={\bf Z},\nonumber\\
\lambda_3&&=-i{\frac {a}{\sqrt {{\rho}^{2}+{\it Mg}}}};
{\bf V}1=[{V^{1}}=0,{V^{2}}={V^{2}},
{V^{3}}=-i{\frac {
\sqrt {{\rho}^{2}+{\it Mg}}}{{\it L^2}\,\rho}}{V^{2}}\,],
{\bf V}1={\bf \bar Z},\nonumber\\
{\rm Type:}&&\{T, Z, \bar Z\}.
\end{eqnarray}
while the electromagnetic energy--momentum tensor amounts to
\begin{eqnarray}
({T^\alpha}_{\beta})&&= \left[ \begin {array}{ccc} -\frac{1}{8\pi}\,{
\frac {{a}^{2}}{\left( {\rho}^{2}+{\it Mg} \right) }}&0&0
\\\noalign{\medskip}0&\frac{1}{8\pi}\,{\frac {{a}^{2}}{\left( {\rho}^{2}+{
\it Mg} \right) }}&0\\\noalign{\medskip}0&0&\frac{1}{8\pi}\,{\frac {{a}^{2}}{
\, \left( {\rho}^{2}+{\it Mg} \right) }}\end {array} \right].
\end{eqnarray}
with eigenvectors
\begin{eqnarray}
\lambda_1&&=-\frac{1}{8\,\pi}{\frac {a^2}{\rho^{2} +\,{\it Mg}}};
{\bf V}1=[{V^{1}}={V^{1}},{V^{2}}=0,{V^{3}}=0 ],\nonumber\\
\,V^{\mu}V_{\mu}&&=-{V^{1}}^2(\rho^{2} +\,{\it Mg}),\,
{\bf V}1={\bf T}1,\nonumber\\
\lambda_2&&=\frac{1}{8\,\pi}{\frac {{a}^{2}}{{\rho}^{2} +
\,{\it Mg}}};
{\bf V}2=[ {V^{1}}=0,
{V^{2}}={V^{2}},{V^{3}}={V^{3}} ],\nonumber\\
V^{\mu}V_{\mu}&&=\frac{{V^{2}}^2}{{\it L^2}}
+\frac{\rho^{2}{\it L^2}}{(\rho^{2} +\,{\it Mg})}{V^{1}}^2,\,
{\bf V}2={\bf S}2,
\nonumber\\
\lambda_3&&=\frac{1}{8\,\pi}{\frac {{a}^{2}}{{\rho}^{2} +
\,{\it Mg}}};
{\bf V}3=[ \tilde{V^{1}}=0,
\tilde{V^{2}}=\tilde{V^{2}},\tilde{V^{3}}=\tilde{V^{3}} ],\nonumber\\
V^{\mu}V_{\mu}&&=\frac{\tilde{V^{2}}^2}{{\it L^2}}
+\frac{\rho^{2}{\it L^2}}{(\rho^{2} +\,{\it Mg})}\tilde{V^{1}}^2,\,
{\bf V}3={\bf S}3,\nonumber\\
{\rm Type:}&&\{T, 2\,S\}.
\end{eqnarray}
The Cotton tensor
\begin{eqnarray}
({C^\alpha}_{\beta})=  \left[
\begin {array}{ccc} 0&0&\frac{1}{2}\,{\frac {{a}^{2}{\it L^2}\,{\rho}^{2}}
{ \left( {\rho}^{2}+{\it Mg} \right) ^{3}}}\\\noalign{\medskip}0&0&0
\\\noalign{\medskip}-\frac{1}{2}\,{\frac {{a}^{2}}{{\rho}^{2}+{\it Mg}}}&0&0
\end {array} \right],
\end{eqnarray}
possesses the following set of eigenvectors
\begin{eqnarray}
&&\lambda_1=0;{\bf V}1=[0,V2,0],V^{\mu}V_{\mu}=\frac{{V^{2}}^2}{{\it L^2}},\,
{\bf V}1={\bf S}1,\nonumber\\&&
\lambda_2=i\,\frac{{a}^{2}}{2}\,{\frac{\rho\,{\it L}}{ \left( {\rho}^{2}+
{\it Mg} \right) ^{2}}};
{\bf V}2=[{ V^{1}}=-i\frac{{\it L}}{{\rho}^{2}+
{\it Mg}}V^{3},V^2=0,V^{3}],{\bf V}2={\bf Z},
\nonumber\\&&
\lambda_3=-i\,\frac{{a}^{2}}{2}\,{\frac {\rho\,{\it L}}{ \left( {\rho}^{2}+
{\it Mg} \right) ^{2}}};
{\bf V}2=[{ V^{1}}=i\frac{{\it L}}{{\rho}^{2}+
{\it Mg}}V^{3},V^2=0,V^{3}],{\bf V}2={\bf \bar Z},
\nonumber\\
&&{\rm Type:}\{T, Z, \bar Z\}.
\end{eqnarray}

\subsection{Hirschman--Welch solution; energy and mass}\label{ModHW}

Accomplishing in the original Peldan solution, Eq. (\ref{PeldanMagstatic}),
the coordinate transformation
$$t\rightarrow t,\,\rho\rightarrow
\sqrt{(\rho^2+r_{+}^2-m\,l^2)/l^2},\phi\rightarrow
\phi\,l^2,\,\chi^2:=a^2\,l^2$$, one obtains the Hirschman--Welch
magnetostatic solution representation~\cite{Hirschmann96}, see
also~\cite{GarciaAnnals09}~(4.33), which is given by the metric
functions
\begin{eqnarray}\label{HWstatic}
ds^2&&=-N(\rho)^2dt^2+\frac{1}{L(\rho)^{2}}\,d\rho^2
+K(\rho)^2[d\phi+W(\rho)dt]^2,\nonumber\\
H(\rho)&=&\frac{\rho^2+r_{+}^2-m\,l^2}{l^2},\,L(\rho)=
\frac{\sqrt{H(\rho)}}{\rho}\,K(\rho) ,\nonumber\\
K(\rho)&=&\sqrt{\rho^2+r_{+}^2+\chi^2\ln{H(\rho)}},\nonumber\\
N(\rho)&=&\sqrt{H(\rho)}, \,W(\rho)=0.
\end{eqnarray}

In the original Hirschman--Welch work~\cite{Hirschmann96}
there is a condition to be
fulfilled by the parameter $r_{+}$, arising from the vanishing
of $K$ at $\rho=0$, namely
\begin{eqnarray}
r_{+}^2+\chi^2\ln(\frac{r_{+}^2}{l^2}-m)=0\rightarrow
(\frac{r_{+}^2}{l^2}-m){\rm e}^{(r_{+}^2/\chi^2)}=1.
\end{eqnarray}
This equation has been used in the quoted publication to
determine the conical angle deficit:
as $\rho\rightarrow 0$ the behavior of $K^2/\rho^2=
1+\chi^2/\rho^2\ln{[(\rho^2+r_{+}^2-m{l^2})/(r_{+}^2-m{l^2})]}$
is given by
$$(K^2/\rho^2)|_{\rho\rightarrow 0}
\rightarrow{\frac{r_{+}^2-m{l^2}+\chi^2}{r_{+}^2-m{l^2}}
=1+\frac{\chi^2}{l^2}\,{\rm e}^{(r_{+}^2/\chi^2)}},$$
hence the spatial sector $(\frac{1}{L^{2}}\,d\rho^2
+K^2\,d\phi^2)_{|\rho\rightarrow 0}$ of the studied
metric behaves as
$${[d\large(\rho\,\frac{{\rm e}^{(r_{+}^2/2\chi^2)}}
{\sqrt{1+\frac{\chi^2}{l^2}\,{\rm e}^{(r_{+}^2/\chi^2)}}}})]^2
+\frac{\rho^2\,{\rm e}^{(r_{+}^2/\chi^2)}}{{1+\frac{\chi^2}{l^2}
\,{\rm e}^{(r_{+}^2/\chi^2)}}}
\,[d\phi\,{{\rm e}^{(-r_{+}^2/2\chi^2)}}{(1+\frac{\chi^2}{l^2}
\,{\rm e}^{(r_{+}^2/\chi^2)})}]^2=:d\tilde\rho^2
+\tilde\rho^2d\tilde\phi^2,$$
hence the angles ranges
$$0\leq \tilde \phi \leq 2\pi
\rightarrow{0\leq \phi \leq 2\pi\,
\frac{{\rm e}^{(r_{+}^2/2\chi^2)}}{1+\frac{\chi^2}{l^2}{\rm e}^{(r_{+}^2/\chi^2)}}
=2\pi\,\frac{l\,\sqrt{r_{+}^2-m{l^2}}}{r_{+}^2-m{l^2}+\chi^2}},$$
thus, the conical singularity at $\rho=0$, as pointed out in the HW paper, arises in
$\phi$ with the period $T_{\phi}=2\pi\nu:=2\pi\,
{{\rm e}^{(r_{+}^2/2\chi^2)}}/
(1+\frac{\chi^2}{l^2}{\rm e}^{(r_{+}^2/\chi^2)}),$
consequently the
angle deficit is  $\delta T_{\phi}=2\pi(1-\nu)$  as
reported also in~\cite{Dias-Lemos-JHEP02}.\\

\subsection{ Mass, energy and momentum for the HW solution}\label{HWMag}

For this electromagnetic field solution the surface energy density is given by
\begin{eqnarray}
\epsilon(\rho,\epsilon_{0})&=&-\frac{1}{\pi\,l}\,
\frac{\rho^2+r_{+}^2-ml^2+\chi^2}{\sqrt{\rho^2+r_{+}^2
+\chi^2\ln{H}}\sqrt{\rho^2+r_{+}^2-l^2m}}-\epsilon_{0},\nonumber\\
\end{eqnarray}
while the integral energy and mass amount to
\begin{eqnarray}\label{den-HW}
E(\rho,\epsilon_{0})&=&-\frac{2}{l}\,
\frac{\rho^2+r_{+}^2-ml^2+\chi^2}{\sqrt{\rho^2
+r_{+}^2-l^2m}}-2\pi\,K\,\epsilon_{0},\nonumber\\
M(\rho,\epsilon_{0})&=&-\frac{2}{l^2}
(\rho^2+r_{+}^2-ml^2+\chi^2)-2\pi\,N\,K\epsilon_{0}
\end{eqnarray}
The evaluation of the  above functions independent of $\epsilon_{0}$ behave at
infinity according to
\begin{eqnarray}\label{enerEpsZ}
\epsilon({\rho\rightarrow\infty},\epsilon_{0}=0)&\approx&-\frac{1}{\pi\,l}+
\frac{ml^2-2\chi^2}{2\pi\,l\,\rho^2}
+\frac{\chi^2}{\pi\,l\,\rho^2}\ln{(\frac{\rho}{l})},\nonumber\\
E({\rho\rightarrow\infty},\epsilon_{0}=0)&\approx&
\,-2\frac{\rho}{l}+\frac{m\,l^2-r_{+}^2-2\chi^2}{l\,\rho}
,\nonumber\\
M({\rho\rightarrow\infty},\epsilon_{0}=0)&=
&-\frac{2}{l^2}(\rho^2+r_{+}^2-ml^2+\chi^2).
\end{eqnarray}
Using in the expressions (\ref{den-HW})
the energy density for the
anti--de Sitter solution counterpart, namely
$\epsilon_{0}=-\frac{1}{\pi\,l^2}\rho/\sqrt{M_{0}+\frac{\rho^2}{l^2}}$,
which at the spatial infinity behaves as
$\epsilon_{0\mid\infty}(M_{0})\approx
-\frac{1}{\pi\,l}+\frac{l\,M_{0}}{2\pi\,\rho^2}$, the series
expansions of the corresponding quantities at
$\rho\rightarrow{\infty}$ result in
\begin{eqnarray}
\epsilon({\rho\rightarrow\infty},\epsilon_{0\mid\infty}(M_{0})) &\approx&
-\frac{l\,M_{0}}{2\pi\,\rho^2}+\frac{ml^2-2\chi^2}{2\pi\,l\rho^2}
+\frac{\chi^2}{\pi\,l\,\rho^2}\ln{(\frac{\rho}{l})},\nonumber\\
E({\rho\rightarrow\infty},\epsilon_{0\mid\infty}(M_{0}))&\approx
&\,-\frac{l\,M_{0}}{\rho}+\frac{ml^2-2\chi^2}{l\,\rho}
+2\frac{\chi^2}{l\,\rho}\ln{(\frac{\rho}{l})},\nonumber\\
M({\rho\rightarrow\infty},\epsilon_{0\mid\infty}(M_{0}))&\approx&
m-M_{0}-2\frac{\chi^2}{l^2}+2\frac{\chi^2}{l^2}\ln{(\frac{\rho}{l})}.
\end{eqnarray}
Therefore, comparing with the energy characteristics of the BTZ
solution, one concludes that the mass logarithmically diverges at
spatial infinity, and that the role of mass is played by $m$.

\subsection{Field, energy-momentum and Cotton tensors
for the generalized--via $SL(2,R)$ transformations--Peldan solution}

Under $SL(2,R)$  transformations of the form
\begin{eqnarray}
t&&=\alpha\,T+\beta\,\Phi ,\,\,\alpha\delta-\beta\gamma=1,\nonumber\\
\phi&&=\gamma\,T+\delta\,\Phi ,
\end{eqnarray}
the metric transforms into
\begin{eqnarray}
\left({g_{\alpha}}_{\beta}\right)= \left[ \begin {array}{ccc} -{ {
  {{\alpha}}^{2}}{\, \left( {\rho}^{2}+{\it Mg} \right) }}
  +{\gamma}^{2}{\frac {
{\it L^2}\,{\rho}^{2}}{\, \left( {\rho}^{2}+{\it
Mg} \right) }}&0&-{ {  \alpha\beta\,}{\, \left( {\rho}^{2}+{
\it Mg} \right) }}+\gamma\,\delta\,{\frac {{\it L^2}\,{\rho}^{2}}{
\, \left( {\rho}^{2}+{\it Mg} \right) }}\\\noalign{\medskip}0&
\frac{1}{{
\it L^2}}&0\\\noalign{\medskip}-{ {  \alpha\beta\,}{
\, \left( {\rho}^{2}+{\it Mg} \right) }}
+\gamma\,\delta\,{\frac {{\it
L^2}\,{\rho}^{2}}{\, \left( {\rho}^{2}+{\it Mg} \right) }}&0
&-{ {  {\beta}^{2}}{\, \left( {\rho}^{2}+{\it Mg} \right) }}+{\delta}^{2}{
\frac {{\it L^2}\,{\rho}^{2}}{\, \left( {\rho}^{
2}+{\it Mg} \right) }}\end {array} \right] ,
\end{eqnarray}
and the field tensor becomes
\begin{eqnarray}
({F^\alpha}_{\beta})= \left[
\begin {array}{ccc} 0&{\frac {\beta\,a}
{ {} {\it L^2}\,\rho}}&0
\\\noalign{\medskip}{\frac {\gamma\,{\it L^2}\,\rho\,a}
{\left( {\rho}^{2}+{\it Mg} \right) }}&0&{\frac {\delta\,{\it
L^2}\,\rho\,a}{\left( {\rho}^{2}+{\it Mg} \right) }
}\\\noalign{\medskip}0&-{\frac {{\alpha}\,a}{  {\it L^2}\,\rho}}&0
\end {array} \right],
\end{eqnarray}
\begin{eqnarray}
\lambda_1&&=0;
{\bf V}1=[{V^{1}}={V^{1}},{V^{2}}=0,
{V^{3}}=-{\frac {\,
\gamma}{\delta}}{V^{1}} ],\nonumber\\
&&V^{\mu}V_{\mu}=-{V^{1}}^2({\rho}^{2}+{\it Mg})/\delta^2,\,
{\bf V}1={\bf T}1,\nonumber\\
\lambda_2&&=i{\frac {a}{\sqrt {{\rho}^{2}+{\it Mg}}}};
{\bf V}2=[ {V^{1}}=-\frac{\beta}{\alpha}{V^{3}},
{V^{2}}=-i{\frac { {\it L^2}\,\rho\,{V^{3}}}{\alpha\,
\sqrt {{\rho}^{2}+{\it Mg}}}},{V^{3}}={V^{3}}],
{\bf V}2={\bf Z},\nonumber\\
\lambda_3&&=-i{\frac {a}{\sqrt {{\rho}^{2}+{\it Mg}}}};
{\bf V}3=[ {V^{1}}=-\frac{\beta}{\alpha}{V^{3}},
{V^{2}}=-i{\frac { {\it L^2}\,\rho\,{V^{3}}}{\alpha\,
\sqrt {{\rho}^{2}+{\it Mg}}}},{V^{3}}={V^{3}} ],
{\bf V}3={\bf \bar Z},\nonumber\\
\end{eqnarray}
and the energy tensor
\begin{eqnarray}
({T^\alpha}_{\beta})= \left[
\begin {array}{ccc} \frac{1}{8} \,{\frac {{a}^{2}
\left( \beta\,\gamma+{\alpha}\,\delta \right) }
{{\rho}^{2}\pi }}&0&-\frac{1}{4}\,
{\frac {\gamma\,\alpha\,{a}^{2}
}{{\rho}^{2}\pi }}
\\\noalign{\medskip}0&\frac{1}{8}\,
{\frac {{a}^{2}}{\pi \,{\rho}^{2}}}&0
\\\noalign{\medskip}\frac{1}{4} \,
{\frac {\delta\,\beta\,{a}^{2}}
{{\rho}^{2}\pi }}&0&-\frac{1}{8}\,{
\frac{{a}^{2}
\left( \beta\,\gamma+\alpha\,\delta \right) }
{{\rho}^{2}\pi }}
\end {array} \right] .
\end{eqnarray}
allows for the following eigenvectors
\begin{eqnarray}
\lambda_1&&=-\frac{1}{8\,\pi}{\frac {a^2}{\rho^{2} +\,{\it Mg}}};
{\bf V}1=[{V^{1}}={V^{1}},{V^{2}}=0,{V^{3}}
=-\frac {\gamma}{\delta}{V^{1}} ],\nonumber\\
&&V^{\mu}V_{\mu}=-{V^{1}}^2({\rho}^{2}+{\it Mg})/\delta^2,\,
{\bf V}1={\bf T}1,\nonumber\\
\lambda_2&&=\frac{1}{8\,\pi}{\frac {{a}^{2}}{{\rho}^{2} +
\,{\it Mg}}};
{\bf V}2=[ {V^{1}}=-\frac{\beta}{\alpha}{V^{3}},
{V^{2}}={V^{2}},{V^{3}}={V^{3}} ],\nonumber\\
&&V^{\mu}V_{\mu}=\frac{{V^{2}}^2}{{\it L^2}}
+\frac{{\it L^2}\rho^2}{{\rho}^{2}+{\it Mg}}\frac{{V^{3}}^2}{\alpha^2},\,
{\bf V}2={\bf S}2,\nonumber\\
\lambda_3&&=\frac{1}{8\,\pi}{\frac {{a}^{2}}{{\rho}^{2}
+
\,{\it Mg}}};
{\bf V}3=[ \tilde{V^{1}}=-\frac{\beta}{\alpha}
\tilde{V^{3}},
\tilde{V^{2}}=\tilde{V^{2}},\tilde{V^{3}}=
\tilde{V^{3}} ],\nonumber\\
&&V^{\mu}V_{\mu}=\frac{\tilde{V^{2}}^2}{{\it L^2}}
+\frac{{\it L^2}\rho^2}{{\rho}^{2}+{\it Mg}}\frac{\tilde{V^{3}}^2}{\alpha^2},\,
{\bf V}3={\bf S}3,\nonumber\\
&&{\rm Type:}\{T, 2\,S\}.
\end{eqnarray}
The transformed Cotton tensor is given by
\begin{eqnarray}
({C^\alpha}_{\beta})=\left[ \begin {array}{ccc} -\frac{1}{2}\,{
\frac {\alpha\,\beta\,{a}^{2}}{ \left( {\rho}^{2}+{\it Mg}
 \right)}}-\frac{1}{2}\,{
\frac {{a}^{2}{\it L^2}\,{\rho}^{2}\gamma\,\delta}{ \left( {\rho}^
{2}+{\it Mg} \right) ^{3}  }}&0&
-\frac{1}{2}\,{\frac {{a}^{2}{\beta}^{2}}{ \left( {\rho}^{2}+{
\it Mg} \right)  {} }}-
\frac{1}{2}\,{\frac {{a}^{2}
{\it L^2}\,{\rho}^{2}{\delta}^{2}}{ \left( {
\rho}^{2}+{\it Mg} \right) ^{3}}}\\
\noalign{\medskip}0&0&0\\
\noalign{\medskip}\frac{1}{2}\,{
\frac {{a}^{2}{\alpha}^{2}}{ \left( {\rho}^{2}
+{\it Mg} \right)}}+\frac{1}{2}\,{\frac {{a}
^{2}{\it L^2}\,{\rho}^{2}{\gamma}^{2}}{ \left( {\rho}^{2}
+{\it Mg}\right) ^{3}}}&0&\frac{1}{2}
\,{\frac {\alpha\,\beta\,{a}^{2}}{ \left( {\rho}^{2}+{\it Mg}
 \right)  {} }}+\frac{1}{2}\,
 {\frac {{a}^{2}{\it L^2}\,
 {\rho}^{2}\gamma\,\delta}{ \left( {\rho}^
{2}+{\it Mg} \right) ^{3}  }}\end {array} \right],
\end{eqnarray}
and has the eigenvectors
\begin{eqnarray}
&&\lambda_1=0;{\bf V}1=[0,V2,0],V^{\mu}V_{\mu}=\frac{{V^{2}}^2}{\it L^2},\,
{\bf V}1={\bf S}1,\nonumber\\&&
\lambda_2=i\,\frac{1}{2}\,{\frac {\rho\,{a}^{2}{\it L}}{ \left( {\rho}^{2}+
{\it Mg} \right) ^{2}}};\nonumber\\&&
{\bf V}2=[{ V^{1}}=-{\frac {{ V^{3}}\, \left[ {\beta}^{2} \left( {\rho}^{2}
+{\it Mg} \right) ^{2}+{\it L^2}\,{\rho}^{2}{\delta}^{2} \right]
}{\beta\,{\alpha}\, \left( {\rho}^{2}+{\it Mg} \right) ^{2}-i
{\it L}\rho\, \left( {\rho}^{2}+{\it Mg} \right)  +{\it L^2}\,{\rho}^{2}
\gamma\,\delta}},V^{2}=0,V^{3}],\,{\bf V}2={\bf Z}\nonumber\\&&
\lambda_3=-i\,\frac{1}{2}\,{\frac {\rho\,{a}^{2}{\it L}}{ \left( {\rho}^{2}+
{\it Mg} \right) ^{2}}};\nonumber\\&&
{\bf V}3=[{ V^{1}}=-\frac {{ V^{3}}\, \left[ {\beta}^{2} \left( {\rho}^{2}
+{\it Mg} \right) ^{2}+{\it L^2}\,{\rho}^{2}{\delta}^{2} \right]
}{\beta\,{\alpha}\, \left( {\rho}^{2}+{\it Mg} \right) ^{2}+i
{\it L}\rho\, \left( {\rho}^{2}+{\it Mg} \right)  +{\it L^2}\,{\rho}^{2}
\gamma\,\delta},V^{2}=0,V^{3}],\,{\bf V}3={\bf \bar Z},\nonumber\\
&&{\rm Type:}\{S, Z,\bar Z\}.
\end{eqnarray}

\subsection{False Stationary Peldan magnetostatic solution}

The magnetostatic
solution with a negative cosmological constant is determined by the
metric
\begin{equation*}
ds^2=-N(\rho)^2dt^2+\frac{1}{L(\rho)^{2}}\,d\rho^2
+K(\rho)^2[d\phi+W(\rho)dt]^2,
\end{equation*}
\begin{eqnarray}
&&h(\rho):=\rho^2+M_{+},\nonumber\\
&&K(\rho)=\frac{1}{l}\sqrt{K_{0}+{\rho^2}+a^2{l^2}\ln{h(\rho)}},\nonumber\\
&&L(\rho)=\frac{1}{\rho\,l}\sqrt{(K_{0}+{\rho^2}
+a^2{l^2}\ln{h(\rho)})h(\rho)},\nonumber\\
&&N(\rho)=\sqrt{h(\rho)},\,\,W(\rho)=-2J_{0}.
\end{eqnarray}
The surface energy and momentum densities are given by
\begin{eqnarray}
&&\epsilon(\rho)=-\frac{1}{\pi\,l^2\,K(\rho)
\,N(\rho)}({\rho^2}+M_{+}+{a^2}{l^2})-
\epsilon_{0},\nonumber\\
&&j(\rho)=0.
\end{eqnarray}

Since the momentum $j(\rho)$ is zero, the stationarity
of this solution is fictitious, as a matter of fact
one is dealing with a static metric.

\section{Dias--Lemos solution}
Subjecting the static Hirschman--Welch metric (\ref{HWstatic})
to the $SL(2,R)$ transformation $$t
\rightarrow{}\sqrt {1+{\omega}^{2}}t-\omega\,l\phi,
\,\rho\rightarrow{}\rho,\phi\rightarrow{}-{
\frac {\omega}{l}\,t}+\sqrt {1+{\omega}^{2}}\phi,
$$
one arrives at
the Dias--Lemos solution~\cite{Dias-Lemos-JHEP02},
see also~\cite{GarciaAnnals09}~(11.41), determined by the metric
\begin{eqnarray}\label{DLmetric}
ds^2&&=-N(\rho)^2dt^2+\frac{1}{L(\rho)^{2}}\,d\rho^2
+K(\rho)^2[d\phi+W(\rho)dt]^2,\nonumber\\
H(\rho)&:&=(\rho^2+r_{+}^2-ml^2)/{l^2},\nonumber\\
L(\rho)&=&\frac{\sqrt{H(\rho)}}{\rho}\,\sqrt{\rho^2
+r_{+}^2+\chi^2\ln{H(\rho)}},\nonumber\\
K(\rho)&=&\sqrt{\rho^2+r_{+}^2+\omega^2l^2m
+(1+\omega^2)\chi^2\ln{H(\rho)}},\nonumber\\
N(\rho)&=&\rho\frac{\sqrt{L(\rho)}}{\sqrt{K(\rho)}},\,\,
W(\rho)=-\frac{\omega\sqrt{1+\omega^2}}{l}
\frac{[ml^2+\chi^2\ln{H(\rho)}]}{K(\rho)^2}.
\end{eqnarray}
Notice that this metric, in the case of vanishing charge $\chi=0$, yields
to an alternative coordinate representation of the
rotating BTZ solution, with parameter $\omega$,
namely
\begin{eqnarray}
ds^2&&=-\left({\frac {{\rho}^{2}+r_{+}^2}{{l}^{2}}}
-\left(1 +{\omega}^{2} \right) m\,l^2\right)dt^2
+\frac{{\rho}^{2}l^2}{\left({\rho}^{2}+{ r_{+}}^{2} \right)\left({\rho}^{2}+
r_{+}^{2}-m\,l^2 \right)}\,d\rho^2\nonumber\\
&&-2\,\omega\,m\,l\,\sqrt {1+{\omega}^{2}}d\phi\,dt
+({\rho}^{2}+{{r_{+}}}^{2}+{l}^{2}{\omega}^{2}){d\phi}^2,
\end{eqnarray}
which differs from the standard BTZ solution
representations. \noindent\\
Accomplishing in the above
metric the transformations
\begin{eqnarray*}\label{DLtrans}
t\rightarrow l^2\,t,\,\,\rho\rightarrow
\sqrt {{{\rho}^{2}}/{{l}^{2}}-m{l}^{2}{\omega}^{2}
-{{\it r_{+}}}^{2}+M}
\end{eqnarray*}
and identifying the parameters according to
\begin{eqnarray*}\label{DLparam}
m&&={\frac {M}{1+2\,{\omega}^{2}}},
M^2\frac{(1+\omega^2)\omega^2\,l^2}{(2\omega^2+1)^2}
=J^2/4,\rightarrow
\nonumber\\&& {\omega}^2=\frac{1}{2}\,
{\frac {Ml\pm\sqrt {{M}^{2}{l}^{2}-{J}^{2}}}{\sqrt {{M}^{2}{l}^{2}-
{J}^{2}}}},\,m={\frac {\sqrt {{M}^{2}{l}^{2}-{J}^{2}}}{l^3}},
\end{eqnarray*}
one arrives at the BTZ solution counterpart representation (\ref{metricSBTZcount})
\begin{eqnarray}
ds^2=-{\rho}^{2}dt^2+\left(\frac{{\rho}^{2}}{{l}^{2}}+M
+\frac{J^2}{4{\rho}^{2}}\right)^{-1}\,d\rho^2
-J\,d\phi\,dt+\left(\frac{{\rho}^{2}}{{l}^{2}}+M\right){d\phi}^2.
\end{eqnarray}
\noindent\\
On the other hand, by replacing $\rho\rightarrow
\sqrt {{\rho}^{2}-m\,{l}^{2}{\omega}^{2}-{{\it r_{+}}}^{2}} $
one arrives at the middle of the road metric
\begin{eqnarray}
ds^2&&=-\left({\frac {{\rho}^{2}}{{l}^{2}}}
-\left(1 +2\,{\omega}^{2} \right) m\right)dt^2
+\left({{\frac {{\rho}^{2}}{{l}^{2}}}-(1+2\,{\omega}^{2})m+\frac{{l}^{2}{m}^{2}{
\omega}^{2}(1+{
\omega}^{2})}{\rho^2}}\right)^{-1}\,d\rho^2\nonumber\\
&&-2\,\omega\,m\,l\,\sqrt {1+{\omega}^{2}}d\phi\,dt+\rho^2{d\phi}^2,
\end{eqnarray}
which, identifying
\begin{eqnarray*}
m&&={\frac {M}{1+2\,{\omega}^{2}}},\,{l}^{2}{m}^{2}{
\omega}^{2}(1+{
\omega}^{2})=J^2/4\rightarrow \nonumber\\
{\omega}^2&&=\frac{1}{2}\,
{\frac {Ml\pm\sqrt {{M}^{2}{l}^{2}-{J}^{2}}}{\sqrt {{M}^{2}{l}^{2}-
{J}^{2}}}},\,m=
{\frac {\sqrt {{M}^{2}{l}^{2}-{J}^{2}}}{l}},\,2\,
\sqrt {1+{\omega}^{2}}\omega\,ml\,\rightarrow J,
\end{eqnarray*}
gives rise to the standard description of the stationary BTZ
black hole metric (\ref{BTZmetric}).\\

Therefore, as the vacuum limit of the DL metric (\ref{DLmetric})
one may consider the rotating BTZ solution counterpart, and
consequently one may think  it of as the reference vacuum solution in
the evaluation of the quasi local energy, momentum and mass.

\subsection{ Mass, energy and momentum for the DL solution}\label{DLMag}

The surface energy and momentum densities are given by
\begin{eqnarray}
\epsilon(\rho,\epsilon_{0})&=&-\rho\,\frac{L}{\pi\,l^2\,K^2\,{H}}\,
(\rho^2+r_{+}^2-ml^2+(1+\omega^2)\chi^2)-\epsilon_{0}\nonumber\\
j(\rho,\epsilon_{0})&=&\omega\sqrt{1+\omega^2}\frac{\rho}{\pi\,l}\frac{L}{N\,K^2}
[m\,l^2-\chi^2+\chi^2\ln{H}],
\end{eqnarray}
while the integral quantities amount to
\begin{eqnarray}\label{den-DL}
J(\rho,\epsilon_{0})&&=2\omega\sqrt{1+\omega^2}\,\frac{\rho}{l}\frac{L}{N
\,K}[m\,l^2-\chi^2+\chi^2\ln{H}]
=\frac{2}{l}\omega\sqrt{1+\omega^2}\,[m\,l^2-\chi^2+\chi^2\ln{H}],\nonumber\\
E(\rho,\epsilon_{0})&&=-2\frac{\rho}{l^2}\frac{L}{K\,{H}}
\,P(\rho)-2\pi\,K\epsilon_{0},\nonumber\\
M(\rho,\epsilon_{0})&&=-2\frac{\rho}{l^2}\frac{N L}{H
K}\,P(\rho)-W\,J
-2\pi\,N\,K\epsilon_{0}\nonumber\\
&&=-\frac{2}{l^2}(\rho^2+\chi^2+r_{+}^2-ml^2)+2\frac{\omega^2}
{l^2}[ml^2-\chi^2+\chi^2\ln{H}]-2\pi\,N\,K\epsilon_{0},\nonumber\\
P(\rho)&&:=\rho^2+r_{+}^2-ml^2+(1+\omega^2)\chi^2.
\end{eqnarray}
The evaluation of the main parts of above functions, i.e., the
corresponding functions independent of $\epsilon_{0}$ behave at
infinity according to
\begin{eqnarray}
j({\rho\rightarrow\infty})&\approx&\frac{\omega}{l\pi\rho}\sqrt{1+\omega^2}
[m\,l^2-\chi^2+2{\chi^2}\ln{(\frac{\rho}{l})}],\nonumber\\
J({\rho\rightarrow\infty})&\approx&2\frac{\omega}{l}\sqrt{1+\omega^2}
[m\,l^2-\chi^2+2{\chi^2}\ln{(\frac{\rho}{l})}],\nonumber\\
\epsilon({\rho\rightarrow\infty},\epsilon_{0}=0)&\approx&-\frac{1}{\pi\,l}
+\frac{ml^2-2\chi^2}{2\pi\,l\,\,\rho^2}+\frac{\chi^2}{\pi\,l\,\rho^2}
\ln{(\frac{\rho}{l})}\nonumber\\
&&{}+\omega^2[\frac{m\,l^2-\chi^2}
{\pi\,l\,\,\rho^2}+2\frac{\chi^2}{\pi\,l\,\rho^2}
\ln{(\frac{\rho}{l})}],\nonumber\\
E({\rho\rightarrow\infty},\epsilon_{0}=0)&\approx&-\frac{2\rho}{l}
+\frac{ml^2-r_{+}^2-2\chi^2}{l\,\,\rho}+\frac{\omega^2}{l\,\rho}
[m\,l^2-2\chi^2+2{\chi^2}\ln{(\frac{\rho}{l})}]
,\nonumber\\
M({\rho\rightarrow\infty},\epsilon_{0}=0)&
\approx&\,2m-2\frac{1}{l^2}(\rho^2+r_{+}^2+\chi^2)+2\frac{\omega^2}{l^2}
[m\,l^2-\chi^2+2{\chi^2}\ln{(\frac{\rho}{l})}].
\end{eqnarray}
Using in the expressions (\ref{den-DL}) as reference energy
density the quantity
$\epsilon_{0}=-\frac{1}{\pi\rho}\sqrt{\frac{\rho^2}{l^2}-M_{0}}$,
which at the spatial infinity behaves as
$\epsilon_{0\mid\infty}(M_{0})\approx
-\frac{1}{\pi\,l}+\frac{lM_{0}}{2\pi\,\rho^2},$ the series
expansions of the corresponding quantities at
$\rho=$ infinity
result in
\begin{eqnarray}
\epsilon({\rho\rightarrow\infty},
\epsilon_{0\mid\infty}(M_{0}))&\approx&
\frac{l}{2\pi\,\rho^2}(m-M_{0})
-\frac{1}{2\pi\,l\,\rho^2}({\chi^2}-\chi^2
\ln{(\frac{\rho}{l})})\nonumber\\
&{}&+\frac{\omega^2}{\pi\,l\,\rho^2}[ml^2-\chi^2
+2\chi^2\ln{(\frac{\rho}{l})}],\nonumber\\
E ({\rho\rightarrow\infty},
\epsilon_{0\mid\infty}(M_{0}))&
\approx&\,\,\frac{l}{\rho}(m-M_{0})
-2\frac{\chi^2}{l\,\,\rho}
+2\frac{\chi^2}{l\,\rho}\ln{(\frac{\rho}{l})}\nonumber\\
&&{}+2\frac{\omega^2}{{l\,\rho}}[{m\,l^2-\chi^2}
+2{\chi^2}\ln{(\frac{\rho}{l})}],\nonumber\\
M ({\rho\rightarrow\infty},\epsilon_{0\mid\infty}(M_{0}))
&\approx&\,\,m
-M_{0}-2\frac{\chi^2}{l^2}+2\frac{\chi^2}{l^2}
\ln{(\frac{\rho}{l})}
+2\frac{\omega^2}{l^2}[m\,l^2-\chi^2
+2\chi^2\ln{(\frac{\rho}{l})}].\nonumber\\
\end{eqnarray}
Therefore, comparing with the energy characteristics of the BTZ
solution, one concludes that the mass logarithmically diverges at
spatial infinity. For vanishing rotation parameter $\omega$ one
recovers the static solution in the representation of
Hirschman--Welch and certainly the corresponding energy quantities.
The parameter $m$ can be considered as the BTZ mass.

\subsection{Field, energy and Cotton tensors for the Dias--Lemos solution}

To determine the algebraic types of the electromagnetic field,
energy--momentum, and Cotton tensors it is more convenient
to work with the DL metric
in the form
\begin{eqnarray}
g= \left[
\begin {array}{ccc} -{\it h}-{\omega}^{2}\left({\it h}-{\it L^2}\right)
&0& l\,\sqrt {1+{\omega}^{2}}\omega\, \left( {\it h}-{\it
L^2} \right) \\\noalign{\medskip}0&{\frac {{\rho}^{2}}{{\it h
}\,{\it L^2}\,{l}^{2}}}&0\\\noalign{\medskip} l\,\sqrt {1+{\omega}^
{2}}\omega\, \left( {\it h}-{\it L^2} \right)&0& {l}^{2}\,\left( -{
\omega}^{2}{\it h}+{\it L^2}+{\it L^2}\,{\omega}^{2}
 \right) \end {array} \right] ,
\end{eqnarray}
where
\begin{eqnarray}
{\it L^2}=\frac {{\chi}^{2}\ln \left( {\it h} \right)
+{\rho}^{2}}{{l}^{2}},
h=\frac{{\rho}^{2}+{\it Mg}}{{l}^{2}},\,{\it Mg}=r_{+}^{2}-m{l}^{2}.
\end{eqnarray}
In his representation, the electromagnetic field tensor
becomes
\begin{eqnarray}
({F^\alpha}_{\beta})= \left[
\begin {array}{ccc} 0&-{\frac {\rho\,\chi\,\omega}{{l}^{3}{\it
L^2}\,{\it h}}}&0\\\noalign{\medskip}-{\frac {\omega\,\chi\,{
\it L^2}}{l\rho}}&0&{\frac {\chi\,\sqrt {1+{\omega}^{2}}{\it
L^2}}{\rho}}\\\noalign{\medskip}0&-{\frac {\rho\,\chi\,\sqrt {1+
{\omega}^{2}}}{{\it L^2}\,{l}^{4}{\it h}}}&0\end {array}
 \right],
\end{eqnarray}
with the following eigenvalues and their corresponding eigenvectors
\begin{eqnarray}
\lambda_1&&=0;{\bf V}1=[{V^{1}}={V^{1}},{V^{2}}=0,{V^{3}}={\frac {\omega}{
l\,\sqrt {1+{\omega}^{2}}}}\,{V^{1}}],V_\mu\,V^\mu=-\frac{h}{1+\omega^2}{V^{1}}^2
\nonumber\\
&&{\bf V}1={\bf T}1,\nonumber\\
\lambda_2&&=-i{\frac {\chi}{\sqrt {{\it h}}{l}^{2}}}
;{\bf V}2=[{V^{1}}=i{\frac {\omega\,\rho}{l{\it
L^2}\,\sqrt {{\it h}}}}\,{V^{2}},{V^{2}}={V^{2}},{V^{3}}=i{\frac {\sqrt {1+{
\omega}^{2}}\rho}{{\it
L^2}\,{l}^{2}\sqrt {{\it h}}}}{V^{2}}],\nonumber\\
&&{\bf V}2={\bf Z},\nonumber\\
\lambda_3&&=i{\frac {\chi}{\sqrt {{\it h}}{l}^{2}}};
{\bf V}3=[{V^{1}}=-i{\frac {\omega\,\rho}{l{\it
L^2}\,\sqrt {{\it h}}}}\,{V^{2}},{V^{2}}={V^{2}},{V^{3}}=-i{\frac {\sqrt {1+{
\omega}^{2}}\rho}{\,{\it
L^2}\,{l}^{2}\sqrt {{\it h}}}}\,{V^{2}}],\nonumber\\
&&{\bf V}3={\bf \bar Z},\nonumber\\
&&{\rm Type}:\{T,Z,\bar Z\}.
\end{eqnarray}
As far as to the electromagnetic energy momentum tensor is concerned, its matrix
is given by
\begin{eqnarray}
({T^\alpha}_{\beta})=  \left[
\begin {array}{ccc} -\frac{1}{8\,\pi }\,{\frac {{\chi}^{2} \left( 1+2\,{\omega}^{2}
 \right) }{{l}^{2} \left( {\rho}^{2}+{\it Mg} \right)}}&0&\frac{1}{4\,\pi}\,{
\frac {\omega\,{\chi}^{2}\sqrt {1+{\omega}^{2}}}{l \left( {\rho}^{2}+{
\it Mg} \right)}}\\\noalign{\medskip}0&
\frac{1}{8\,\pi}\,{\frac {{\chi}^{2}}{{l
}^{2} \left( {\rho}^{2}+{\it Mg} \right) }}&0
\\\noalign{\medskip}-
\frac{1}{8\,\pi}\,
{\frac {\omega\,{\chi}^{2}\sqrt {1+{\omega}^{2}}}{{l}^{3} \left( {
\rho}^{2}+{\it Mg} \right)  }}&0&
\frac{1}{8\,\pi}\,{\frac {{\chi}^{2} \left( 1+2
\,{\omega}^{2} \right) }{{l}^{2} \left( {\rho}^{2}+{\it Mg} \right)
 }}\end {array} \right],
\end{eqnarray}
with the following eigenvalues and their corresponding eigenvectors
\begin{eqnarray}
\lambda_1&&=-\frac{1}{8\,\pi}\,
{\frac {{\chi}^{2}}{{l}^{2} \left( {\rho}^{2}+{\it Mg} \right)
 }};{\bf V}1=[{V^{1}},0,{
\frac {\omega}{l\,\sqrt {1+{\omega}^{2}}}}\,{V^{1}}],\nonumber\\
&&
V_\mu\,V^\mu=-\frac{h}{1+\omega^2}{V^{1}}^2,
{\bf V}1={\bf T}1,\nonumber\\
\lambda_2&&=\frac{1}{8\,\pi}\,
{\frac {{\chi}^{2}}{{l}^{2} \left( {\rho}^{2}+{\it Mg} \right)
}}
;{\bf V}2=[{V^{1}}={\frac {\omega\,l}
{\sqrt {1+{\omega}^{2}}}}\,{V^{3}},{
V^{2}}={V^{2}},{V^{3}}={V^{3}}],\nonumber\\
&&
\,V^{\mu}V_{\mu}=\frac {{l}^{2}{\it L^2}}
{1+{\omega}^{2}}{V^{3}}^{2}+{
\frac {{\rho}^{2}}{{\it h}\,
{\it L^2}\,{l}^{2}}}{{V^{2}}}^{2},\,
{\bf V}2={\bf S}2,
\nonumber\\
\lambda_3&&=\frac{1}{8\,\pi}\,
{\frac {{\chi}^{2}}{{l}^{2} \left( {\rho}^{2}+{\it Mg} \right)
 }};{\bf V}3=[{\frac {\omega
\,\,l}{\sqrt {1+{\omega}^{2}}}}{ \tilde V^{3}},
{ \tilde V^{2}},{\it \tilde V^{3}}
],\nonumber\\
&&V^{\mu}V_{\mu}=\frac {{l}^{2}{\it L^2}}
{1+{\omega}^{2}}({{\tilde V}^{3}})^{2}+{
\frac {{\rho}^{2}}{{\it h}\,{\it L^2}\,{l}^{2}}}
({{\tilde V^{2}}})^{2},\,{\bf V}3={\bf S}3,\nonumber\\
&&{\rm Type}:\{T,2\,S\}.
\end{eqnarray}
This tensor structure corresponds to that one describing a
perfect fluid energy momentum tensor, but this time for the state
equation: $\it {energy = pressure}$. Again, the solutions generated from this
metric by using coordinate transformations  possesses this perfect fluid
feature because the invariance of the eigenvalues.

The Cotton tensor for stationary cyclic symmetric gravitational field is
given by
\begin{eqnarray}
({C^\alpha}_{\beta})=\left[
\begin {array}{ccc} -\frac{{\chi}^{2}}{2}
\,{\frac {\omega\,\sqrt {1+{\omega}^{2}} \left( {\it h}
+{\it L^2} \right) }{{{\it h}}^{2}{l}^{5}}}
&0&\frac{{\chi}^{2}}{2}\,\frac { \left( {\omega}^{2}{\it h}+{\it L^2}
+{\it L^2}\,{\omega}^{2} \right)}{{\it h}^{2}{l}^{4}}
\\\noalign{\medskip}0&0&0\\
\noalign{\medskip}-\frac{\chi^2}{2}\,\frac{\left( {\it h}
+\omega^2{\it h}+{\it L^2}\,\omega^2\right)}
{{\it h}^2\,l^6}&0&
\frac{\chi^2}{2}\,\frac{\omega\,\sqrt{1+{
\omega}^2}\left( {\it h}+{\it L^2} \right)}{{\it h}^2\,{l}^{5}}
\end {array}
\right].
\end{eqnarray}
Searching for its eigenvectors, one arrives at
\begin{eqnarray}
&&\lambda_{1}=0;{\bf V}1=[{V^{1}}=0,{V^{2}}={V^{2}},{V^{3}}=0],
V_\mu\,V^\mu={\frac {{\rho}^{2}}{{\it h
}\,{\it L^2}\,{l}^{2}}}{V^{1}}^2,
{\bf V}1={\bf S},\nonumber\\
&&\lambda_2=\frac{i}{2}\,{\frac {{\chi}^{2}\,{\it L}}{{{\it h}}^{3/2}
{l}^{5}}};\nonumber\\
&&
{\bf V}2=[{V^{1}}={\frac {{V^{3}}\,l \left( -i\,{\it
L}\sqrt {{\it h}}+\omega\,
\sqrt {1+{\omega}^{2}}{\it h}+\omega\,\sqrt {1+{
\omega}^{2}}{\it L^2} \right) }{{\it h}+{\omega}^{2}{\it h}+{
\it L^2}\,{\omega}^{2}},{V^{2}}=0,{V^{3}}={V^{3}}}],
\,{\bf V}2={\bf Z},\nonumber\\
&&\lambda_3=-\frac{i}{2}\,{\frac {{\chi}^{2}\,{\it L}}{{{\it h}}^{3/2}
{l}^{5}}};\nonumber\\
&&{\bf V}3=[{V^{1}}={\frac {{V^{3}}\,l \left( i\,{\it
L}\sqrt {{\it h}}+\omega\,
\sqrt {1+{\omega}^{2}}{\it h}+\omega\,\sqrt {1+{
\omega}^{2}}{\it L^2} \right) }{{\it h}+{\omega}^{2}{\it h}+{
\it L^2}\,{\omega}^{2}}},{V^{2}}=0,{V^{3}}={V^{3}}],\,
{\bf V}3={\bf \bar Z},\nonumber\\
&&{\rm Type}:\{T,Z,\bar Z\}.
\end{eqnarray}
The eigenvectors ${\bf V}2$ and ${\bf V}3$ are complex
conjugated while the vector ${\bf V}1$, associated to the zero
eigenvalue, occurs to be the only physically meaningful spacelike
direction in this case.

\section{Matyjasek-Zaslavski solution}

The uniform electrostatic solution~\cite{Matyjasek-cqg04},
see also~\cite{GarciaAnnals09}~(5.20), is given by the metric functions
\begin{eqnarray}
ds^2&&=-N(\rho)^2dt^2+\frac{1}{L(\rho)^{2}}\,d\rho^2
+K^2[d\phi+Wdt]^2,\nonumber\\
L(\rho)&=&N(\rho)=\sqrt{\frac{2}{l^2}\rho^2+4c_{1}\rho+c_{0}},\,K=1,\, W=0.\nonumber\\
\end{eqnarray}
\subsection{ Vanishing mass, energy and momentum of the MZ solution}\label{MZMag}
Since the surface energy density $\epsilon$ occurs proportional to
$\epsilon_{0}$, $\epsilon=-\epsilon_{0}$, consequently all the
energy--mass quantities are given through it
\begin{eqnarray}\label{den-catgx}
\epsilon=-\epsilon_{0},\, M(\rho,\epsilon_{0})= -2\pi\,N(\rho)\epsilon_{0},\,
E(\rho,\epsilon_{0})=-2\pi\epsilon_{0}.
\end{eqnarray}
Thus, for the natural choice of a vanishing reference energy density
 $\epsilon_{0}=0$ all the energy quantities vanish:
 $\epsilon=0, M(\rho,0)=0=E(\rho,0)$. On the other hand, if the reference energy is the
 one corresponding to the anti--de Sitter spacetime,
 $\epsilon_{0}=-\frac{1}{\pi\,\rho}\sqrt{\frac{\rho^2}{l^2}-M_{0}}$, the energies
 $M(\rho,\epsilon_{0})$ and $E(\rho,\epsilon_{0})$ will
 be again expressed through  $\epsilon_{0}$.

Metric
\begin{eqnarray}
g=  \left[
\begin {array}{ccc} -{N}^{2}&0&0\\\noalign{\medskip}0&{N}^{-2}&0
\\\noalign{\medskip}0&0&1\end {array} \right],{N(\rho)}^{2}
={2\,{\frac {{\rho}^{2}}{{l}^{2}}}+{\it c\_1}\,\rho+{\it c\_0}}
\end{eqnarray}

\subsection{Cotton, field, and energy--momentum tensors}

As far as the eigenvalue--vector properties of this solution one
establishes straightforwardly that the Cotton tensor ought to vanish
because of uniform character of the electromagnetic field, hence the
2+1 Matyjasek-Zaslavski gravitational field is conformally flat,
${C^\alpha}_{\beta}=0$. On the other hand the electromagnetic field
tensor
\begin{eqnarray}
({F^\alpha}_{\beta})=\left[ \begin {array}{ccc}0
&{\frac {1}{{N}^{2}l}}&0\\
\noalign{\medskip}{\frac {{N}^{2}}{l}}&0&0
\\\noalign{\medskip}0&0&0\end {array} \right],
\end{eqnarray}
allows for the eigenvectors
\begin{eqnarray}
\lambda_{1}&&=0;{\bf V}1=\left[{V^{1}}=0,{V^{2}}=0,{V^{3}} \right ],
 V_{\mu}V^\mu ={{V^{3}}}^{2},\,{\bf V}1={\bf S}1,\nonumber\\
\lambda_2&&=\frac{1}{l};
{\bf V}2=\left[{V^{1}}={V^{1}},{V^{2}}={N}^{2}{V^{1}},{V^{3}}=0 \right
],\nonumber\\&&
\,V^{\mu}V_{\mu}=0,\,{\bf V}2={\bf  N}2,\nonumber\\
\lambda_{3}&&=-\frac{1}{l};
{\bf V}3=\left[{V^{1}}={V^{1}},{V^{2}}=-{N}^{2}{V^{1}},{V^{3}}=0 \right
],\,\nonumber\\&&
V^{\mu}V_{\mu}=0,\,{\bf V}3={\bf N}3,
\end{eqnarray}
consequently its type is
$$\{S,N,N\}.$$
For the electromagnetic energy--momentum tensor we have
\begin{eqnarray}
({T^\alpha}_{\beta})=  \left[ \begin {array}{ccc}
-\,{\frac {1}{8{l}^{2}\pi }}&0&0
\\\noalign{\medskip}0&-\,{\frac {1}{8{l}^{2}\pi }}&0
\\\noalign{\medskip}0&0&\,{\frac {1}{8{l}^{2}\pi }}
\end {array} \right]
,
\end{eqnarray}
with eigenvectors
\begin{eqnarray}
\lambda_1&&=\frac {1}{8\,\pi \,{l}^{2}};{\bf V}1=\left[{V^{1}}=0,
{V^{2}}=0,{V^{3}}={V^{3}} \right],
V_{\mu}V^\mu ={V^{1}}^{2},{\bf V}1={\bf S}1,\nonumber\\
\lambda_{2,3}&&=-\frac {1}{8\,\pi \,{l}^{2}};
{\bf V}2,3=\left[{V^{1}}={V^{1}},{V^{2}}={V^{2}},{V^{3}}=0\right],\nonumber\\&&
V_{\mu}V^\mu =-{\frac { \left( {N}^{2}{V^{1}}-{V^{2}} \right)
 \left( {N}^{2}{V^{1}}+{V^{2}} \right) }{{N}^{2}}}\nonumber\\
&&\,{\bf V}2={\bf T}2,{\bf S}2,{\bf N}2,\,\,{\bf V}3={\bf T}3,{\bf S}3,{\bf N}3,
\end{eqnarray}
and therefore it allows for the types
$$\{S,2T\},\{S,2N\},\{S,2S\}.$$

\section{Cataldo solution}

The structural functions of the
Cataldo static solution~\cite{Cataldo02},
see~\cite{GarciaAnnals09}~Eq.~(4.46), are given by
\begin{eqnarray}
ds^2&&=-N(\rho)^2dt^2+\frac{1}{L(\rho)^{2}}\,d\rho^2
+K(\rho)^2[d\phi+W(\rho)dt]^2,\nonumber\\
N(\rho)&=&\rho^{(1/2-\sqrt{\alpha}/2)}
(\rho^2/l^2-M)^{(1/4+\sqrt{\alpha}/4)},\nonumber\\
L(\rho)&=&(\rho^2/l^2-M)^{(1/2)},\nonumber\\
K(\rho)&=&\rho^{(1/2+\sqrt{\alpha}/2)}
(\rho^2/l^2-M)^{(1/4-\sqrt{\alpha}/4)}.
\end{eqnarray}

\subsection{ Mass, energy and momentum for
the Cataldo solution}\label{CataldoHIB}

The corresponding surface densities occur to be
\begin{eqnarray}\label{den-catgCP1}
\epsilon(\rho,\epsilon_{0})&=&-\frac{1}{\pi
l\rho}\frac{1}{\sqrt{\rho^2-Ml^2}}
[\rho^2-Ml^2\frac{(1+\sqrt{\alpha})}{2}]
-\epsilon_{0},\nonumber\\
j_{\phi}(\rho)&=&0=J(\rho),
\end{eqnarray}
while the integral quantities amount to
\begin{eqnarray}\label{den-catgCP2}
E(\rho,\epsilon_{0})&=&-{l}^{(\sqrt{\alpha}/2-3/2)}
{(\rho^2-Ml^2)}^{(-\sqrt{\alpha}/4-1/4)}
\rho^{(\sqrt{\alpha}/2-1/2)}[2\rho^2-(1+\sqrt{\alpha})Ml^2]
\nonumber\\
&-&2\pi\epsilon_{0}{l}^{(\sqrt{\alpha}/2-1/2)}
{(\rho^2-Ml^2)}^{(-\sqrt{\alpha}/4+1/4)}
\rho^{(\sqrt{\alpha}/2+1/2)},\nonumber\\
M(\rho,\epsilon_{0})&=&
-2\frac{\rho^2}{l^2}+(1+\sqrt{\alpha})M
-\frac{2\pi}{l}\epsilon_{0}\rho\sqrt{\rho^2-Ml^2}
\end{eqnarray}
The evaluation of energy and mass functions independent of
$\epsilon_{0}$ behave at infinity as
\begin{eqnarray}\label{enerBTZapproxEpsZ}
\epsilon({\rho\rightarrow\infty},\epsilon_{0}=0)&\approx&-\frac{1}{\pi\,l}
+\frac{l\,\sqrt{\alpha}M}{2\pi\,\rho^2},\nonumber\\
E({\rho\rightarrow\infty},\epsilon_{0}=0)&\approx&
\,l^{\frac{1}{2}(1+\sqrt{\alpha})}[-2\frac{\rho}{l^2}
+\frac{(1+\sqrt{\alpha})M}{2\rho}]
,\nonumber\\
M({\rho\rightarrow\infty},\epsilon_{0}=0)&\approx&
\,M(1+\sqrt{\alpha})-2\frac{\rho^2}{l^2}.
\end{eqnarray}
Using in the expressions (\ref{den-catgCP1})
and (\ref{den-catgCP2}) as the reference energy
density the quantity
$\epsilon_{0}=-\frac{1}{\pi\rho}\sqrt{-M_{0}
+\frac{\rho^2}{l^2}}$, which
at the spatial infinity behaves as
$\epsilon_{0\mid\infty}(M_{0})\approx
-\frac{1}{\pi\,l}+\frac{M_{0}}{2\pi\,\rho^2}$ the series expansions of
the corresponding quantities at $\rho\rightarrow$ infinity result in
\begin{eqnarray}
\epsilon({\rho\rightarrow\infty},
\epsilon_{0\mid\infty}(M_{0}))&\approx&\, \frac{l}{2\pi
\,\rho^2}(-M_{0}+\sqrt{\alpha}M),\nonumber\\
E({\rho\rightarrow\infty},\epsilon_{0\mid\infty}(M_{0}))&\approx&
\,\frac{(-M_{0}+\sqrt{\alpha}M)}{\rho}\,
l^{\frac{1}{2}(1+\sqrt{\alpha})},\nonumber\\
M({\rho\rightarrow\infty},
\epsilon_{0\mid\infty}(M_{0}))&\approx& -M_{0}+\sqrt{\alpha}M.
\end{eqnarray}
Therefore, comparing with the energy characteristics of the BTZ
solution, one concludes that the mass parameter at spatial infinity is
determined by the product $\sqrt{\alpha}M$, although the mass
function diverges at infinity as fast as $1/\rho^2$, a similar behavior is
exhibited by the energy density in that spatial region.\\

\subsection{Field, energy--momentum, and Cotton tensors}

The electromagnetic field of this solution is given by
\begin{eqnarray}
({F^\alpha}_{\beta})&&=
\frac{l\,M(1-\alpha)^{1/2}}{(\rho^2-Ml^2)^{1/2}\,\rho}\times\nonumber\\
&&\left[ \begin {array}{ccc} 0&0&
-\frac{1}{2}\,{l}^{\sqrt {\alpha}}
\left( {\rho}^{2}-M{l}^{2} \right) ^{1/2\,
\sqrt {\alpha}}{\rho}^{-\sqrt {
\alpha}}\\\noalign{\medskip}0&0&0
\\\noalign{\medskip}-\frac{1}{2}\,{l}^{-
\sqrt {\alpha}} \left( {\rho}^{2}-M{l}^{2} \right) ^{-1/2\,\sqrt {
\alpha}}{\rho}^{\sqrt {\alpha}}&0&0\end {array} \right],
\end{eqnarray}
and it is characterized by the following eigenvalues and eigenvectors
\begin{eqnarray}
&&\lambda_1=0;{\bf V}1=(0,V^{2},0),\,
V^{\mu}V_{\mu}=(V^{2})^2g_{\rho \rho},\,
{\bf V}1={\bf S}1,\nonumber\\
&&\lambda_2=-1/2\,{\frac {Ml\sqrt {1-\alpha}}{\sqrt {{\rho}^{2}-M{l}^{2}}\rho}};
{\bf V}2=[{l}^{\sqrt {\alpha}}
\left( {\rho}^{2}-M{l}^{2} \right) ^{-1/2\,\sqrt {
\alpha}}{\rho}^{\sqrt {\alpha}}{V^{3}}
,0,{V^{3}}],\nonumber\\&&V^{\mu}V_{\mu}=0,\,
{\bf V}2={\bf N}2,\nonumber\\&&
\lambda_3=1/2\,{\frac {Ml\sqrt {1-\alpha}}{\sqrt {{\rho}^{2}-M{l}^{2}}\rho}};
{\bf V}3=[-{l}^{\sqrt {\alpha}} \left( {\rho}^{2}-M{l}^{2} \right) ^{-1/2\,
\sqrt {
\alpha}}{\rho}^{\sqrt {\alpha}}{V^{3}}
,0,{V^{3}}],\nonumber\\&&V^{\mu}V_{\mu}=0,\,{\bf V}3={\bf N}3,\nonumber\\&&
{\rm Type}:\{S,N,N\}.
\end{eqnarray}

On the other hand, the energy--momentum tensor, having the structure
 \begin{eqnarray}
({T^\alpha}_{\beta}) =-\frac{1}{32} \,{\frac {{M}^{2}{l}^{2} \left(1- \alpha
 \right) }{{\rho}^{2} \left( {\rho}^{2}-M{l}^{2} \right) \pi }}\left[
\begin {array}{ccc} 1&0&0
\\\noalign{\medskip}0&-1&0
\\\noalign{\medskip}0&0&1
\end {array} \right],
\end{eqnarray}
allows for the eigenvalues $\lambda_1=\frac{1}{32}
\,{\frac {{M}^{2}{l}^{2} \left(1- \alpha
 \right) }{{\rho}^{2}
 \left( {\rho}^{2}-M{l}^{2} \right) \pi }}$
 and
 the other one, of multiplicity two,
 $\lambda_2=\lambda_3=-\frac{1}{32}
 \,{\frac {{M}^{2}{l}^{2} \left(1- \alpha
 \right) }{{\rho}^{2} \left( {\rho}^{2}-M{l}^{2} \right) \pi }}$
with the corresponding eigenvectors
\begin{eqnarray}
\lambda_1=\frac{1}{32} \,{\frac {{M}^{2}{l}^{2} \left(1- \alpha
 \right) }{{\rho}^{2} \left( {\rho}^{2}-M{l}^{2} \right) \pi }}
 ;{\bf V}1&&=(0,V^{2},0),\,V_{\mu}=V^{2}g_{\rho\rho}\delta^\rho_{\mu},
 \,V^{\mu}V_{\mu}=(V^{2})^2g_{\rho\rho},\,{\bf V}1={\bf S}1,\nonumber\\
\lambda_2=-\frac{1}{32} \,{\frac {{M}^{2}{l}^{2} \left(1- \alpha
 \right) }{{\rho}^{2} \left( {\rho}^{2}-M{l}^{2} \right) \pi }}
 ;{\bf V}2&&=(V^{1},0,V^{3}),\,
 V^{\mu}V_{\mu}=(V^{1})^2g_{t \,t}+(V^{3})^2g_{\phi \,\phi},\,\nonumber\\
{\bf V}2&&={\bf T}2, {\bf S}2, {\bf N}2,\nonumber\\
\lambda_3=-\frac{1}{32} \,{\frac {{M}^{2}{l}^{2} \left(1- \alpha
 \right) }{{\rho}^{2} \left( {\rho}^{2}-M{l}^{2} \right) \pi }};
 {\bf V}3&&=(\tilde V^{1},0,\tilde V^{3}),\,
 V^{\mu}V_{\mu}=({\tilde V}^{1})^2g_{t \,t}
 +({\tilde V}^{3})^2g_{\phi \,\phi},\,\nonumber\\
{\bf V}3&&={\bf T}3, {\bf S}3, {\bf N}3.\nonumber\\
\end{eqnarray}
For ${\bf V}2$ and ${\bf V}3$, the character of the vector depends
on the sing of its norm; for instance, by choosing
\begin{eqnarray*}
{ V}^{1}&&=s\,\sqrt{g_{\phi \,\phi}}/\sqrt{|g_{t \,t}|}\,{ V}^{3},
\,s=const.,\,{{\bf V}1}^{\mu}{{\bf V}1}_{\mu}=(1-s^2)g_{\phi \,\phi}\,(V^{3})^2;
\nonumber\\&&
s>1\rightarrow{{\bf V}1={\bf T}},\,
s=\pm 1\,\rightarrow{{\bf V}1={\bf N}},\,s<1\rightarrow{{\bf V}1={\bf S}}.
\end{eqnarray*}
Recall that in 3+1 gravity the eigenvectors of the
electromagnetic energy--momentum tensor (and at the
same time of the electromagnetic field tensor) are
null in pairs, i.e. they exhibit double coincidence.
Hence in the 2+1 case under study one may
think of the alignments $\{{\bf S},{2\bf N}\}$
or $\{{\bf N},{\bf S},{\bf N}\}$ as the corresponding
reductions of electromagnetic field eigen--directions of 3+1 gravity.

To complete the characterization of this solution, it is reasonable to add
some comments about the conformal Cotton tensor, which is given by
\begin{eqnarray}
 ({C^\alpha}_{\beta})=&&\frac{l^3\sqrt {\alpha}
 \left( \alpha-1 \right){M}^{3} }{8\left( {\rho}^{2}
 -M{l}^{2} \right)^{3/2}{\rho}^{3}}\nonumber\\&&\times
 \left[ \begin {array}{ccc} 0&0
&{l}^{-\sqrt {\alpha}}{\rho}^{\sqrt {\alpha}}
 \left( {\rho}^{2}-M{l}^{2} \right) ^{-\sqrt {\alpha}/2}
 \\\noalign{\medskip}0&0&0
\\\noalign{\medskip}-{l}^{\sqrt {\alpha}}
\left( {\rho}^{2}-M{l}^{2} \right) ^{\sqrt {\alpha}/2}{\rho}^{-\sqrt {\alpha}}
&0&0
\end {array} \right],
\end{eqnarray}
with eigenvectors
\begin{eqnarray}
\lambda_1=0;{\bf V}1&&=(0,V^{2},0),\,
V^{\mu}V_{\mu}=(V^{2})^2/L^2,\,{\bf V}1={\bf S},\nonumber\\
\lambda_2={-\frac {\,i\sqrt {\alpha}{l}^{3}
\left( 1-\alpha \right) {M}^{3}}{8
 \left( {\rho}^{2}-M{l}^{2} \right) ^{3/2}{\rho}^{3}}}
;\nonumber\\
{\bf V}2&&=[{V^{1}}=-i{\rho}^{\sqrt {\alpha}}
 \left( {\rho}^{2}-M{l}^{2} \right) ^{-\,\sqrt {\alpha}/2}\,{
l}^{-\sqrt {\alpha}}{V^{3}},0,{V^{3}}],\,{\bf V}2={\bf Z},\nonumber\\
\lambda_3={\frac {\,i\sqrt {\alpha}{l}^{3} \left( 1-\alpha \right) {M}^{3}}{8
 \left( {\rho}^{2}-M{l}^{2} \right) ^{3/2}{\rho}^{3}}}
;\nonumber\\
{\bf V}3&&=[{V^{1}}=i{\rho}^{\sqrt {\alpha}}
 \left( {\rho}^{2}-M{l}^{2} \right) ^{-\,\sqrt {\alpha}/2\,{
l}^{-\sqrt {\alpha}}}{V^{3}},0,{V^{3}}],\,{\bf V}3={\bf \bar Z},\nonumber\\
{\rm Type}:\{S,Z,\bar Z\}.
\end{eqnarray}

\section{Stationary generalization of the Cataldo static solution}

The structural functions corresponding to the stationary
generalization of the Cataldo static solution, via $SL(2,R)$
transformations, are given by
\begin{eqnarray}
ds^2&&=-N(\rho)^2dt^2+\frac{1}{L(\rho)^{2}}\,d\rho^2
+K(\rho)^2[d\phi+W(\rho)dt]^2,\nonumber\\
K(\rho)^2&=&\frac{\rho(\rho^2-Ml^2)^{1/2}}
{l\,{\Delta}}\left[\delta_{0}^2l^{\sqrt{\alpha
}}\rho^{\sqrt{\alpha}}
(\rho^2-Ml^2)^{-\sqrt{\alpha}/2}-
\beta_{0}^2l^{-\sqrt{\alpha}}\rho^{-\sqrt{\alpha}}
(\rho^2-Ml^2)^{\sqrt{\alpha}/2}\right],\nonumber\\
N(\rho)^2&=&
\frac{\rho^2}{l^2}\frac{\rho^2-Ml^2}{K(\rho)^2},
\,\,{\Delta}:={(\alpha_{0}\delta_{0}-\beta_{0}\gamma_{0})}\neq 0,
\nonumber\\
W(\rho)&=&-\frac{1}{\Delta}
\frac{\rho}{l}\frac{\sqrt{\rho^2-Ml^2}}{K(\rho)^2}
[\alpha_{0}\beta_{0}\,{l}^{-\sqrt{\alpha}}\rho^{-\sqrt{\alpha}}
(\rho^2-Ml^2)^{\sqrt{\alpha}/2}
-\gamma_{0}\delta_{0}\,l^{\sqrt{\alpha}}\rho^{\sqrt{\alpha}}
(\rho^2-Ml^2)^{-\sqrt{\alpha}/2}],
\nonumber\\
L(\rho)^2&=&\frac{\rho^2}{l^2}-M.
\end{eqnarray}
\subsection{Mass, energy and momentum for the
generalized Cataldo solution}\label{CataldoHIBmass}

The corresponding surface densities occur to be
\begin{eqnarray}\label{den-catgC1}
\epsilon(\rho,\epsilon_{0})&=&-
\frac{\left[\beta_{0}^2(1-\frac{Ml^2}{\rho^2})^{\frac{\sqrt{\alpha}}{2}}
(2-(1-\sqrt{\alpha})\frac{Ml^2}{\rho^2})
-\delta_{0}^2l^{2\sqrt{\alpha}}
(1-\frac{Ml^2}{\rho^2})^{-\frac{\sqrt{\alpha}}{2}}
(2-(1+\sqrt{\alpha})\frac{Ml^2}{\rho^2})\right]}{2\pi
l\sqrt{1-Ml^2/\rho^2}}\nonumber\\
&\times&
\left[\beta_{0}^2(1-\frac{Ml^2}{\rho^2})^{\frac{\sqrt{\alpha}}{2}}
-\delta_{0}^2\,l^{2\sqrt{\alpha}}
(1-\frac{Ml^2}{\rho^2})^{-\frac{\sqrt{\alpha}}{2}}
\right]^{-1}
-\epsilon_{0},\nonumber\\
j(\rho)&=&\beta_{0}\delta_{0}
\frac{\sqrt{\alpha}M}{\pi\Delta\,K(\rho)},\,
\end{eqnarray}
therefore the product $\beta_{0}\delta_{0}$ is
related with the rotation properties of the considered solution.
The integral quantities amount to
\begin{eqnarray}\label{den-catgC2}
&&J(\rho)=2\,\pi\,K(\rho)\,j(\rho)=
2\beta_{0}\delta_{0}\,\frac{\sqrt{\alpha}M}{\Delta},\nonumber\\
&&E(\rho)=2\,\pi\,K(\rho)\,\epsilon(\rho),\nonumber\\
&&M(\rho)=N(\rho)\,E(\rho)-W(\rho)\,J(\rho).
\end{eqnarray}

The evaluation of energy and mass functions independent of
$\epsilon_{0}$ behave at infinity as
\begin{eqnarray}
\epsilon({\rho\rightarrow\infty},\epsilon_{0}=0)&
\approx&-\frac{1}{\pi\,l}
+\frac{\sqrt{\alpha}l\,M}{2\pi\rho^2}\frac{\delta_{0}^2l^{2\sqrt{\alpha}}
+\beta_{0}^2}
{\delta_{0}^2l^{2\sqrt{\alpha}}-\beta_{0}^2},\nonumber\\
E({\rho\rightarrow\infty},\epsilon_{0}=0)&
\approx&-\,l^{1/2-\sqrt{\alpha}/2}\,
\frac{M}{2\rho\sqrt{\Delta}}
\frac{[(\sqrt{\alpha}-1)\beta_{0}^4+
2\beta_{0}^2\delta_{0}^2l^{2\sqrt{\alpha}}
-(\sqrt{\alpha}+1)\delta_{0}^4\,l^{4\sqrt{\alpha}}
]}{(\delta_{0}^2l^{2\sqrt{\alpha}}-\beta_{0}^2)^{3/2}}\nonumber\\
&&-2\rho\,l^{-3/2-\sqrt{\alpha}/2}\frac{\sqrt{\delta_{0}^2l^{2\sqrt{\alpha}}
-\beta_{0}^2}}{\sqrt{\Delta}},
\nonumber\\
M({\rho\rightarrow\infty},\epsilon_{0}=0)&=&-2\frac{\rho^2}{l^2}
+{M}\left[1+\sqrt{\alpha}\frac{\delta_{0}^2l^{2\sqrt{\alpha}}+\beta_{0}^2}
{\delta_{0}^2l^{2\sqrt{\alpha}}-\beta_{0}^2}+
2\sqrt{\alpha}\frac{\beta_{0}\delta_{0}}{\Delta}
\frac{\alpha_{0}\beta_{0}-\gamma_{0}\delta_{0}l^{2\sqrt{\alpha}}}
{\delta_{0}^2l^{2\sqrt{\alpha}}-\beta_{0}^2}\right].\nonumber\\
\end{eqnarray}
Using in the expressions (\ref{den-catgC1})--(\ref{den-catgC2})
as reference energy
density the quantity
$\epsilon_{0}=-\frac{\rho}{\pi\,l^2}/\sqrt{M_{0}+\frac{\rho^2}{l^2}}$, which
at the spatial infinity behaves as $\epsilon_{0\mid\infty}(M_{0})\approx
-\frac{1}{\pi\,l}+\frac{l\,M_{0}}{2\pi\,\rho^2}$, the series expansions of
the corresponding quantities at $\rho\rightarrow$ infinity result in
\begin{eqnarray}
\epsilon({\rho\rightarrow\infty},\epsilon_{0\mid\infty}(M_{0}))
&\approx&\frac{l}{2\pi
\,\rho^2}(\sqrt{\alpha}M-M_{0})-\frac{l\,\sqrt{\alpha}M}{\pi
\,\rho^2}\frac{\beta_{0}^2}{\delta_{0}^2l^{2\sqrt{\alpha}}
-\beta_{0}^2},\nonumber\\
E({\rho\rightarrow\infty},\epsilon_{0\mid\infty}(M_{0}))&\approx&\,
{l^{1/2-\sqrt{\alpha}/2}}\frac{\sqrt{\delta_{0}^2l^{2\sqrt{\alpha}}
-\beta_{0}^2}}{\rho\,\sqrt{\Delta}}
(\sqrt{\alpha}M-M_{0})+2l^{1/2-\sqrt{\alpha}/2}\frac{\beta_{0}^2\sqrt{\alpha}M}
{\rho\,\sqrt{\delta_{0}^2l^{2\sqrt{\alpha}}-\beta_{0}^2}},\nonumber\\
M ({\rho\rightarrow\infty},\epsilon_{0\mid\infty}(M_{0}))&\approx&
\sqrt{\alpha}M-M_{0}
-\frac{\sqrt{\alpha}M\beta_{0}}{\Delta}
\frac{\gamma_{0}\delta_{0}^2l^{2\sqrt{\alpha}}
-2\alpha_{0}\beta_{0}\delta_{0}+\beta_{0}^2\gamma_{0}}
{\delta_{0}^2l^{2\sqrt{\alpha}}-\beta_{0}^2}.
\end{eqnarray}
Therefore, comparing with the energy characteristics of the BTZ
solution, one concludes that role of the mass
parameter is played by the product $\sqrt{\alpha}M$.
At spatial infinity the mass function occurs to be
finite, the energy density and global energy approach to infinity
as fast as $1/\rho^2$ and $1/\rho$ correspondingly.

\section{Ayon--Cataldo--Garcia hybrid solution}

The metric
defining this kind of stationary electromagnetic
solution~\cite{AyonCG04}, see also~\cite{GarciaAnnals09},~Eq.(8.15), can be
given in the standard form as
\begin{eqnarray}\label{metriCAG2}
\bm{g}&=&-N^2\bm{d\,t}^2+\frac{1}{L^{2}}\,\bm{d\rho}^2
+K^2[\bm{d\phi}+W\bm{d\,t}]^2=-\frac{\rho^2\,f}{H}\bm{d\,t}^2
+\frac{\bm{d\rho}^2}{f}
+H(\bm{d\phi}+W\bm{d\,t})^2,\nonumber\\
f(\rho)&=&\frac{\rho^2}{l^2}-M+\frac{J^2}{4\rho^2},\nonumber\\
H(\rho)&=&\frac{1}{4K_{1}\sqrt{\alpha}S_{M}}
\sqrt{2\rho^2-lR_{-}}\sqrt{2\rho^2-lR_{+}}
\left[J^2K_{1}^2({2\rho^2-lR_{-}})^{-\sqrt{\alpha}/2}
({2\rho^2-lR_{+}})^{\sqrt{\alpha}/2}
\right.\nonumber
\\&&
\left.-({2\rho^2-lR_{-}})^{\sqrt{\alpha}/2}
({2\rho^2-lR_{+}})^{-\sqrt{\alpha}/2}\right],\nonumber\\
W(\rho)&=&-\frac{R_{-}}{J\,l}
\left[({2\rho^2-lR_{+}})^{\sqrt{\alpha}}(2\sqrt{\alpha}S_{M}+R_{-})R_{-}
-({2\rho^2-lR_{-}})^{\sqrt{\alpha}}J^2\right]
\times\nonumber\\
&&\left[({2\rho^2-lR_{+}})^{\sqrt{\alpha}}R_{-}^2
-({2\rho^2-lR_{-}})^{\sqrt{\alpha}}J^2\right]^{-1},
\nonumber\\
R_{\pm}:&=&M\,l\pm\sqrt{M^2l^2-J^2},
\,K_{1}:=-\frac{R_{-}}{J^2},\,S_{M}:=\sqrt{M^2\,l^2-J^2}.
\end{eqnarray}
The structural functions appearing in the definitions of the energy
and momentum quantities are expressed as
\begin{eqnarray}\label{metriCAG1}
N(\rho)&=&\sqrt{ \frac{\rho^2\,f(\rho)}{H(\rho)}},\,
L(\rho)=\sqrt{f(\rho)},\,K(\rho)=\sqrt{H(\rho)},\, W(\rho)=W(\rho).
\end{eqnarray}
The corresponding electromagnetic tensors are given as
\begin{eqnarray}
F_{\mu\,\nu}&=&-\frac{\sqrt{1-\alpha}}{\,l}\sqrt{M^2\,l^2-J^2}
{\delta_{[\mu}}^{t}{\delta_{\nu]}}^{\phi},\nonumber\\
8\,\,\pi\,{T_{\mu}}^{\nu}&=
&\frac{\alpha-1}{4\,l^2}\frac{M^2\,l^2-J^2}{\rho^2\,f(\rho)}
[{\delta_{\mu}}^{T}{\delta_{T}}^{\nu}-{\delta_{\mu}}^{\rho}{\delta_{\rho}}^{\nu}
+{\delta_{\mu}}^{\Phi}{\delta_{\Phi}}^{\nu}].
\end{eqnarray}
When the electromagnetic field is turned off, $\alpha=1$, the above
metric components reduce to
$$g_{TT}=M-\frac{\rho^2}{l^2}, g_{T\Phi}=-\frac{J}{2}, g_{\Phi\Phi}=\rho^2,
 g_{\rho\rho}=\left(\frac{\rho^2}{l^2}-M+\frac{J^2}{4\rho^2}\right)^{-1},$$
which correspond to the BTZ ones.

\subsection{ Mass, energy and momentum for the ACG solution}\label{ACataldoGHIB}

In terms of the structural metric functions the momentum quantities
they allow for very simple expressions
\begin{eqnarray}\label{MomAG}
j(\rho)&=&\frac{J}{2\pi}\frac{1}{\sqrt{H(\rho)}}, J(\rho)=J,
\end{eqnarray}
while the energy and mass characteristics become
\begin{eqnarray}\label{EndMas1}
\epsilon(\rho,\epsilon_{0})=-\frac{1}{2\pi}\frac{\sqrt{f(\rho)}}
{H(\rho)}\frac{d}{d\rho}H(\rho)-\epsilon_{0},
\end{eqnarray}
\begin{eqnarray}\label{EndMas2}
E(\rho,\epsilon_{0})=-\frac{\sqrt{f(\rho)}}{\sqrt{H(\rho)}}\frac{d}
{d\rho}H(\rho)-2\pi\epsilon_{0}\sqrt{H(\rho)},
\end{eqnarray}
\begin{eqnarray}\label{EndMas3}
M(\rho,\epsilon_{0})=-\rho\frac{{f(\rho)}}{H(\rho)}\frac{d}{d\rho}H(\rho)
-J\,W(\rho)-2\rho\pi\epsilon_{0}\sqrt{f(\rho)},
\end{eqnarray}
Because of the involved dependence of the metric functions upon the
$\rho$ coordinate, the evaluation of the energy quantities will de
done in the approximation of the spatial infinity.
 The momentum density at infinity becomes
\begin{eqnarray}\label{MomBTZapproxEpsZ1}
j({\rho\rightarrow\infty})\approx\frac{J}{2\pi}\frac{{\alpha}^{1/4}}{\rho},
\end{eqnarray}
while the global momentum remains constant in the whole space
\begin{eqnarray}\label{MomBTZZ1}
J(\rho)=J.
\end{eqnarray}
It becomes apparent then that the role of the momentum parameter
is played and coincides with $J$.

\noindent
The approximated at $\rho\rightarrow\infty$ surface energy
density, global energy and mass, for zero base energy density
$\epsilon_{0}$ are given by
\begin{eqnarray}
\epsilon({\rho\rightarrow\infty},\epsilon_{0}=0)&
\approx&-\frac{1}{\pi\,l}
+\frac{l\,M\sqrt{\alpha}}{2\pi\,\rho^2},\nonumber\\
E({\rho\rightarrow\infty},\epsilon_{0}=0)&
\approx&\,-2\frac{\rho}{l\,{\alpha}^{1/4}}
+\frac{l}{2\,\rho{\alpha}^{1/4}}(1+\sqrt{\alpha})M
,\nonumber\\
M({\rho\rightarrow\infty},\epsilon_{0}=0)&
\approx&\,-2\frac{\rho^2}{l^2}+2M
+\frac{\sqrt{\alpha}-1}{l}\sqrt{M^2\,l^2-J^2}.
\end{eqnarray}

Using in the expressions (\ref{EndMas1})--(\ref{EndMas3}) as reference energy
density the energy corresponding to the anti--de--Sitter metric,
$\epsilon_{0}=-\frac{1}{\pi\rho}\sqrt{\frac{\rho^2}{l^2}-M_{0}}$,
$\epsilon_{0\mid\infty}(M_{0})\approx
-\frac{1}{\pi\,l}+\frac{l\,M_{0}}{2\pi\,\rho^2}$, the series
expansions of the global energy and mass quantities at $\rho\rightarrow\infty$
result in
\begin{eqnarray}
\epsilon({\rho\rightarrow\infty},
\epsilon_{0\mid\infty}(M_{0})) &\approx&\,\frac{l}{2\pi
\,\rho^2}(\sqrt{\alpha}M-M_{0}),\nonumber\\
E({\rho\rightarrow\infty},\epsilon_{0\mid\infty}(M_{0}))&\approx&
\,l\,\frac{(\sqrt{\alpha}M-M_{0})}{\rho\,{\alpha}^{1/4}},
\nonumber\\
M({\rho\rightarrow\infty},\epsilon_{0\mid\infty}(M_{0}))&\approx&
M-M_{0}+\frac{\sqrt{\alpha}-1}{l}\sqrt{M^2\,l^2-J^2}.
\end{eqnarray}
Comparing with the energy characteristics of the BTZ
solution, the mass parameter occurs to be an involved
quantity depending on $M$, the momentum $J$, and the charge $\alpha$,
namely $M+\frac{\sqrt{\alpha}-1}{l}\sqrt{M^2\,l^2-J^2}$,
although the mass function is finite at spatial infinity.
 On the other hand, if $M\,l>>J$ then $\sqrt{\alpha}M$
 becomes the mass parameter. The energy density and
 global energy are proportional at infinity
 to $1/\rho^2$ and $1/\rho$ correspondingly.

\subsection{Field, energy--momentum, and Cotton tensors}

The coordinate system $\{t, r,\phi\}$ occurs to be more
adequate in the derivation of the eigenvalue--vector
characteristics of the considered
solution~\cite{GarciaAnnals09},~Eq.(8.11).
The metric in $\{t, r,\phi\}$--coordinates is given by
\begin{eqnarray}
g= \left[
\begin {array}{ccc} -{F }/{H }+H\,W^{2}&0&H W \\
\noalign{\medskip}0&  1/F&0\\
\noalign{\medskip}H W &0&H \end {array}
 \right] ,
\end{eqnarray}
with structural functions
\begin{eqnarray}
F(r)&&=4\,{\frac { \left( r-{\it r1} \right)  \left( r-{\it r2} \right) }{{l}
^{2}}},\nonumber\\
H(r)&&=\frac{l}{2}\,\frac{\sqrt{\left( r-{\it r1}\right)\left( r-{\it r2} \right)} }
 {{\it K1}\,\sqrt {\alpha} \left( {\it r2}-{\it r1}\right)}
 \left[  \left( {\frac {r-{\it r1}}{r-{\it r2}}} \right) ^{1/2\,\sqrt
{\alpha}}-{J}^{2}{{\it K1}}^{2} \left( {\frac {r-{\it r1}}{r-{\it r2}}
} \right) ^{-1/2\,\sqrt {\alpha}} \right]  ,\nonumber\\
W(r)&&={\it W0}-2\,{\it K1}\,J \left( {\frac {r-{\it r1}}{r-{\it r2}}}
 \right) ^{-1/2\,\sqrt {\alpha}}\frac{\sqrt { \left( r-{\it r1} \right)
 \left( r-{\it r2} \right) }}{l \,H(r) },\,\frac{d\,W}{dr}=-\frac{J}{H^2}.
 \nonumber\\
\end{eqnarray}
The electromagnetic field tensor is given by
\begin{eqnarray}
\left({F^\alpha}_{\beta}\right)&&=\left[
\begin {array}{ccc} -{ c\frac {H\,W}{F }}&0&-c{\frac {\,H }{F}}\\
\noalign{\medskip}0&0&0\\
\noalign{\medskip}{-c\frac {\left( F - H^{2} W^{2} \right) }{H F }}
&0&{c\frac {H W }{F }}\end {array} \right],\,c=
{\frac { \left( {\it r2}-{\it r1} \right)
\sqrt {1-\alpha}}{{l}^{2}}}.
\end{eqnarray}
In the search of its eigenvectors, one arrives at
\begin{eqnarray}
\lambda_{1}&&=0;{\bf V}1=[{V^{1}}=0,{V^{2}}={V^{2}},{V^{3}}=0],
 V_{\mu}V^\mu =\frac{{V^{2}}^2}{F},\,{\bf V}1={\bf S}1,\nonumber\\
\lambda_2&&=\frac {c}{\sqrt {F}};
\nonumber\\
{\bf V}2&&=[{V^{1}}={V^{1}},{V^{2}}=0,{V^{3}}
=-{\frac { \left( H W  +\sqrt {F } \right) {
V^1}}{H }}],\,V^{\mu}V_{\mu}=0,\,{\bf V}2
={\bf N}2,\nonumber\\
\lambda_{3}&&=-\frac {c}{\sqrt {F}};
\nonumber\\
{\bf V}3&&=[{V^{1}}={V^{1}},{V^{2}}=0,{V^{3}}
={\frac { \left(- H W  +\sqrt {F } \right) {
V^1}}{H }}],\,V^{\mu}V_{\mu}=0,\,{\bf V}3={\bf N}3,
\end{eqnarray}
hence this tensor is of the type
$$\{S,N1,N2\}.$$
For the energy--momentum tensor
\begin{eqnarray}
\left({T^\alpha}_{\beta}\right)=  \left[
\begin {array}{ccc} -\frac{1}{8\pi}\,{\frac {{c}^{2}}{F }}&0
&0\\\noalign{\medskip}0&\frac{1}{8\pi}\,{\frac {{c}^{2}}{F }
}&0\\\noalign{\medskip}0&0&-\frac{1}{8\pi}\,{\frac {{c}^{2}}{F
 }}\end {array} \right],
\end{eqnarray}
the eigenvectors are
\begin{eqnarray}
\lambda_1&&=\frac{1}{8\pi}\,{\frac {{c}^{2}}{F }};
{\bf V}1=[{V^{1}}=0,{V^{2}}={V^{2}},{V^{3}}=0],
V_{\mu}V^\mu =\frac {{V^{2}}^{2}}{ F },{\bf V}1={\bf S}1,\nonumber\\
\lambda_{2,3}&&=-\frac{1}{8\pi}\,{\frac {{c}^{2}}{F }};
{\bf V}2,3=[{V^{1}}={V^{1}},{V^{2}}=0,{V^{3}}={V^{3}}],
\,\nonumber\\&&
V_{\mu}V^\mu =-{\frac {{{V^{1}}}^{2}F}{H }}+
 \left( {V^{1}}\,W  +{V^{3}} \right) ^{2}H =
 {\frac {{{V^{1}}}^{2}F \left( Z^2-1 \right) }{H }},
 \nonumber\\
&&\,{\bf V}2=\{{\bf T}2,{\bf N}2,{\bf S}2\},
\,\,{\bf V}3=\{{\bf T}3,{\bf N}3,{\bf S}3\},\nonumber\\
\end{eqnarray}
hence it allows for the types:
$$\{S,2T\},\{S,2N\},\{S,2S\}.$$

The Cotton tensor
\begin{eqnarray}
\left({C^\alpha}_{\beta}\right)&&= \left[
\begin {array}{ccc} {{C^1}_{1}}&0&{{C^1}_{3} }\\\noalign{\medskip}0&
{0}&0\\\noalign{\medskip}{{C^3}_{1} }&0&-{{C^1}_{1}}\end {array}
 \right],\nonumber\\
{C^1}_{1}&&=-{C^3}_{3}=-\frac{{c}^{2} }{32\pi F^{2}}\,
\left( W\,{\mathcal{Q}}+2\,{\frac {F
 J}{H }}\right),\nonumber\\
{C^1}_{3}&&=-\frac{{c}^{2} }{32\pi F^{2}}\,{\mathcal{Q}},
\nonumber\\
{C^3}_{1}&&=-\frac{{c}^{2} }{32\pi F^{2}H^2}\left[ -W  H^{2}
 \left( W {\mathcal{Q}}+4\,{\frac {F
 J}{H }} \right) -F {\mathcal{Q}} \right],\nonumber\\
\mathcal{Q}:&&=2\,F\frac {d}{dr}\,H - H\frac {d}{dr}
\,F;\nonumber\\
\mathcal{Q}{}&&=-\frac{2}{l{\it K1}}\,
\sqrt { \left( r-{\it r1} \right)  \left( r-{\it r2} \right) }
 \left[ \left( {\frac {r-{\it r1}}{r-{\it r2}}} \right) ^{1/2\,\sqrt
{\alpha}}+{J}^{2}{{\it K1}}^{2} \left( {\frac {r-{\it r1}}{r-{\it r2}}
} \right) ^{-1/2\,\sqrt {\alpha}} \right].\nonumber\\
\end{eqnarray}
possesses the following set of eigenvectors
\begin{eqnarray}
\lambda_{1}&&=0;
{\bf V}1=[{V^{1}}=0,{V^{2}}={V^{2}},{V^{3}}=0],
\,V_{\mu}V^\mu ={V^{2}}^2/F,\,{\bf V}1={\bf S}1,
\nonumber\\
\lambda_{2}&&=\frac{1}{ 4}i\,
{\frac {\sqrt {\alpha}{c}^{2} \left( {\it r2}-{\it r1} \right)
}{\pi \, F ^{3/2}{l}^{2}}};\nonumber\\&&
{\bf V}2=[{V^{1}}=-{c}^{2}{V^{3}}\,{\mathcal{Q}} {\left( {c}^{2}W{\it
\mathcal{Q}}+2\,{\frac {{c}^{2}J\,F}{H }}
+32\,\lambda_{2}\,  F^{2}\pi  \right)^{-1}},{V^{2}}=0
,{V^{3}}={V^{3}}],{\bf V}2={\bf Z},
\nonumber\\
\lambda_{3}&&=-\frac{1}{4}i\,
{\frac {\sqrt {\alpha}{c}^{2} \left( {\it r2}-{\it r1} \right)
}{\pi \,  F^{3/2}{l}^{2}}};\nonumber\\&&
{\bf V}3=[{V^{1}}=-{c}^{2}{V^{3}}\,
{\mathcal{Q}} {\left( {c}^{2}W{\it
\mathcal{Q}}+2\,{\frac {{c}^{2}J\,F}{H }}
+32\,\lambda_{3}\,  F^{2}\pi  \right)^{-1}},{V^{2}}=0
,{V^{3}}={V^{3}}],{\bf V}3={\bf {\bar Z}}.\nonumber\\
\end{eqnarray}
therefore its type is
$$\{S, Z, {\bar Z}\}.$$

\section{Concluding remarks}

In the framework of the Einstein--Maxwell theory with negative cosmological
constant different families of exact solutions for cyclic symmetric
stationary (static) metrics have been studied,
evaluated their energy--momentum densities and
global energy--momentum-mass quantities using the Brown--York approach. As reference
characteristics there have been used the ones corresponding to the stationary or
static BTZ--AdS with parameter $M_{0}$--solutions. The electric
and magnetic solutions, and their generalizations
through $SL(2,R)$--transformations exhibit at the spatial infinity
$\rho\rightarrow \infty$ the following generic behavior
\begin{eqnarray*}
J(\rho\rightarrow\infty)&\approx&\alpha_{J}\,J+
\beta_{J}\,\ln{\rho},\nonumber\\
\epsilon(\rho\rightarrow\infty,
\epsilon_{0\mid\infty}(M_{0}))&
\approx&\,\frac{l\,}{2\pi\,\rho^2}(\alpha_{M}M-\alpha_{M_{0}}M_{0})
+\alpha_{Q}\frac{Q^2}{2\pi\,\rho^2}+\frac{a\,}{2\pi\,\rho}J
+\frac{A}{\pi\,\rho^2}\ln{\rho},\nonumber\\
E(\rho\rightarrow\infty,
\epsilon_{0\mid\infty}(M_{0}))&\approx&\,\frac{l}{\rho}(\beta_{M}M-\beta_{M_{0}}M_{0})
+\beta_{2}\frac{Q^2}{\rho}
+\frac{b\,}{\rho}J+2\frac{B}{\rho}\ln{\rho},\nonumber\\
M(\rho\rightarrow\infty,
\epsilon_{0\mid\infty}(M_{0}))&
\approx&\, \gamma_{M}M-\gamma_{M_{0}}M_{0}
+\gamma_{Q}Q^2 + c\,J+C\,\ln{\rho},
\end{eqnarray*}
where $\alpha_{J}$, $\beta_{M}$,...,
$\gamma_{Q}$ are constant numerical factors related to the
physical parameters: $J$ momentum, $M$ mass,..., $Q$ electromagnetic charge.

The momentum, energy and mass of the hybrid solutions behaves at spatial infinity
$\rho\rightarrow \infty$ as follows
\begin{eqnarray*}
J(\rho\rightarrow\infty)&\approx&\alpha_{J}\,J
+\beta_{J},\nonumber\\
\epsilon(\rho\rightarrow\infty,
\epsilon_{0\mid\infty}(M_{0}))&
\approx&\,\frac{l\,}{2\pi\,\rho^2}(\alpha_{M}M-\alpha_{M_{0}}M_{0})
+\alpha_{Q}\frac{Q^2}{2\pi\,\rho^2}+\frac{a\,}{2\pi\,\rho^2}J,
\nonumber\\
E(\rho\rightarrow\infty, \epsilon_{0\mid\infty}(M_{0}))&
\approx&\,\frac{l}{\rho}(\beta_{M}M-\beta_{M_{0}}M_{0})
+\beta_{Q}\frac{Q^2}{\rho}
+\frac{b\,}{\rho}J,\nonumber\\
M(\rho\rightarrow\infty,
\epsilon_{0\mid\infty}(M_{0}))&
\approx&\, \gamma_{M}M-\gamma_{M_{0}}M_{0}+\gamma_{Q}Q^2 + c\,J,
\end{eqnarray*}
where the charge $Q$ is related with the electromagnetic
parameter $\alpha$.

 Moreover, the eigenvectors
for their  electromagnetic field, energy--momentum and Cotton tensors have
been explicitly determined; the static and stationary
Peldan electric
classes, the Martinez--Teitelboim--Zanelli and  the
Clement solutions exhibit the following algebraic
types:$${\rm Field}:\{S,N,N\},\,{\rm Energy}
:\{S,2T\},\,\{S,2N\},\{S,2S\};\,{\rm Cotton}:\{S,Z,\bar Z\},$$
while the static and stationary Peldan magnetic
families, the Hirschmann--Welch and the
Dias--Lemos solutions exhibit the following
algebraic types:$${\rm Field}:\{T,Z,\bar Z\},\,
{\rm Energy}:\{T,2S\};\,{\rm Cotton}:\{S,Z,\bar Z\}.$$
The Garcia solution allows for the set of
types:$${\rm Field}:\{T,Z,\bar Z\},\,\{S,Z,\bar Z\},
\,\{T,2N\},\,\{S,2N\},\,\{3N\},$$ $${\rm Energy}:
\{T,2T\},\{T,2S\},\{T,2N\}.\{T,2N\},\{S,2T\},
\{S,2S\},\{S,2N\}.\{3N\},$$$$\,{\rm Cotton}:\{S,Z,\bar Z\}.$$
The hybrid Cataldo and Ayon--Cataldo--Garcia
solutions fall into the types:
$${\rm Field}:\{S,N,N\},{\rm Energy}:\{S,2T\},
\{S,2S\},\{S,2N\},\,{\rm Cotton}:\{S,Z,\bar Z\}.$$
The Kamata--Koikawa belongs to types
$${\rm Field}:\{3S\},{\rm Energy}:\{3S\},\{3N\},
\,{\rm Cotton}:\{3S\},\{3N\}.$$
Finally, the Matyjasek--Zaslavski solution
exhibits the types
$${\rm Field}:\{S,N,N\},{\rm Energy}
:\{S,2T\},\{S,2S\},\,\{S,2N\},\,{\rm Cotton}:\{0\}.$$ Recall that
algebraic structures $\{T,2S\}$ are thought of as perfect fluids.
though

\section{Acknowledgments}
This work has been partially supported by UC MEXUS--CONACyT for
Mexican and UC Faculty Fellowships 09012007, and Grant CONACyT 82443
and 178346.

\appendix
\section{Summary of the Brown--York approach to
${\Sigma}$--hyper\-surfaces and ${B}$--three--boundary}

Following the Brown--York
formulation~\cite{BrownY-prd93}, for a
$4$--dimensional spacetime with metric $g_{\mu\,\nu}$ a foliation
$\Sigma$ determines and is determined by a congruence of timelike
unit normal vectors $u_{\mu}$ proportional to the gradient of a
scala field $t$ labeling the hypersurfaces, i. e.,
$u_{\mu}=-N\,t_{,\mu}$, where $N$ is the lapse function. The spatial
metric tensor field $h_{\mu\nu}$ is defined on $\Sigma$ by
$h_{\mu\nu}=g_{\mu\nu}+u_{\mu}\,u_{\nu}$.The covariant derivative
$\nabla_{\mu}$ with respect to $g_{\mu\nu}$ induces on $\Sigma$ a
covariant derivative $D_{\mu}$ by means of the projection
$D_{\mu}=h^{\alpha}_{\mu}\,\nabla_{\alpha}$. For any spatial tensor
field $T_{\mu}^{\nu}$, $T_{\mu}^{\nu}\,u^{\mu}=0$, the spatial
covariant derivative is defined as
$D_{\mu}T_{\nu}^{\lambda}=h^{\alpha}_{\mu}\,
h^{\lambda}_{\gamma}\,h^{\beta}_{\nu}\,\nabla_{\alpha}T_{\beta}^{\gamma}$.
The extrinsic curvature of $\Sigma$ occurs to be
$K_{\mu\nu}=-h^{\alpha}_{\mu}\,\nabla_{\alpha}u_{\nu}=-D_{\mu}u_{\nu}$,
which is a symmetric spatial tensor because the normal timelike
vector $u^{\mu}$ is a gradient and consequently possesses vanishing
rotation, $\omega_{\mu\nu}=(u_{\mu;\nu}-u_{\nu;\mu})/2
+(u_{\mu;\alpha}u^\alpha\,u_{\nu}-u_{\nu;\alpha}u^\alpha\,u_{\mu})/2=0$,
which implies
$h^{\alpha}_{\mu}\nabla_{\alpha}u_{\nu}-h^{\alpha}_{\nu}\nabla_{\alpha}u_{\mu}=0.$

\noindent Following Brown--York paper, for convenience, coordinates
adapted to the foliation are introduced by choosing $t$ as the time
coordinate while $x^i, i=1,2,3,$ lie in the surface $\Sigma$,
consequently $\frac{\partial}{\partial\,x^i}$ are spacelike vectors.
Accordingly, the spacetime metric can be written as
$$ds^2=g_{\mu\nu}dx^\mu dx^\nu=(h_{\mu\nu}-u_{\mu}u_{\nu})dx^\mu dx^\nu\\
=-Ndt^2+h_{ij}(dx^i+V^idt)(dx^j+V^jdt),$$ where $N$ and
$V^i=h^i_{0}=-N\,u^i$ are correspondingly the shift function and the
shift vector. It follows that the spatial tensor
$h^{\mu\nu}=2h^{ij}\delta_{i}^{(\mu}\delta_{j}^{\nu)}$, where
$h^{ij}$ form a matrix inverse to the metric components $h_{ij}$ of
the hypersurface, $h^{is}h_{sj}=\delta^i_{j}$. Spatial vector fields
$T^\mu$ possess vanishing contravariant time components $T^0=0$,
therefore their space components are lowered and raised by means of
$h_{ij}$ and $h^{ij}$, $T_{i}=g_{i\mu}T^{\mu}=h_{ij}T^{j}$, and
$T^{i}=g^{i\mu}T_{\mu}=h^{ij}T_{j}$. In particular, the spacetime
tensors $D_{\mu}f$, $D_{\mu}T^\alpha$, for a spatial vector
$T^\alpha$, and $K_{\mu\nu}=-D_{\mu}u_\nu$ are spatial tensors, then
$D_{i}f$, $D_{i}T^j$, and $K_{ij}$ are tensors on $\Sigma$ with
indices raised and lowered by $h^{ij}$ and $h_{ij}$.

\noindent With respect to this coordinate decomposition, the space
components of the extrinsic curvature $K_{ij}$ occurs to be
\begin{equation}
K_{ij}=-\frac{1}{2\,N}\left[ \frac{\partial }{\partial
t}\,h_{ij}-2D_{(i}V_{j)}\right],
\end{equation}
hence the momentum $P^{ij}$ for the hypersurfaces $\Sigma$ is
defined as
\begin{equation}
P^{ij}=\frac{1}{2\,\kappa}\sqrt{\det(h_{ij})}\left[
K\,h^{ij}-K^{ij}\right],
\end{equation}
which is appropriate if the matter fields are minimally coupled to
gravity, i. e., those fields do not contain derivatives of
$g_{\mu\nu}$.

The extrinsic geometry of a three--boundary $\stackrel{3}{B}$ is
defined in a way similar to the one exhibited above for the
hypersurfaces $\Sigma$. Nevertheless, as pointed in
Ref.~\cite{BrownY-prd93}, the three--boundary is not
thought of as a member of a foliation of the whole spacetime $M$
since the extension of $\stackrel{3}{B}$ throughout all $M$ could be
forbidden by the topology of $M$. Denoting by $n^\mu$ the outward
pointing spacelike normal to the three--boundary $\stackrel{3}{B}$,
$n^\mu\,n_\mu=1$, the projection metric on $\stackrel{3}{B}$ is
given by $\gamma_{\mu\nu}=g_{\mu\nu}-n_\mu\,n_\nu$. The extrinsic
curvature is defined by
$\Theta_{\mu\nu}=-\gamma_{\mu}^{\lambda}\nabla_{\lambda}n_\nu=-\mathcal{D}_{\mu}n_{\nu}$,
where $\mathcal{D}_{\mu}=\gamma_{\mu}^{\lambda}\nabla_{\lambda}$
denote the induced covariant derivative for tensors that are tangent
to the three--boundary $\stackrel{3}{B}$.

\noindent Introducing intrinsic coordinates on $\stackrel{3}{B}$
adapted to the choice of the spacelike normal
$n_{\mu}=\mathcal{N}\delta^3_{\mu}$ by means of $x^i, i=0,1,2$, the
spacetime metric becomes
$$ds^2=(\gamma_{\mu\nu}+n_{\mu}n_{\nu})dx^\mu dx^\nu\\
=\mathcal{N}^2(dx^3)^2+\gamma_{ij}(dx^i+N^idx^3)(dx^j+N^jdx^3).$$
Likewise, the components of tensors tangent to the three--boundary
$\stackrel{3}{B}$ are raised and lowered by the intrinsic metric
tensors $\gamma^{ij}$ and $\gamma_{ij}$, in particular this hold for
the components of the intrinsic curvature $\Theta_{ij}$.

\noindent The definition of the boundary momentum amounts to
\begin{equation}
\pi^{ij}=-\frac{1}{2\,\kappa}\sqrt{-\det(\gamma_{ij})}\left[
\Theta\,\gamma^{ij}-\Theta^{ij}\right].
\end{equation}

{\it Two dimensional subspace $V_{2}$ immersed in a spacetime $M$}\\

According to the theory of $V_{n}$ embedded in a $V_{m}, m\geq n$,
in  the specific case of $n=2$ and
$m=4$ there exist two orthogonal vector fields, say $\xi_{1}$ and
$\xi_{2}$, normal to two three--dimensional manifolds $\Sigma_{1}$
and $\Sigma_{2}$. The set of points where these "hypersurfaces"
intersect gives rise to a 2-subspace having as outward normals the
vectors $\xi_{1}$ and $\xi_{2}$. This 2-subspace can be
characterized by its metric and extrinsic curvature.

From the Brown--York paper it follows that the intersections of the
families of hypersurfaces $\Sigma$ and the three--boundary
$\stackrel{3}{B}$, such that for their normals $u$ and $n$ the
orthogonality condition $(u\cdot n)|_{\stackrel{3}{B}}=0$ holds,
determine the two--boundaries $B$ on which the metric
$\sigma_{\mu\nu}$, tangent to both $u$ and $n$ fields at the
intersection, is defined by
\begin{equation}
\sigma_{\mu\nu}=g_{\mu\nu}+u_{\mu}u_{\nu}-n_{\mu}n_{\nu}=h_{\mu\nu}-n_{\mu}n_{\nu}
=\gamma_{\mu\nu}+u_{\mu}u_{\nu},
\end{equation}
fulfilling $\sigma_{\mu\nu}u^\mu=0=\sigma_{\mu\nu}n^\mu$; notice
that $\gamma_{\mu\nu}u^{\mu}=u_{\nu}$ and
$h_{\mu\nu}n^{\mu}=n_{\nu}$ because of the tangent property of the
vectors involved. The spacetime metric can be given as
$$ds^2=(\sigma_{\mu\nu}+n_{\mu}n_{\nu}
-u_{\mu}u_{\nu})dx^\mu dx^\nu=-(u_{\mu}dx^\mu)^2+
(n_{\nu}dx^\nu)^2+\sigma_{\mu\nu}dx^\mu dx^\nu.
$$
The metric element on $\stackrel{3}{B}$, constructed via the space
metric $\gamma_{ij}$, $i=0,1,2$ through the metric $\sigma_{ab}$,
$a,b=1,2$, of the two boundary $B$,  and the components of
$u_{\mu}=-N\delta^0_{\mu}$ can be given as
\begin{equation}
\gamma_{ij}dx^i\,dx^j=-N^2\,dt^2+\sigma_{ab}(dx^a
+V^adt)(dx^b+V^bdt).
\end{equation}
If we were using the decomposition of the spacetime metric with
respect to a normal spacelike vector
$n_{\mu}=\mathcal{N}\delta^3_{\mu}$ one would arrive at
$$g_{\mu\nu}=\gamma_{\mu\nu}+n_{\mu}n_{\nu}, g_{ij}
=\gamma_{ij}, g_{3j}=\gamma_{3j}=-N^i \gamma_{ij},
g_{33}=\mathcal{N}^2+N^iN^j\gamma_{ij},$$ where $i,j=0,1,2$. Hence
the four--dimensional metric would be written as
$$ds^2=(\gamma_{\mu\nu}+n_{\mu}n_{\nu})dx^\mu dx^\nu\\
=\mathcal{N}^2(dx^3)^2+\gamma_{ij}(dx^i-N^idx^3)(dx^j-N^jdx^3).$$
Restricting oneself to lie on the surface $x^3=\rm const.$, the
three--boundary $\stackrel{3}{B}$ metric reduces to
$$ds^2|_{\stackrel{3}{B}}=\gamma_{ij}dx^i\,dx^j.$$
This projection metric $\gamma_{ij}$, $i,j=0,1,2$, can be considered
as spanned by the two-dimensional metric $\sigma_{ab}$ of the
two--boundary $B$ and the vector field $u_{i}=-N\delta^0_{i}$
orthogonal to the hypersurface $\Sigma$. I consider illustrative to
give certain details about the derivation of the metric of
$\stackrel{3}{B}$ in terms of the tensor components $\sigma_{ab}$
and the components of $u^i$, $\gamma_{ij}=\sigma_{ij}-u_{i}u_{j}$,
$i,j=0,1,2$. Since $u_{i}=-N\delta^0_{i}$, $u^j=\gamma^{ji}u_{i}$,
$u_{j}=\gamma_{ji}u^{i}$, then $u^j=-N\gamma^{j0}$, and from the
unit condition $u^\mu\,u_{\mu}=-1$ one gets $\gamma^{00}=-1/N^2$,
consequently $u^0=1/N$ and $u^a=-N\gamma^{a0}=:N^a/N$, which yields
$N^a=-N\,u^a$. On the other hand, from the condition of
orthogonality $\sigma_{ij}u^j=0$ one has
$\sigma_{i0}u^0+\sigma_{ia}u^a=0$, the substitution of $u^0$ and
$u^a$ yields $\sigma_{i0}=\sigma_{ia}N^a$, or explicitly by
components $\sigma_{b0}=\sigma_{ab}N^a$,
$\sigma_{00}=\sigma_{b0}N^b=\sigma_{ab}N^a\,N^b$. Gathering these
results one has
\begin{eqnarray}
&{}&\gamma_{00}=\sigma_{00}-N^2=\sigma_{ab}N^a\,N^b-N^2,\nonumber\\
&{}&\gamma_{0b}=\sigma_{0b}=\sigma_{ab}N^a,\nonumber\\
&{}&\gamma_{ab}=\sigma_{ab},\nonumber\\
\end{eqnarray}
consequently
\begin{eqnarray}
\gamma_{ij}dx^i\,dx^j&=&
\gamma_{00}dt^2+2\gamma_{0a}dt\,dx^a+\gamma_{ab}\,dx^a\,dx^b\nonumber\\
&=&-N^2\,dt^2+\sigma_{ab}N^a\,N^b\,dt^2+\sigma_{ab}\,N^a\,dx^b\,dt+
\sigma_{ab}\,dx^a\,dx^b\nonumber\\
&=&-N^2\,dt^2+\sigma_{ab}(dx^a+N^a\,dt)(dx^b+N^b\,dt).\nonumber\\
\end{eqnarray}
The three--boundary unit normal $n^\mu$ is orthogonal to both the
three--boundary $\stackrel{3}{B}$ embedded in the spacetime $M$ and
the two--boundary $B$ as embedded in the hypersurface $\Sigma$. The
extrinsic curvature of the two--boundary $B$ as embedded in the
hypersurface $\Sigma$ is defined by
\begin{eqnarray}\label{intcurB}
k_{\mu\nu}=-\sigma^\alpha_{\mu}D_{\alpha}n_{\nu}
=-\sigma^\alpha_{\mu}h^\beta_{\alpha}h^\lambda_{\nu}\nabla_{\beta}n_{\lambda}
=-\sigma_{\mu\alpha}h^{\alpha\beta}n_{\lambda;\beta}h^\lambda_{\nu},
\end{eqnarray}
where
$D_{\alpha}n_{\nu}=h^{\beta}_{\alpha}n_{\lambda;\beta}h^\lambda_{\nu}$
is the covariant derivative on $\Sigma$ of $n_\alpha$ obtained by
projecting with the tensor $h^{\beta}_{\alpha}$ the spacetime
covariant derivative of the spacelike vector $n_\alpha$. The tensors
$\sigma_{\alpha\beta}$ and $k_{\mu\nu}$ are defined only on
$\stackrel{3}{B}$.

The relevance of this splitting of the spacetime $M$ into
hypersurfaces $\Sigma$, the three--boundary $\stackrel{3}{B}$, and
the two--boundary space $B$ in the Brown--York approach resides in
its use in the formulation of the quasilocal energy and conserved
charges of gravitational and matter fields minimally coupled to
gravity in a spatially bounded region given in detail in
Ref.~\cite{BrownY93}.

For the choice of the timelike field
$u_{\alpha}=-N\delta^{0}_{\alpha}=-\delta^{0}_{\alpha}/\sqrt{-g^{00}}$
normal to the hypersurface $\Sigma$ one establishes that the
contravariant components $u^{\alpha}=-N\,g^{0\alpha}$ are given in
terms of the metric components as
$u^{\alpha}=-g^{0\alpha}/\sqrt{-g^{00}}$, and the shift function
is $N=1/\sqrt{-g^{00}}$, while the shift vector components amount
to $N^{i}=-N\,u^{i}=-g^{i0}/g^{00}$. Hence the projection tensor
can be given as:
\begin{eqnarray}
h_{\alpha\,\beta}&=&\left[
\begin {array}{cc}
h_{a\,0}=g_{ab}N^{b}=-g_{ab}g^{b0}\frac{1}{g^{00}}&
h_{a\,b}=g_{a\,b}
\\\noalign{\medskip}
g_{ab}N^{a}N_{b}
=g_{ab}g^{a0}g^{b0}\frac{1}{(g^{00})^2}&h_{0\,b}=g_{bs}N^{s}
=-g_{ab}g^{a0}\frac{1}{g^{00}}
\\\noalign{\medskip}
\end {array} \right],\nonumber\\
h^{\alpha\,\beta}&=&\left[
\begin {array}{cc}0&
h^{ab}=g^{a\,b}-g^{a0}g^{b0}\frac{1}{(g^{00})^3}
\\\noalign{\medskip}
0&0
\\\noalign{\medskip}
\end {array} \right],\nonumber\\
h_{\alpha}^{\beta}&=&\left[
\begin {array}{cc} 0&
h_{a}^{b}=\delta_{a}^{b}
\\\noalign{\medskip}0&
h_{0}^{b}=N^{b}=-g^{b0}\frac{1}{g^{00}}
\\\noalign{\medskip}
\end {array} \right],
\end{eqnarray}
where Latin letters run $a,b,...,i=1,2,3$.

If one chose the spacelike normal $n_{\alpha}$ to the
three--boundary $\stackrel{3}{B}$ as pointing along the coordinate
$x^3$ then $n_{\alpha}=\mathcal{N}\delta^{3}_{\alpha}$. One would
have that the contravariant components
$n^{\alpha}=\mathcal{N}\,g^{3\alpha}$ are given in terms of the
metric components as $n^{\alpha}=g^{3\alpha}/\sqrt{g^{33}}$, and the
function is $\mathcal{N}=1/\sqrt{g^{33}}$, while the vector
components amount to
$\mathcal{N}^{A}=\mathcal{N}\,n^{A}=g^{A3}/g^{33}$. Moreover, the
following decomposition will hold:
$$g_{\mu\nu}=\gamma_{\mu\nu}+n_{\mu}n_{\nu}, g_{AB}
=\gamma_{AB}, g_{3B}=\gamma_{3B}=-\mathcal{N}^A\,g_{AB},
g_{33}=\mathcal{N}^2+\mathcal{N}^A\,\mathcal{N}^B\,g_{AB},$$ where
capital Latin letters run $A,B,...,J=0,1,2$ have been introduced to
avoid misunderstanding.

In the definition of the intrinsic curvature $k_{\mu\nu}$, Eq.
(\ref{intcurB}), of the two--surface $B$ enter the metric tensor
$\sigma{\mu\nu}$, the covariant derivative on $\Sigma$ and the
normal $n$ to $\stackrel{3}{B}$. The metric on $\stackrel{3}{B}$,
i.e., the spacetime metric on $x^3=\rm const.$ is given by
$$ds^2|_{\stackrel{3}{B}}=g_{AB}dx^A\,dx^B=\gamma_{AB}dx^A\,dx^B.$$
One can relate $\gamma_{AB}$ to  $\sigma_{\mu}{\nu}$ and the
timelike normal vector $u^\mu$ to the hypersurface $\Sigma$ by
defining $\sigma_{AB}=\gamma_{AB}+u_{A}u_{B}$. Since
$u_{A}=-N\,\delta_{A}^{0}$, $u^{\mu}=g^{\mu A}u_{A}=-N\,g^{\mu 0}$,
and the unit condition
$u_{\mu}u^{\mu}=-1\rightarrow{N^2g^{00}=-1}\rightarrow{N=1/\sqrt{-g^{00}}}$.
Moreover, due to $(u\cdot n)|_{\stackrel{3}{B}}=0$, therefore
$u^\mu\, n_{\mu}=0\rightarrow{u^3=0}$, hence
$u^\mu\rightarrow{u^A}$. Consequently $u^{B}=g^{BA}u_{A}=-g^{B
0}/\sqrt{-g^{00}}$. On the other hand $\sigma_{\mu\nu}u^\mu=0$, thus
$u^A=\gamma^{AB}u_{B}$, $u_A=\gamma_{AB}u^{B}$, in
particular,$u^A=-N\,\gamma^{A0}=-\gamma^{A0}/\sqrt{-g^{00}}$. The
unit condition yields $N^2\gamma^{00}=-1$, thus
$u^0=\sqrt{-g^{00}}$, and
$u^{\imath}=-N\gamma^{\imath\,0}=:-N^{\imath}/N$, where
$\imath,\jmath=1,2$. The condition $\sigma_{\mu\nu}u^\mu=0$ yields
$\sigma_{\mu0}u^0+\sigma_{\mu1}u^1+\sigma_{\mu2}u^2
=0\rightarrow{\sigma_{\mu0}u^0+\sigma_{\mu\imath}u^\imath}=0$, using
the expressions above of $u^0$ and $u^\imath$ one obtains
$\sigma_{\mu\,0}=\sigma_{\mu\,\imath}N^\imath$, or explicitly
$\sigma_{\jmath\,0}=\sigma_{\jmath\,\imath}N^\imath$,
$\sigma_{00}=\sigma_{0\,\imath}N^\imath=\sigma_{\jmath\,\imath}N^\imath\,N^\imath$.
Moreover, from
$\sigma^{\mu\nu}u_{\nu}=0\rightarrow{\sigma^{\mu\,0}=0=\sigma^{0}_{\mu}}$,
thus $\sigma^{i
j}=\gamma^{ij}+u^iu^j=\gamma^{ij}-g^{0i}g^{0j}/g^{00}$, and
$\gamma^{0\mu}=g^{0\mu}$. Therefore
$$\sigma_{\imath\,\jmath}=\gamma_{\imath\,\jmath},\,\sigma_{0\,\jmath}
=\gamma_{0\,\jmath}=\sigma_{\imath\,\jmath}N^\imath
=\gamma_{\imath\,\jmath}N^\imath,\,N^\imath=-g^{\imath\,0}/g^{0\,0},
$$
$$\sigma_{00}=\gamma_{00}+N^2
=\sigma_{\imath\,\jmath}N^\imath\,N^\jmath
=\gamma_{\imath\,\jmath}N^\imath\,N^\jmath,\,N^2=-1/g^{0\,0}.
$$
\begin{eqnarray}
k_{\nu}^{\mu}&=&-\sigma^{\mu\alpha}D_{\alpha}n_{\nu}=-\sigma^{\mu\alpha}h_{\alpha}^\beta
\,h^{\lambda}_{\nu}\nabla_{\beta}n_{\lambda}\nonumber\\
&=&-\sigma^{\mu\alpha}h_{\alpha}^0\,h^{\lambda}_{\nu}
\nabla_{0}n_{\lambda}-\sigma^{\mu\alpha}h_{\alpha}^i\,h^{\lambda}_{\nu}
\nabla_{i}n_{\lambda}=-\sigma^{\mu\alpha}h_{\alpha}^i\,h^{\lambda}_{\nu}
\nabla_{i}n_{\lambda}\nonumber\\
&=&-\sigma^{\mu\,0}h_{0}^i
h^{\lambda}_{\nu}\nabla_{i}n_{\lambda}-\sigma^{\mu\,j}h_{j}^i\,h^{\lambda}_{\nu}
\nabla_{i}n_{\lambda}=-\sigma^{\mu\,j}\,h^{\lambda}_{\nu}
\nabla_{j}n_{\lambda}
\end{eqnarray}
\begin{eqnarray}
k_{\nu}^{0}&=&0,\,k_{\nu}^{i}=-\sigma^{i\,j}
\,h^{\lambda}_{\nu}\nabla_{j}n_{\lambda}
\end{eqnarray}
It remains still to give the expression of $\sigma^{ij}$ in terms of
the metric components $g^{\mu\nu}$. Since $u_{\alpha}n^{\alpha}=0$
on $B^3$, because of the choice of coordinates along $u_{\alpha}$
and $n_{\alpha}$, one has $u^3=0$ and $n^0=0$, hence ${g^{03}=0}$.
Consequently, $\sigma^{ij}$ reduces to the components $i, j=1, 2$,
i.e., with $\imath, \jmath=1, 2$, namely
\begin{eqnarray}
\sigma^{\imath\,\jmath}=g^{\imath\,\jmath}-n^{\imath}n^{\jmath}
+u^{\imath}u^{\jmath}
=g^{\imath\,\jmath}-g^{\imath3}g^{\jmath3}/g^{33}-g^{\imath0}g^{\jmath0}/g^{00},
\end{eqnarray}
while $\sigma^{i3}=\sigma^{3i}=0, i=1, 2, 3.$

In this coordinate frame the intrinsic curvature of $B$ becomes
\begin{eqnarray}
k^{\imath}_{\nu}=-\sigma^{\imath\,\jmath}
\,h^{\lambda}_{\nu}\nabla_{\jmath}(\mathcal{N}\delta^3_{\lambda})
=-\sigma^{\imath\,\jmath}h^{\lambda}_{\nu}\left(\delta^3_{\lambda}
\,\frac{\partial\,\mathcal{N}}{\partial
x^\jmath}-\mathcal{N}{\Gamma^3}_{\jmath\,\lambda}\right).
\end{eqnarray}

\subsection{Energy and momentum surface densities, and spatial stress}

The energy surface density $\epsilon$, the momentum surface density
$j_{a}$, and the spatial stress $s^{ab}$ are defined as normal and
tangential projections of
$\tau^{ij}=2/\sqrt{-\gamma}(\pi^{ij}_{cl}-\pi^{ij}_{0})$ on the
two-surface $B$
\begin{eqnarray} \epsilon&=&
u_{i}u_{j}\tau^{ij}=-\frac{1}{\sqrt{\sigma}}\frac{\delta
S_{cl}}{\delta\,N},\nonumber\\
j_{a}&=&-\sigma_{ai}u_{j}\tau^{ij}=\frac{1}{\sqrt{\sigma}}\frac{\delta
S_{cl}}{\delta\,V^a},\nonumber\\
s^{ab}&=&\sigma^a_{i}\sigma^a_{j}\tau^{ij}=\frac{2}{\sqrt{-\gamma}}\frac{\delta
S_{cl}}{\delta\,\sigma_{ab}}.
 \end{eqnarray}
The subscript $cl$ stands for classical. The surface
stress--energy--momentum tensor $\tau^{ij}$ includes contributions
from both the gravitational and the matter fields. These quantities
are tensors with respect to the metric $\sigma_{ab}$ defined on the
two--surface $B$ and physically represent the corresponding
quantities associated with matter and gravitational fields on the
hypersurface $\Sigma$ with boundary $B$. In particular, the total
energy for $\Sigma$ is obtained by integrating $\epsilon$ on the
boundary $B$, $E=\int_{B}d^2x\sqrt{\sigma}\epsilon$. \underline{}

These tensors can be given as
\begin{eqnarray} \epsilon&=&
\frac{1}{\kappa}k|_{cl}+\frac{1}{\sqrt{\sigma}}\frac{\delta
S^{0}}{\delta\,N},\nonumber\\
j_{a}&=&-2\left(\frac{1}{\sqrt{h}}\,\sigma_{a\underline{i}}
n_{\underline{k}}P^{\underline{i}\,\underline{k}} \right)|_{cl}
-\frac{1}{\sqrt{\sigma}}\frac{\delta
S^{0}}{\delta\,V^a},\nonumber\\
s^{ab}&=&\frac{1}{\kappa}\left[k^{ab}+(n\cdot\, a-k)\sigma^{ab}
\right]|_{cl}-\frac{2}{\sqrt{-\gamma}}\frac{\delta
S^{0}}{\delta\,\sigma_{ab}},
\end{eqnarray}
$|_{cl}$ means evaluation for the classical solution, i.e.,
evaluation for a particular spacelike hypersurface $\Sigma$ in the
spacetime, indices that refer to coordinates on $B^3$ are denoted by
$i, j$, tensor indices that refer to coordinates on $\Sigma$ are
underlined $\underline{i}, \underline{k}$, indices that refer to
coordinates on $B$ are denoted by $a, b.$

There exist an ambiguity in the choice of the reference term $S^0$,
which is eliminated in the Brown--York paper by demanding that
$\epsilon$, and $j_{a}$ of a particular $\Sigma$ should depend only
on the canonical  variables $h_{ij}$, and $P^{ij}$ defined on
$\Sigma$. In particular, with this aim in mind, a possible choice is
$S^0$ as a linear functional of the lapse function $N$ and shift
vector $V^a$ on the two--boundary $B$
\begin{equation}
S^0=-\int_{B^3}d^3x\left(\sqrt{\sigma}N\,A+2\sqrt{\sigma}V^a\,B_{a}\right)
\end{equation}
where $A$ and $B_{a}$ are arbitrary functions of the two--metric
$\sigma_{ab}$

In Ref.~\cite{BrownY-prd93} a comment in extenso follows the choice of
$B_{a}=\left(\frac{1}{\sqrt{h}}\,
\sigma_{a\underline{i}}n_{\underline{k}}P^{\underline{i}\,\underline{k}}
\right)|_{0}$ and $A=k|_{0}$ by choosing a reference space--a fixed
spacelike slice of some fixed spacetime--and then consider a
two--surface in the slice whose induced two--metric is
$\sigma_{ab}$. If the two--surface exists then one evaluates the
specific functions $A$ and $B$ above yielding the sought functions
of $\sigma_{ab}$. For such choice of $A$ and $B$ one gets
\begin{eqnarray} \epsilon&=&
\frac{1}{\kappa}k|^{cl}_{0},\nonumber\\
j_{a}&=&-2\left(\frac{1}{\sqrt{h}}\,\sigma_{a\underline{i}}
n_{\underline{k}}P^{\underline{i}\,\underline{k}} \right)|^{cl}_{0}.
\end{eqnarray}
In particular, such functions $A$ and $B$ are uniquely
determined by the flat reference space, at least for all positive
curvature two--metrics with two--sphere topology. Moreover, for a
flat slice of flat spacetime $\left(\frac{1}{\sqrt{h}}\,
\sigma_{a\underline{i}}n_{\underline{k}}P^{\underline{i}\,\underline{k}}
\right)|_{0}=0$ because $P^{\underline{i}\,\underline{k}}=0$
identically.

The total quasilocal energy is defined over the two--surface B as
\begin{eqnarray}
E=\int_{B}d^2x\sqrt{\sigma}\epsilon.
\end{eqnarray}
For metric allowing the existence of symmetries associated with a
Killing field $\bf {\xi}$ on the boundary $\stackrel{3}{B}$ a
conserved charge can be defined as
\begin{eqnarray}
Q_{\xi}=\int_{B}d^2x\sqrt{\sigma}(\epsilon\,u^i+j^i)\,\xi_{i}.
\end{eqnarray}
If the Killing field $\xi$ is timelike, then the negative of the
corresponding charge defines a conserved mass $M:=-Q_{\xi_{\rm
timelike}}$. If the Killing field $\xi$ is a rotational symmetry on
$\stackrel{3}{B}$, then the corresponding conserved charge defines
the angular momentum $J:= Q_{\xi_{\rm rotational}}$; if the surface
$B$ contains the orbits of $\xi_{\rm rotational}$, then the angular
momentum can be determined by
\begin{eqnarray}
J=\int_{B}d^2x\sqrt{\sigma}j^i\,\xi_{i}.
\end{eqnarray}
As pointed out in Ref.~\cite{BrownCM-prd94}, the distinction between
mass $M$ and energy $E$ is relevant for spacetimes that are
asymptotically anti--de Sitter due to the divergent character at
spatial infinity of the magnitude of the timelike Killing vector in
that case; the timelike Killing vector does not approach the unit
normal to the stationary time slices at spatial infinity,
consequently $E$ and $M$ do not coincide.

The reduction of the above theory to (2+1)--dimensional spacetime is straightforward.

\section{Symmetries of the stationary and
static cyclic symmetric BTZ solutions}\label{appendixSym}

The study of the symmetries of the stationary and static cyclic
symmetric BTZ families and AdS classes of solutions starts  with
the stationary metric for the standard BTZ solution
\begin{eqnarray}\label{BTZsymm}
\bm{g}&=&-F(r)^2\,{d\,t}^2+\frac{{dr}^2}{F(r)^2}
+{r}^{2}\left[{d\phi}+V(r){d\,t}\right]^2,\nonumber\\
&&{F(r)}^{2}=\frac{r^2}{l^2}-M+\frac{J^2}{4\,r^2},\,V(r)=-\frac{J}{2\,r^2}.
\end{eqnarray}
The covariant components of the Killing  vectors $V_{\alpha}$ are denoted by
${\it v{\alpha}}$, namely
\begin{eqnarray}
V_{1}( t,r,\phi)={\it v1},\,V_{2}( t,r,\phi)={\it v2},
\,V_{3}( t,r,\phi)={\it v3} .
\end{eqnarray}
The Killing equations $V_{\mu;\nu}+V_{\nu;\mu}=0$ amount explicitly to
\begin{subequations}\label{BTZ_KEq}
\begin{eqnarray}\label{BTZ_KEq11}
EQ11={\frac {
\partial }{\partial t}}{\it v1}-{\frac { r{F(r)}^2 }{{l}^{2}}}{\it v2} ,
\end{eqnarray}
\begin{eqnarray}\label{BTZ_KEq12}
2EQ12={\frac {\partial }{\partial r}}{\it v1}  +
{\frac {\partial }{\partial t}}{\it v2}  -
\frac{J}{{l}^{2}{r}{F(r)}^2}\,{\it v3}
 -2\,\frac{r}{{l}^{2}{F(r)}^2}{\it v1},
\end{eqnarray}
\begin{eqnarray}\label{BTZ_KEq13}
2EQ13={\frac {\partial }{\partial \phi}}{\it v1}
+{\frac {\partial }{\partial t}}{\it v3} ,
\end{eqnarray}
\begin{eqnarray}\label{BTZ_KEq22}
EQ22=F(r){\frac {\partial }{\partial r}}{\it v2}
+{\it v2} \,{\frac {d }{d r}}F(r)
\rightarrow v2={\frac {{\it F_{1}} \left( t,\phi \right) }{2lF(r)}},
\end{eqnarray}
\begin{eqnarray}\label{BTZ_KEq23}
EQ23&&={\frac {\partial }{\partial \phi}}{\it v2}
+{\frac {\partial }{\partial r}}{\it v3}  +\frac{J}{{r}{F(r)}^2}\,{\it v1}
+2\,\frac { ( M{l}^{2}-{r}^{2} ) }{{r}{l}^{2}{F(r)}^2}{\it v3} ,
\end{eqnarray}
\begin{eqnarray}\label{BTZ_KEq33}
EQ33={\frac {\partial }{\partial \phi}}{\it v3}
 +r\,{F(r)}^2\, {\it v2},
\end{eqnarray}
\end{subequations}
where ${\it F_{i}}\left( t,\phi \right), i=1,2,3,$ are integration functions.\\
Isolating ${\it v1}$ from (\ref{BTZ_KEq23}) in terms
of ${\it v2}$ and ${\it v3}$ and their derivatives one gets
\begin{eqnarray}\label{KBTZV1}
r{l}^{2}J{\it v1} &&=2\,r \left( {r}^{2}-M{
l}^{2} \right) {\it v3}-r^2\,l^2\,F(r)^2\left({\frac {\partial }{\partial
\phi}}{\it v2} + \frac {\partial }{\partial r}{\it v3}\right).
\end{eqnarray}
Next, substituting ${\it v1}$ from above and the first integral
of ${\it v2}$ from (\ref{BTZ_KEq22}) into equation (\ref{BTZ_KEq12})
one arrives at a linear second
order equation for ${\it v3}$ with integrals
\begin{eqnarray}\label{KV3Fs}
{\it v3}=\frac{r^{2}}{2} {\it F_{2}} \left( t,\phi \right)+{\it F_{3}} \left( t,\phi
 \right) -\frac{1}{2}\frac{r\,l\,F(r)}{{M}^{2}{l}^{2}-{J}^{2}}
 \left( \,J\frac {\partial }{\partial t}{\it  F_{1}} \left( t,\phi \right)
+  M\,\frac {\partial }{\partial \phi}{\it F_{1}} \left( t,\phi\right) \right)
\end{eqnarray}
which substituted, together with ${\it v2}$
from (\ref{BTZ_KEq22}), into Eq. (\ref{KBTZV1}) for ${\it v1}$ gives
\begin{eqnarray}\label{KV1Fs}
{\it v1}&&=
2\,{\frac { \left( {r}^{2}-M{l}^{2} \right) }{J{l}^{2}}}{\it F_{3}} \left( t,\phi
 \right) -\frac{1}{4}\,J\,{\it F_{2}} \left( t,\phi \right) \nonumber\\
 &&
 +\frac{1}{2}\,
\frac {r\,l\,F(r)}{ \left(l^2\, M^2- J^2\right)  }\left(\frac{J}{{l}^{2}}\frac {
\partial }{\partial \phi}{\it F_{1}} \left( t,\phi \right)+M\,\frac {\partial }{\partial t}
{\it F_{1}} \left( t,\phi \right)\right).
\nonumber\\
 \end{eqnarray}
The dependence of the Killing vector components on the $r$ variable has been established;
it remains still
to determine their dependence on the $t$ and $\phi$ variables hiding in
the  ${\it F_{1}}\left( t,\phi \right), {\it F_{2}}\left( t,\phi \right)$
and ${\it F_{3}}\left( t,\phi \right)$ functions. Substituting the expressions
of ${\it v1}$ from (\ref{KV1Fs}), ${\it v3}$ from (\ref{KV3Fs}),
and ${\it v2}$ from (\ref{BTZ_KEq22}) one arrives at the independent equations
\begin{subequations}\label{BTZ_KEqF}
\begin{eqnarray}
{\it F_{1}} \left( t,\phi \right) {J}^{2}+{l}^{2}J{\frac {\partial ^{2}
}{\partial t\partial \phi}}{\it F_{1}} \left( t,\phi \right) +{l}^{2}
 \left( {\frac {\partial ^{2}}{\partial {\phi}^{2}}}{\it F_{1}} \left(
t,\phi \right)  \right) M-{\it F_{1}} \left( t,\phi \right) {M}^{2}{l}^{2}=0,
\end{eqnarray}
\begin{eqnarray}
{\it F_{1}} \left( t,\phi \right) {J}^{2}+{l}^{2}J{\frac {\partial ^{2}
}{\partial t\partial \phi}}{\it F_{1}} \left( t,\phi \right) +M \left(
{\frac {\partial ^{2}}{\partial {t}^{2}}}{\it F_{1}} \left( t,\phi
 \right)  \right) {l}^{4}-{\it F_{1}} \left( t,\phi \right) {M}^{2}{l}^{2}=0,
\end{eqnarray}
\end{subequations}
with integral
\begin{eqnarray}
{\it F_{1}} \left( t,\phi \right) ={\it C_{1}}\,
{{\rm e}^{{\frac {\sqrt {Ml-J} \left( l\phi+t
 \right) }{{l}^{3/2}}}}}+{\it C_{2}}\,{{\rm e}^{{\frac {\sqrt {Ml+J}
 \left( l\phi-t \right) }{{l}^{3/2}}}}}+{\it C_{3}}\,
 {{\rm e}^{{-\frac {\sqrt {Ml+J} \left( l\phi-t \right) }{{l}
^{3/2}}}}}+{\it C_{4}}\,{{\rm e}^{-{
\frac {\sqrt {Ml-J} \left( l\phi+t \right) }{{l}^{3/2}}}}}.
\end{eqnarray}
Furthermore,there have to be solved constraints on ${\it F_{2}} \left( t,\phi \right)$
and ${\it  F_{2}} \left( t,\phi \right)$, namely
\begin{eqnarray}
&&{\frac {\partial }{\partial t}}{\it F_{3}} \left( t,\phi \right) =0,
{\frac {\partial }{\partial \phi}}{\it F_{3}} \left( t,\phi \right) =0,
{\it F_{3}} \left( t,\phi \right) ={J\,\it C_{6}}=\rm const.,
\nonumber\\
&&{\frac {\partial }{\partial t}}{\it  F_{2}} \left( t,\phi \right) =0,
{\frac {\partial }{\partial \phi}}{\it  F_{2}} \left( t,\phi \right) =0,
{\it  F_{2}} \left( t,\phi \right) ={\it C_{5}}=\rm const.
\end{eqnarray}
where the integration constants are denoted  through ${\it C_{i}},\, i=1,...,6$.
\noindent
Finally, the covariant Killing vector components are
\begin{subequations}\label{BTZ_KVFin}
\begin{eqnarray}
V_{1}&&=
\frac{1}{2}\,\frac{r\,F(r)}{{l}^{3/2}\sqrt{ {M}^{2}{l}^{2}-{J}^{2}}}
\left(
{\it C_{1}}\,\sqrt {Ml+J}{\rm \exp}({\frac {\sqrt {Ml-J}  }{l^{3/2}}}( l\phi+t ))
\right.\nonumber\\&&\left.
-{\it C_{2}}\,\sqrt {Ml-J}{\rm \exp}({\frac {\sqrt {J+Ml}  }{l^{3/2}}}
( l\phi-t))
+{\it C_{3}}\,\sqrt {Ml-J}{\rm \exp}
(-\frac {\sqrt{J+Ml}}{l^{3/2}}( l\phi-t ))\right.\nonumber\\&&\left.
-{\it C_{4}}\,\sqrt {Ml+J}{\rm \exp}({-\frac {\sqrt {Ml-J} }{l^{3/2}}} ( l\phi+t ))
\right)
-\frac{J}{4}{\it C_{5}}-2\,{\frac { \left( M{l}^{2}-{r}^{2} \right) }{{
l}^{2}}}{\it C_{6}},
\end{eqnarray}
\begin{eqnarray}
V_{2}&&={\frac {
1}{2lF(r)}}
\left(
{\it C_{1}}\,{{\exp}({\frac {\sqrt {Ml-J}  }{{l}^{3/2}}}\left( l\phi
+t \right))}
+{\it C_{2}}\,{{\exp}({\frac {\sqrt {J+Ml}
  }{{l}^{3/2}}}\left( l\phi-t \right))}
 \right.\nonumber\\&&\left.
 +{\it C_{3}}{{\exp}(-{\frac {\sqrt {J+Ml} }{{l}^{3/2}}} \left( l\phi-t \right))}
 +{\it C_{4}}\,{{\exp}(-{\frac
{\sqrt {Ml-J}  }{{l}^{3/2}}}\left( l\phi+t \right))} \right),
\end{eqnarray}
\begin{eqnarray}
 V_{3}&&= -\frac{1}{2{l}^{1/2}}\frac{{r}\,F(r)}{\sqrt{M^2l^2-J^2}}\left(
 {\it C_{1}}\,\sqrt
{M l+J} {\exp({\frac
{\sqrt {Ml-J}  }{{l}^{3/2}}}}\left( l\phi+t \right))\right.\nonumber\\&&\left.
 +{\it C_{2}} \,\sqrt
{Ml-J}\exp({{\frac {
\sqrt {Ml+J}}{{l}^{3/2}}}}\left( l\phi-t \right) )
-{\it C_{3}}\,\sqrt {Ml-J}{\exp(-{\frac {  \sqrt {Ml+J}}{{l}^{3/2}}}}\left( l\phi-t
 \right))\right.\nonumber\\&&\left.
-{\it C_{4}}\,\sqrt {Ml+J}{\exp(-{\frac {\sqrt {Ml-J}  }{{l}^
{3/2}}}}\left( l\phi+t \right))\right)
+\frac{r^2}{2}\,{\it C_{5}}+J\,{\it C_{6}}.
\end{eqnarray}
\end{subequations}

These expressions allow one to determine the Killing
vector $\bf k_{i}$
associated to its corresponding integration constant ${\it C_{i}}$
for each of the possible ${i}$, by means of
$V_{\mu}={\sum}_{{i=1}}^6{k}_{i\mu}$ ,
where  $k_{i\mu}={\it C_{i}}{V}_{i\mu}$,
for each fixed value of $i$. The
contravariant vectors' components ${k_{i}}^{\mu}$,
 $\bm{\partial}_{k_{i}}=
{k_{i}}^{\mu}\frac{\partial}{{\partial{x^{\mu}}}}$,
are derived from the relationship  $V^{\mu}
=V_{\nu}g^{\nu\mu}={\sum}_{{i=1}}^6{k}_{i\nu}g^{\nu\mu}
= {\sum}_{{i=1}}^6{k_{i}}^{\mu}
={\sum}_{{i=1}}^6{\it C_{i}}{V}_{i}^{\mu}$.
Explicitly these Killing vectors are reported
in the main text, Section \ref{SymmBTZa}.

\subsection{\label{sec:symmDeS}Symmetries of
the anti--de Sitter space with positive $M$, $M>0$}

By a straightforward integration of the Killing equations
for the anti--de Sitter metric
\begin{eqnarray}\label{BrownCAGASD}
{g}&=&-F(r)^2\,{d\,t}^2+\frac{{dr}^2}{F(r)^2},
+{r}^{2}{d\phi}^2,\,{F(r)}^2=\frac{r^2}{l^2}-M,
\end{eqnarray}
or as a limiting transition of the previously derived Killing
vectors for the stationary BTZ by setting
$J\rightarrow 0,\,{\it C_{a}}
\rightarrow\,2{\it c_{a}},\,{\it C_{6}}\rightarrow{\it c_{6}}/2$ one arrives at
the covariant components of the Killing vectors;
$V= V_{\mu}dx^{\mu}=C_{a} V_{a\mu}dx^{\mu}$--one vector for
each constant--${\bm{V}}={\it c_{a}} \bm{ V}_{a}$:
\begin{eqnarray}
{V_{1}}
&&=r\frac{\sqrt {{r}^{2}-M{l}^{2}
}}{{l}^{3}\sqrt {M}} \left({{\rm e}^{{\frac {\sqrt {M} \left( l\phi+t
 \right) }{l}}}}{\it c_{1}}-{{\rm e}^{{\frac {\sqrt {M} \left( l\phi-t
 \right) }{l}}}}{\it c_{2}}+ {{\rm e}^{-{\frac {\sqrt {M} \left( l\phi-t
 \right) }{l}}}}{\it c_{3}}-{{\rm e}^{-{\frac {\sqrt {M} \left( l\phi+t
 \right) }{l}}}}{\it c_{4}}\right) \nonumber\\&&
 +{\it c_{6}}\,\frac{{r}
^{2}-M{l}^{2}}{l^2},
\end{eqnarray}
\begin{eqnarray}
{V_{2}}
&&=\frac{1}{\sqrt {{r}^{2}-M{l}^{2}
}}\left({{\rm e}^{{\frac {\sqrt {M} \left( \phi\,l+t \right) }{l}}}}{\it c_{1}}
+{{\rm e}^{{\frac {\sqrt {M} \left( \phi\,l-t \right) }{l}}}}{\it c_{2}}
+{{\rm e}^{-{\frac {\sqrt {M} \left( \phi\,l-t \right) }{l}}}}{\it c_{3}}
+{{\rm e}^{-{\frac {\sqrt {M}
\left( \phi\,l+t \right) }{l}}}}{\it c_{4}}\right),\nonumber\\&&
\end{eqnarray}
\begin{eqnarray}
{V_{3}}
&&=-r\,\frac{\sqrt {{r}^{2}-M{l}^{2}}}{\sqrt {M}{l}^{2} }
\left( {{\rm e}^{{\frac {\sqrt {M} \left(
\phi\,l+t \right) }{l}}}}{\it c_{1}}+{{\rm e}^{{\frac {\sqrt {M}
 \left( \phi\,l-t \right) }{l}}}}{\it c_{2}}-{{\rm e}^{-{\frac {\sqrt {
M} \left( \phi\,l-t \right) }{l}}}}{\it c_{3}}-{{\rm e}^{-{\frac {
\sqrt {M} \left( \phi\,l+t \right) }{l}}}}{\it c_{4}} \right)\nonumber\\&&
 +{r}^{2}{\it c_{5}}.
\end{eqnarray}
From the view point of the group properties more important
are the contravariant ${V_{i}}^{\mu}$ components
of the Killing vectors $\bm{\partial}_{V}=
C_{a}\bm{\partial}_{k_{a}}=C_{a}{k_{a}}^{\mu}\bm{\partial}_{\mu}$.

\subsection{\label{sec:symmBTZ}Symmetries of the
anti--de Sitter metric for negative $M$, $M=-\alpha^2$}

If negative $M$, one equates it to $-\alpha^2$. Moreover,
instead of complex exponential function, it will
be better to use trigonometric sine and cosine
functions. Thus, one can give the  Killing
vector components as $V= V_{\mu}dx^{\mu}=C_{a} V_{a\mu}dx^{\mu}$
\begin{eqnarray}
V_{1}&&=\frac{r\sqrt {\alpha^{2}\,l^{2}+{r}^{2}}}{\alpha\,l^3}\,
\left(
 -{\it C_{1}}\,\sin \left( \alpha\,\phi\right)
 \cos \left( \frac {\alpha\,t}{l} \right)
 +{\it C_{2}}\,\sin \left( \alpha\,\phi \right)
 \sin \left( \frac {\alpha\,t}{l} \right) \right.\nonumber\\&&\left.
-{\it C_{3}}\,\cos \left( \alpha\,\phi \right)
\cos\left( \frac {\alpha\,t}{l} \right)
 +{\it C_{4}}\,\cos \left( \alpha\,\phi \right)
 \sin \left( \frac{\alpha\,t}{l} \right)
\right)
+{\it C_{5}} (\alpha^2{l}^{2}+{r}^{2}),
\end{eqnarray}
\begin{eqnarray}
V_{2}&&=\sqrt {{\alpha}^{2}{l}^{2}+
{r}^{2}}\left(
 {\it C_{1}}\,\sin \left( \alpha\,\phi \right)
 \sin \left( {\frac {\alpha\,t}{l}} \right)
 + {\it C_{2}}\,\sin \left( \alpha\,\phi \right)
 \cos \left( {\frac {\alpha\,t}{l}} \right)
 \right.\nonumber\\&&\left.
+{\it C_{3}}\,\cos \left( \alpha
\,\phi \right) \sin \left( {\frac {\alpha\,t}{l}} \right)
+ {\it C_{4}}\,\cos \left( \alpha\,\phi \right)\cos \left( {\frac {
\alpha\,t}{l}} \right)\right),
\end{eqnarray}
\begin{eqnarray}
V_{3}&&=\frac{r\,\sqrt {{\alpha}^{2}{l}^{2}+{r}^{2}}}{\alpha\,l^2}
\left( {\it C_{1}}\,\cos \left( \alpha\,\phi\right)
\sin \left( {\frac {\alpha\,t}{l}} \right)
 +{\it C_{2}}\,\cos \left( \alpha\,\phi \right)
 \cos \left( {\frac {\alpha\,t}{l}} \right)
 \right.\nonumber\\&&\left.
 -{\it C_{3}}\,\sin \left( \alpha\,\phi \right)
 \sin\left( {\frac {\alpha\,t}{l}} \right)
  -{\it C_{4}}\,\sin \left( \alpha\,\phi \right)
  \cos \left( {\frac {\alpha\,t}{l}} \right) \right)
  +{\it C_{6}}\,r^2,
\end{eqnarray}
correspondingly, the contravariant components
are given in the main text, Section \ref{SymmBTZaneg}.

This anti--de Sitter metric,
(cosmological constant negative--$\lambda=-1/l^2$),
for the coordinates
$\{t,\rho,\phi\}$--merely names--ranging
$-\infty\leq t\leq\infty$, $-\infty\leq \rho\leq\infty$;
$-\infty\leq \phi\leq\infty$ allows for six symmetries,
i.e., six Killing vectors. For these ranges
of determination of the coordinates, the space is maximally symmetric.\\
The same situation takes place if the spatial coordinates are restricted to range\\
$$ 0\leq \rho\leq\infty,\,0\leq \phi\leq\,2\pi, $$
and $\alpha$ is set equal to unity, $\alpha=1=-M$, then
in such case $\rho$ and $\phi$
become polar coordinates with $\phi$ being the angular
coordinate with  period  $2\pi$. This spacetime--the (proper)
anti--de Sitter space (with $M=-1)$--allows for six symmetries,
and as such it is maximally symmetric.

\end{document}